\documentclass[%
reprint,
amsmath,amssymb,
aps,
]{revtex4-2}
\usepackage[bookmarks=true,colorlinks=true,citecolor=blue,linkcolor=blue,urlcolor=magenta]{hyperref}

\usepackage{amssymb}
\usepackage{latexsym}
\usepackage{gensymb} 
\usepackage{bbm}
\usepackage{bm}

\usepackage{xfrac}
\usepackage{calc}
\usepackage[footnotesize]{subfigure}

\usepackage{color}
\usepackage{graphicx}
\usepackage{dcolumn}
\usepackage{multirow}%
\usepackage{rotating}

\usepackage{longtable}
\usepackage{dashrule}

\usepackage{float}
\usepackage{perpage}
\usepackage{morefloats}

\usepackage{comment} 

\DeclareFontFamily{OT1}{pzc}{}
\DeclareFontShape{OT1}{pzc}{m}{it}{<-> s * [0.900] pzcmi7t}{}
\DeclareMathAlphabet{\mathpzc}{OT1}{pzc}{m}{it}

\newcommand\scalemath[2]{\scalebox{#1}{\mbox{\ensuremath{\displaystyle #2}}}}

\begin{document}
\preprint{PRD}
\title{Strings of diquark-quark $(QQ)Q$ baryon\\ before phase transition}
\author{A. S. Bakry}
\email[]{ahmed.bakry@mail.com}
\affiliation{Institute of Modern Physics, Chinese Academy of Sciences, Gansu 730000, China}

\author{M. A. Deliyergiyev}
\email[]{maksym.deliyergiyev@outlook.com}
\affiliation{Department of Nuclear and Particle Physics, University of Geneva, CH-1211, Switzerland}
\affiliation{High Performance Computing Center Stuttgart (HLRS), Universit\"{a}t Stuttgart, 70550 Stuttgart, Germany}

\author{A. A. Galal}
\affiliation{Department Scientific Research, Egyptian Meteorological Authority (E.M.A.), Cairo 11784, Egypt}

\author{M. N.\ Khalil}
\affiliation{Department of Mathematics, Bergische Universit\"at Wuppertal, 42097 Germany}
\affiliation{Department of Physics, University of Ferrara, Ferrara 44121, Italy}
\affiliation{Computation-based science Research center, Cyprus Institute, Nicosia 2121, Cyprus}
\date{\today}
	
\begin{abstract}
	We explore the limit at which the effective baryonic Y-string model of the junction approaches the mesonic stringlike behavior. We calculate and compare the numerical values of the static potential and energy-density correlators of diquark-quark and quark-antiquark configurations. The gauge model is pure Yang-Mills $SU(3)$ lattice gauge theory at coupling $\beta=6.0$ and finite temperature. The diquark setup is approximated as two quarks confined within a sphere of radius $0.1$ fm. The lattice data of the potential and energy show that the string binding the diquark-quark configuration displays an identical behavior to the quark-antiquark confining string. However, with the temperature increase to a small enough neighborhood of the critical point $T_{c}$, the gluonic similarities between the two systems do not manifest neither at short nor intermediate distance scales $R<1.0$ fm. The comparison between the potential and the second moment of the action-density correlators for both systems shows significant splitting. This suggests that subsisted baryonic decoupled states overlap with the mesonic spectrum. The baryonic junction's model for the potential and the profile returns a good fit to the numerical lattice data of the diquark-quark arrangement. However, near the critical point, the mesonic string displays large deviations compared to fits of the corresponding quark-antiquark data.	
\end{abstract}
\maketitle

\section{Introduction}
\label{sec:intro}
Baryons are viewed as the bound states of the three quarks in the conventional quark model, with two relative coordinates as the expressions of the internal space-time degrees of freedom. On the other hand, in the  diquark-quark $(QQ)Q$ model, two of the three quarks are bound together to create a boson (diquark) system, with the third quark assumed to revolve around this boson.

The concept of the diquark is long-established as the quark model itself. Shortly afterward Gell-Mann introduced the notion of the diquark ~\cite{Gell-Mann:1964ewy}, the constituent quark-diquark models for baryons were constructed by Ida-Kobayashi ~\cite{Ida:1966ev} and Lichtenberg et al. ~\cite{Lichtenberg:1967zz, PhysRev.167.1535, Carroll:1969ty} who also explored the electromagnetic characteristics of the baryons within the model's framework.

The diquark model was provoked later to describe several strong interaction phenomena ~\cite{Ono:1972ba, Anisovich:1974fr, Anisovich:1975zd, Schmidt:1977hs, Anselmino:1989vt, Anselmino:1992vg, Goeke:1993zp, Anselmino:1994uj}. In the recent past, there has been reviving interest in the characteristics of diquarks in hadronic systems, since they may debut a key role in the formation of exotic states.

The pertinence between strings and gauge field theory~\cite{HASHIMOTO} of hadrons is a longstanding conjecture~\cite{Nambu:1974zg, Nambu1979372}. The Nambu-Nielsen-Susskind-Goto string~\cite{Goto:1971ce} with point masses at its ends is produced by the interaction of two oppositely (magnetically) charged monopoles with an infinitely massive gauge field~\cite{Balachandran:1975qd}. Another finding is that a convenient limit of non-Abelian gauge theory's would be in agreement with the dual string model ~\cite{tHooft:1973alw}.

It may thus be presumed that a meson can be approximated as two point-quarks bound together by a string, and a baryon as three point-quarks joined up by three strings. Representing the point quarks as Dirac fields constrained to world lines allows for the assignment of appropriate internal quantum numbers \cite{Bars:1975dd, Giles:1975gy}. 
The startling outcome of this approach is that $1+1$ dimensional string model of the gluonic infrared aspects of QCD valid in some approximation~\cite{Callan:1975ps}.

In addition to studies that have investigated the mass of diquarks ~\cite{Hess98,Babich05,Orginos05,Alexandrou:2006cq,Liu06}, and, more recently, the nature of diquark correlations~\cite{Babich07}, now, we have theoretical developments related to hadronic string models in both the mesonic and baryonic configurations ~\cite{ANDREO1976521, Andreev:2015riv,  Solovev:1999ua, Jahn2004700, deForcrand:2005vv, PhysRevD.79.025022}.

In $SU(3)$ color group, a diquark, two quarks in close vicinity to each other, transform according to the conjugate representation $[\bar{3}]$. The scalar-diquark channel is attractive in the spin-singlet rather than the spin-triplet of the axial vector diquark. Hence low-lying diquarks lie within the conjugate representation $[\bar{3}]$ and acquire (+) parity and belong to the color $[\bar{3}]$, and thereof, share common properties to the antiquark.

Owing to the expected formation of flux-tubes of the same energy density and transverse size, the long-range confining force of the $(QQ)Q$ put forth to the same linearly rising potential that of the $Q\bar{Q}$ quark-antiquark~\cite{Bicudo:2014wka, Cardoso:2012aj, Giataganas:2014mla}. The target of this investigation is to scrutinize this interesting conjecture in lattice QCD at various temperature scales~\cite{Bakry:2020flt,Bakry:2020ebo,Bakry:2019cuw,Bakry:2018kpn,Bakry:2017jna,Bakry:2017fii,Bakry:2016aod, Bakry:2016uwt,Bakry:2014gea,Bakry:2014ina,Bakry:2012eq,Bakry:2011vnk,Bakry:2011cn,Bakry:2011kga,Bakry:2010sp, Bakry:2010zt}.

In a previous report, we show the formation of hadronic Y-string systems in static baryon on the lattice at finite temperatures $T$. However,  the properties of the diquark-quark $(QQ)Q$ configuration  and its relevance to the quark-antiquark $Q\bar{Q}$ system yet remains to be fully addressed on the lattice and in particular under extreme conditions of high temperature~\cite{Megias:2018haz, Chen:2021bkc}.

In this work, we extend our investigation of three-quark systems to the scenario in which two quarks interact to create a closely bound state, with a small triangular base length $A=0.2$ fm or a diquark ~\cite{Bakry:2014gea, Bakry:2016uwt, Bakry:2016aod} which then engages in interaction with a third quark to create a baryon.

The paper is organized as follows: We review the free bosonic string theory for the baryon and meson in Sec.~\ref{sec:EffectiveBosonicStringModel}. The simulations setup and lattice measurement operators are described in Sec.~\ref{sec:Operators}. In Sec.~\ref{sec:QQQPotential} we represent and compare the measured numerical data of the potential of the $Q\bar{Q}$ and the $(QQ)Q$ systems. The string model implications for each of these systems are discussed and compared with the numerical data. We display the action density analysis of both the $Q\bar{Q}$ and the $(QQ)Q$ systems in Sec.~\ref{sec:AcDensity_onLattice}. The broadening aspects of these Monte-Carlo data versus the baryonic and mesonic string models are scrutinized at several transverse planes. In the last section, Sec.~\ref{sec:Conclusions} concluding remarks are drawn.  

\section{Hadronic string phenomenology}
\label{sec:EffectiveBosonicStringModel}
It has long been hypothesized that in a pure Yang-Mills (YM) vacuum, a stable stringlike structure can develop, which binds static color charges and results in linearly growing  $Q\bar{Q}$ potential~\cite{Brambilla:2009cd, Brambilla:1995px, Bazavov100}. For example, through the dual Meissner effect, the QCD vacuum confines the color fields into a string that is dual to the Abrikosov line in the dual superconductor model of the QCD vacuum~\cite{Parisi:1974yh, Mandelstam:1974pi}. Further, it was suggested to employ an idealized system of bosonic strings to describe flux tubes transmitting the strong interaction between the color sources~\cite{Nambu1979372, Luscherfr}.

The formation of stringlike topological defects is not exceptional to QCD~\cite{Creutz:1980zw, Creutz:1983ev, Creutz:1980hb, Creutz:1980vq, Fukugita:1983du, Flower:1985gs, Wosiek:1987kx, Sommer:1987uz, DiGiacomo:1989yp, DiGiacomo:1990hc, Bali:1994de, Haymaker:1994fm, Cea:1995zt, Okiharu:2003vt, HASHIMOTO}, and arises in numerous strongly interacting systems \cite{PhysRevB.36.3583,PhysRevB.78.024510,2007arXiv0709.1042K,Nielsen197345,Lo:2005xt}. The classical solution of the string configuration breaks the translational-invariance of the YM vacuum~\cite{Patrascioiu:1974un,Bardeen:1975gx,Bars:1976re,Brandt:2016xsp, Dubovsky:2012sh, NBrambilla}, resulting in the creation of massless Goldstone modes~\cite{Polchinski:1991ax}. The L\"uscher term~\cite{Luscherfr} and logarithmic broadening~\cite{Luscher:1980iy} are the two main predictions of effective string theory that have been confirmed in many lattice simulations~\cite{Juge:2002br, HariDass2008273, Caselle:2016mqu,caselle-2002,Pennanen:1997qm,Brandt:2016xsp,Caselle:1995fh,Bonati:2011nt,HASENBUSCH1994124,Caselle:2006dv,Bringoltz:2008nd,Athenodorou:2008cj,HariDass:2006pq,Giudice:2006hw,Luscher:2004ib,Caselle:2010zs}

At sufficiently high temperatures, the mean-square (MS) width of the string at the middle plane between two quarks is predicted to exhibit a linear broadening pattern~\cite{allais}, which is tested at large quark separation~\cite{allais,Bakry:2010sp,Bakry:2010zt}. In addition to this, the inclusion of higher-order string's self-interactions and other effects~\cite{PhysRevD.27.2944, Aharony:2010cx, Billo:2012da, Caselle:2004er, Caselle:2005xy, Caselle:1994df, Caselle:2004jq, Caselle:1995fh,Gliozzi:2010zv, Gliozzi:2010zt, Athenodorou:2010cs, Athenodorou:2011rx, Ambjorn:2014rwa, Kalaydzhyan:2014tfa, Caselle:2014eka, Brandt:2017yzw, Caselle:2006dv, Dubovsky:2013gi} is found to extend the match between the lattice data and the predictions of the mesonic string model to the intermediate source separation and high temperature~\cite{Caselle:2016iai, Caselle:2016mqu, Khalil:2022gpj, Brandt:2021kvt, Caselle:2021eir, Bonati:2021vbc, Cea:2017ocq, Battelli:2019lkz,Bakry:2020akg}.

In baryonic configurations, the aspects of the confining potential are widely believed to be manifestations of three body forces that emerge due to the formation of three Y-shaped strings connected at a junction. The Y-ansatz emerges in the strong coupling approximation~\cite{Kogut:1974ag, Capstick:1986ter}, and recently, the Y-potential has been derived using two-loop perturbative calculations~\cite{Brambilla:2013vx}. The Y-ansatz, which represents the leading string effect, can adequately characterize the long-distance lattice data of the confining potential at zero temperature~\cite{Takahashi:2000te, Takahashi:2002bw, Alexandrou:2001ip, Alexandrou:2003ip}.

The impact of quantum fluctuations of the Y-string on the $3Q$ potential has indicated a geometrical L\"uscher-like term as a subleading correction to the leading Y-potential ansatz~\cite{Jahn2004700}. Furthermore, the calculation of the MS width of the Y-junction disclosed a logarithmic growth ~\cite{PhysRevD.79.025022} pattern in equilateral string geometry.

The L\"uscher-like corrections and the width of the junction fluctuations are examined in lattice gauge theory~\cite{deForcrand:2005vv,Bakry:2014gea} and revealed indications in favor of the model. That is, we expect no peculiarities thereabout the hypothesis that the fluctuations of an underpinning Y-string system are the origin of the gluonic aspects at a given temperature scale.

The foregoing main points mount the rationale for discussing the lattice gauge theory data of $(QQ)Q$ and $Q\bar{Q}$ versus the string models. This is expected to be  valid, particularly at color charge separation and temperatures where a crossover from the junction behavior to the free mesonic string model would come about.

\subsection{Mesonic string potential}
Physical infrared (IR) features of the string's world sheet ought to compare with what is predicted for a QCD flux tube. The dynamics of the flux tube follow a massless and free-string theory in the IR limit; given large enough color source separation, an effective field theory with infrared action may be described as
\begin{equation}
	\begin{split}
		S[\mathbf{X}]=&S_{\rm{cl}}+S_{0}[\mathbf{X}]\\
		=&\sigma R~L_{T}+\dfrac{1}{2} \int d\zeta_{0} \int d\zeta \left(\dfrac{\partial \mathbf{X}}{\partial \zeta_{\alpha}} \cdot \dfrac{\partial \mathbf{X}}{\partial \zeta_{\alpha}}\right),
		\label{eq:LOaction}
	\end{split}        
\end{equation}
where $S_{cl}$ is the classical configuration or perimeter-area term, the coordinates $\zeta_{0}$ and $\zeta_{1}$ parametrize the world sheet ($\alpha=0,1$), the vector $X^{\mu}(\zeta_{0},\zeta_{1})$ in the transverse gauge describes the fluctuations of the two-dimensional bosonic world sheet relative to the surface of minimal-area/the classical configuration with $(\mu=1,2,...,d-2)$. The above action $S[\mathbf{X}]$, Eq.(\ref{eq:LOaction}), is referred to as the massless free bosonic string action.
\begin{figure}[t]
	\centering
	\includegraphics[scale=0.37]{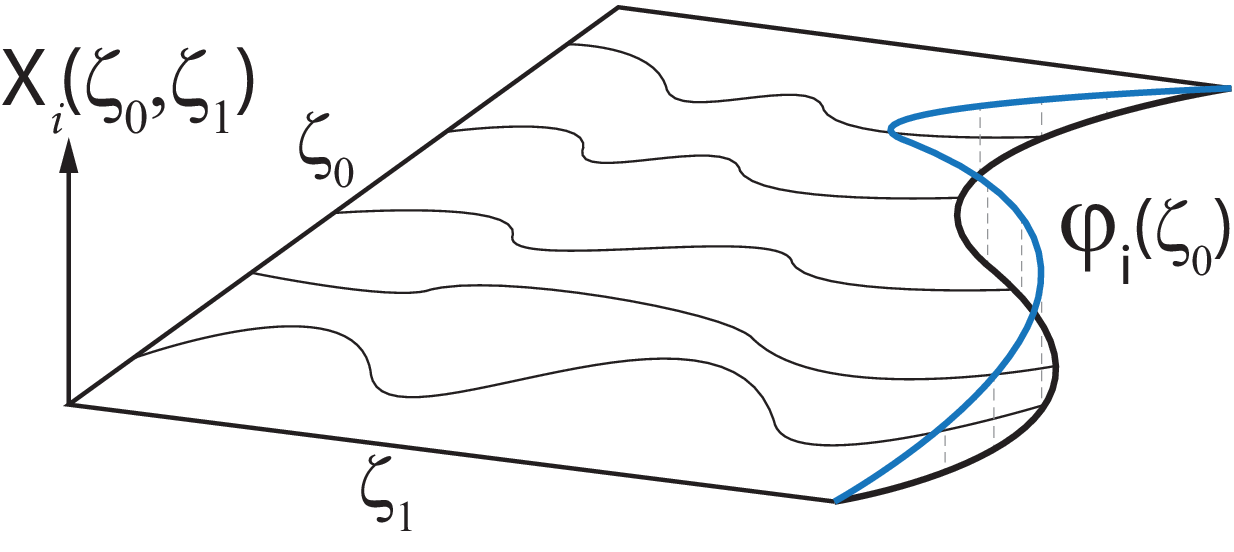}
	\caption{World sheet traced by one of the strings up to the junction position.}
	\label{fig:WorldSheet}
\end{figure}

The Casimir energy is extracted from the string partition function as
\begin{equation}
	V(R,L_{T})=-\dfrac{1}{L_T} \log\left[Z(R,L_{T})\right].
	\label{eq:Casmir}
\end{equation}
The partition function of the free NG model in the physical gauge is a functional integral over all the world sheet configurations swept by the noninteracting string
\begin{equation}
	Z(R,L_{T})= \int_{{\cal C}} [D\, {\mathbf{X}} ] \,\exp(\,-S( {\mathbf{X}} )).
	\label{eq:PI}
\end{equation}  
For a periodic boundary condition (BC) along the time direction and Dirichlet BC at the source's position. The path integral Eq.\eqref{eq:PI} and Eq.\eqref{eq:Casmir} yields the static potential
\begin{equation}
	\label{eq:Pot_NG_LO}
	V^{\rm{NG}}_{\rm{\ell o}}(R,L_{T})= \sigma_{0} R+\frac{2}{L_{T}}\, \log \eta \left(\tau \right)+\mu_c,
\end{equation}  
where $\mu$ is a UV-cutoff and $\eta$ is the Dedekind eta function defined as
\begin{equation}	\eta(\tau)=q_1^{\frac{1}{24}} \prod_{n=1}^{\infty}(1-q_1^{n}),
\end{equation}
where $q_1=e^{i \pi \tau}$ with modular parameter $\tau=\frac{iL_{T}}{R}$. The slope of the linear terms in $R$ defines the renormalized string tension~\cite{Kac,PhysRevD.85.077501} given by
\begin{equation}
	\sigma(T)=\sigma_{0}-\dfrac{\pi}{3} T^{2}+ {\mathcal{O}}(T^4).
	\label{eq:Tension_NG_LO}
\end{equation}
where $T$ is the temperature scale governed by inverse of the temporal extent $T=1/L_{T}$.

\subsection{Baryonic Y-string potential}
\begin{figure}[t]
	\centering	
	\includegraphics[scale=0.18] {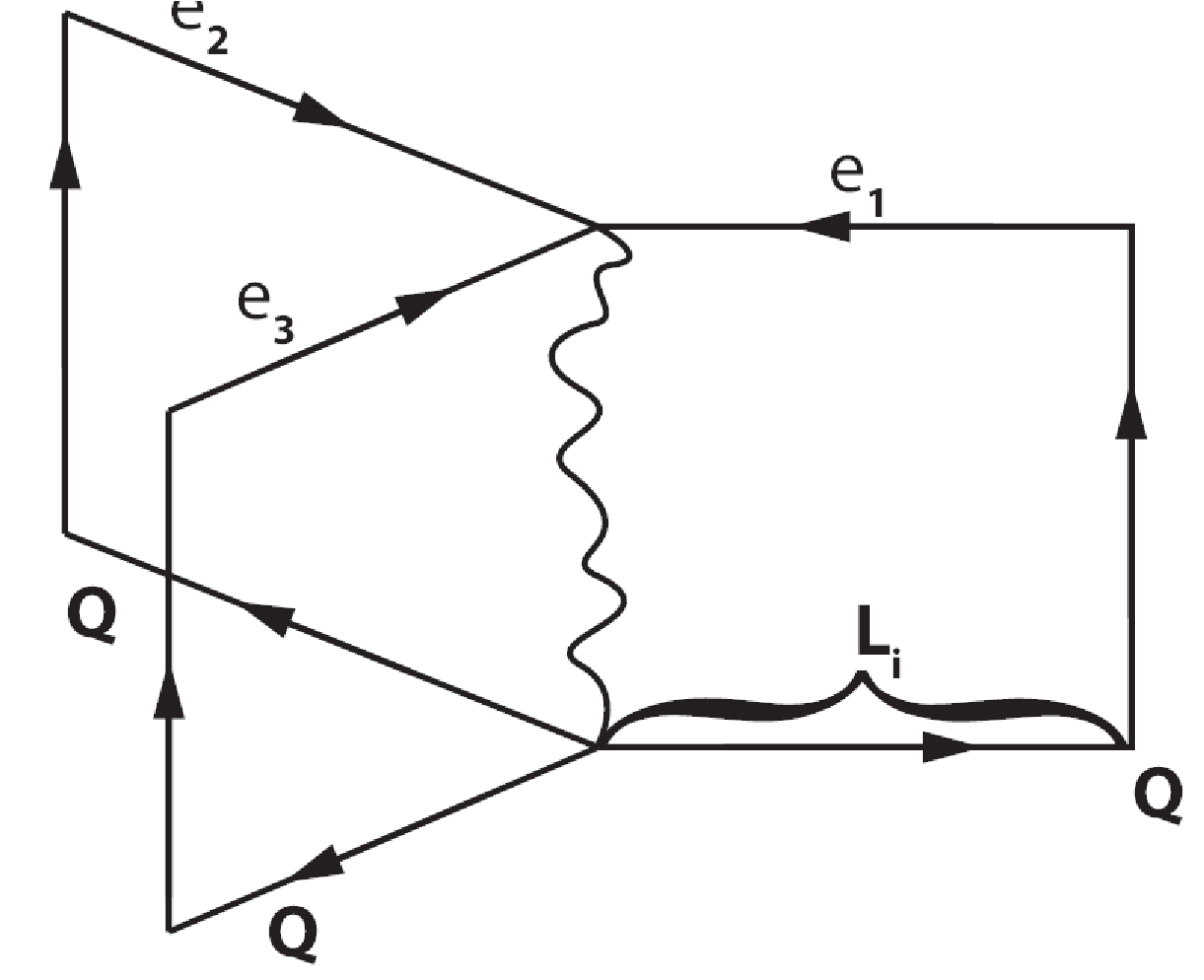}
	\includegraphics[scale=0.18] {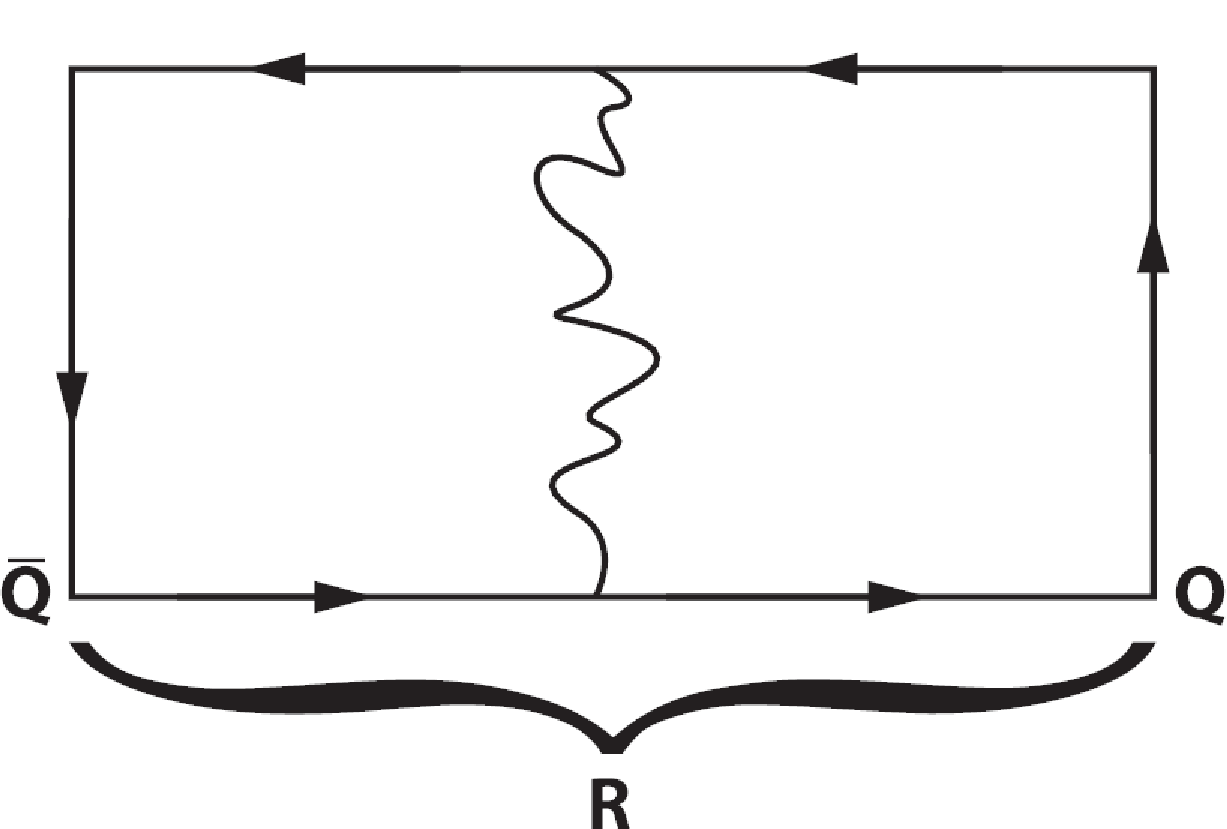}
	\caption{The three and two blade worldsheet systems swept by the fluctuating world lines of the gluonic strings of the baryon and meson. 
	}
	\label{fig:convol}	
\end{figure}

In the Y-string model~\cite{Jahn2004700, PhysRevD.79.025022, Andreev:2008tv, AndreevDiQ} the quarks are connected by three strings that come together at a junction~\cite{Flower:1986ru, Takahashi:2002bw, Ichie:2002mi, Bissey:2006bz, Cornwall:2006, Llanes-Estrada:2011gwu}. The string world sheets' smallest area corresponds to the classical arrangement. The world sheet (blade) of each string is made up of a static quark line and a fluctuating junction world-line (Fig.\ref{fig:WorldSheet} and Fig.\ref{fig:convol}). 
\begin{figure*}[!hptb]
	\centering
	\includegraphics[scale=0.35]{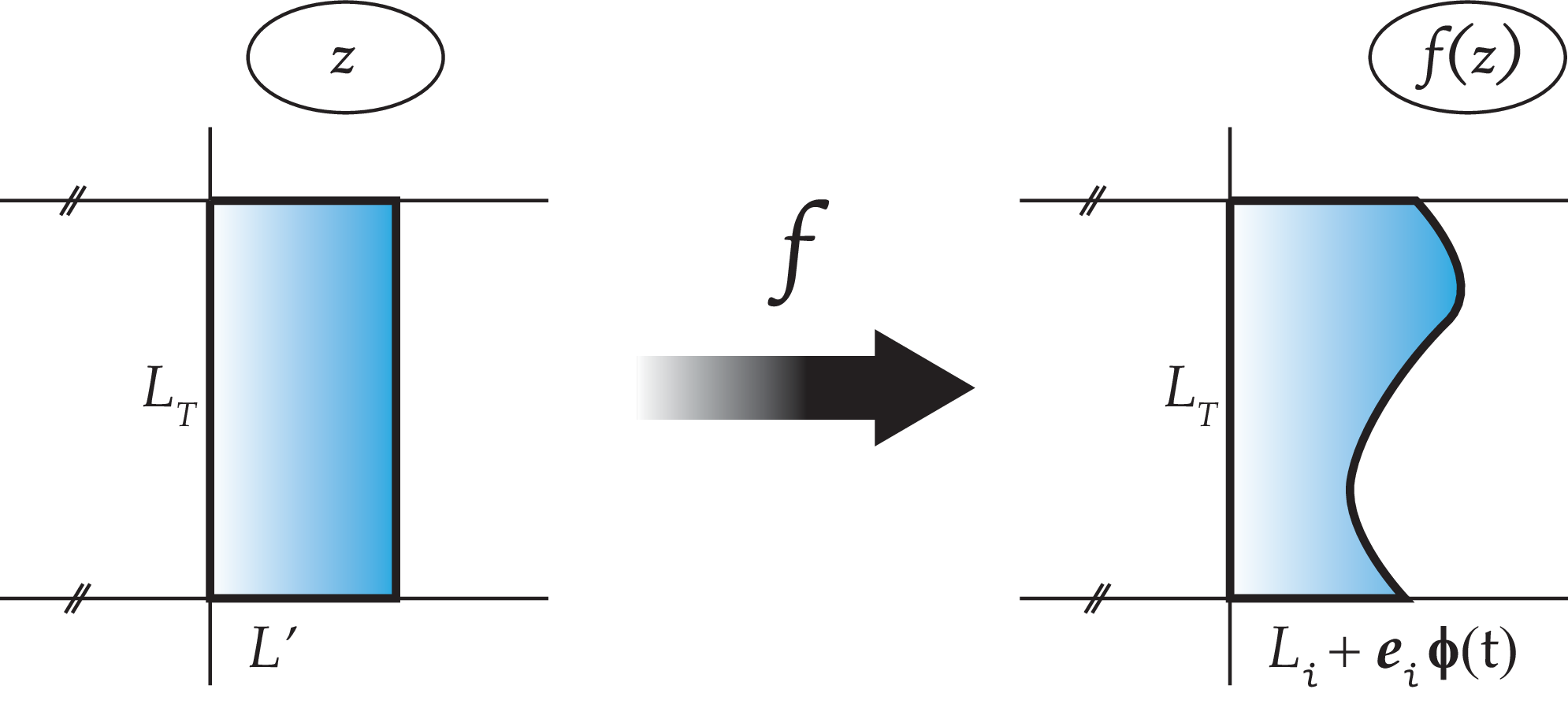} 
	\caption{Conformal map Eq.\eqref{eq:conformal map}  maps the domain $\Theta_{i}$ conformally to a rectangle $L'_{i}\times L_{T}\equiv\tilde\Theta_{i}$.
	}
	\label{fig:conformal_map}
\end{figure*}
The parameter $\zeta_{1}$ and $\zeta_{0}$ (time) label the position on string world-sheet (blade) $i$. The position of the junction is given by $\zeta_{1}=L_{i}+\eta_{i}.\phi(\zeta_{0})$. The transverse fluctuations $\mathbf{X}_i (\zeta_{0},\zeta_{1})$ vanish at the location of the quarks ($\zeta_{1}=0$), and are periodic in the time $\zeta_{0}$, with period $1/L_{T}$ (see Fig.\ref{fig:WorldSheet}).

In addition to the Dirichlet BC at the quark position  we have the BC from the continuity of the transverse fluctuations $X_i (\zeta_{0},\zeta_{1})$
\begin{equation}
	\label{eq:boundarycondition_xi}
	\mathbf{X_i}(\zeta_{0},L_i+\eta_i \cdot \phi(\zeta_{0})) = \phi_{\perp i}(\zeta_{0})\:,
\end{equation}
where $\eta_{i}$ are spatial unit vectors in the direction of the strings such that $\Sigma_{i} \eta_{i}=0$(mixed Dirichlet-Neumann BC~\cite{tHooft,tHooft:2004doe}).

The NG action after gauge-fixing and expanding around the equilibrium configuration yields 
\begin{equation}
	S_{\mathrm{Fluct}}=  \sigma L_Y L_T+\frac{\sigma}{2} \sum_{i} \int_{\Theta_i}\!\!d^2\!{\rm \zeta}\,\, 
	\frac{\partial \mathbf{X}_{i}}{\partial{\rm \zeta}_{\beta}} \cdot \frac{\partial \mathbf{X}_{i}}{\partial{\rm \zeta}_{\beta}},
\end{equation}
where, $L_Y = \sum_i L_i$ above denotes the total string length. In this model ~\cite{Jahn2004700,PhysRevD.79.025022}, the junction is assumed to acquire a self-energy term $m$. This results in an additional boundary term to NG action
\begin{align*}
	S &=  S_{\mathrm{Fluct}}+S_{\mathrm{Boundary}},\\
	\intertext{with a static energy and a kinetic energy terms of junction defined as}
	S_{\mathrm{Boundary}}& = \left(m L_T + \frac {m}{2}\int_0^{L_T}\! d\zeta_{0}\, |\dot{ \boldsymbol\phi}|^2 \right),
\end{align*}
\noindent respectively.

The system's partition function is
\begin{equation}
	\begin{split} 
		\label{eq:partitionfct}
		Z & = e^{-(\sigma L_Y +m)L_T}\\
		& \times \int \, D \boldsymbol\phi~
		\exp\left(-\frac{m}{2}\int \, d\zeta_{0}\,
		|\dot{\boldsymbol\phi}|^2\right)\prod_{i=1}^3 Z_i( \boldsymbol\phi),
	\end{split}  
\end{equation}
\noindent where $Z_i(\phi)$ is the $i$-th partition function of a given blade bounded by the junction worldline $\boldsymbol\phi(\zeta_{0})$ and reads in d-dimension as
\begin{equation}
	Z_i(\phi)=e^{-\frac{\sigma}{2}\int |\partial
		\mathbf{X}_{\min,i}|^2}
	|\det(-\triangle_{\Theta_i})|^{-(d-2)/2}\:,
	\label{eq:Laplace}
\end{equation}
where $\mathbf{X}_{\min,i}$ is the minimal-area solution for a given junction configuration $\mathbf{\phi}(\zeta_{0})$, and $\triangle_{\Theta_i}$ denotes the Laplacian acting on the domain (blade) $\Theta_i$(Fig.\ref{fig:conformal_map}).

The Casimir energy has been computed for the baryonic potential $V_{\rm{3Q}}$ by Jahn and De Forcrand ~\cite{Jahn2004700,deForcrand:2005vv}. In that approach, the  domains of the blades are conformally mapped to rectangles (Fig.\ref{fig:conformal_map}) prior to evaluating the Laplacian's determinant in Eq.\eqref{eq:Laplace}.

The conformal map~\cite{Jahn2004700} of the string's blade $\Theta_i$ to a rectangle~\cite{Jahn2004700} $\tilde\Theta_i$ takes the form
\begin{equation}
	f_i(z)=z+\frac{1}{\sqrt L_{T}} \sum_{\omega \neq 0} \frac{\boldsymbol\eta_i \cdot \boldsymbol\phi_{i,\omega} }{\sinh(\omega L_i)} e^{\omega z},
	\label{eq:conformal map}
\end{equation}
then the determinant in Eq.\eqref{eq:Laplace} is obtained after taking into account the change in the Laplacian~\cite{Luscherfr, PhysRevD.79.025022} by the conformal map
\begin{equation}
	{\rm{ln}}\frac{\det (-\triangle_{\Theta_{i}})} {\det (-\triangle_{\tilde\Theta_{i}}) } = \frac{1}{12\pi}\sum_{\omega} \omega^3 | \boldsymbol\eta_i \cdot \boldsymbol\phi_{i,\omega}|^2 \coth(\omega L_{i}).
	\label{eq:ratio}
\end{equation}

Further, conformally mapping the rectangle $L'_i\times L_{T}\equiv\tilde\Theta_{i}$ into a circle, the determinant of the Laplacian with respect to the blade $i$ would then read
\begin{equation}
	\label{eq:barypotgaussian}
	\scalemath{0.85}{
		\det(-\triangle_{\Theta_i}) = \eta^2\!\left( \frac{i L_{T}}{2 L'_i}\right) 
		\exp\left(- \frac{1}{12\pi}\sum_{\omega} \omega^3 \coth (\omega L_i)| \boldsymbol\eta_i \cdot \boldsymbol\phi_{i,\omega}|^2\right)\: 
	}.
\end{equation}

The sum over all eigenmodes formally result in the baryonic potential ${\rm V}_{3Q}$ which reads as
\begin{equation}
	\label{eq:vba1}
	V_{3Q}(L_i)= \sigma L_Y + V_{\parallel} +2 V_{\perp} + \mathcal{O}(L^{-2}_i),
\end{equation}
such that $L_{\rm Y} \equiv L_1+L_2+L_3$ is  the sum of the lengths of the three strings, and
\begin{equation}
	\begin{split}
		V_{\parallel}(L_i) & = \sum_i \frac{1}{L_T}\eta\left(\frac{i L_T}{2L_{i}}\right) \\ 
		& + \sum_{\omega=1} \frac{1}{L_T}\ln\left[\frac 1 3 \sum_{i<j} \coth(\omega L_i)\coth(\omega L_j)\right],
	\end{split}
\end{equation}
is the potential component owing to the in-plane fluctuations and  
\begin{equation}
	\begin{split} 
		\scalemath{0.85}{
			V_{\perp}(L_i) = \sum_i \frac{1}{L_T}\eta\left(\frac{i L_T}{2L_{i}}\right) + \sum_{\omega=1} \frac{1}{L_T} \ln\left[ \frac 1 3 \sum_i \coth(\omega L_i)\right] },
	\end{split}  
\end{equation}
\noindent is that due to the perpendicular components.

The corresponding mesonic limit would read 
\begin{equation}
	\begin{split} 
		V_{\perp} &= \frac{1}{L_T}\eta \left(\frac{iL_T}{2L_{1}}\right)+ \frac{1}{L_T}\eta \left(\dfrac{i L_T}{2 L_{2}}\right) \\
		& + \sum_{\omega=1} \frac{1}{L_T} \ln\left[ \frac 1 2 (\coth(\omega L_1)+\coth(\omega L_2)) \right].\:
	\end{split}  
\end{equation}
The quark-antiquark ($Q\bar{Q}$) potential is then
\begin{equation}
	V_{Q\bar{Q}}=\sigma (L_1+L_2)+ \frac{2}{L_T}\ln\left[\eta \left(\dfrac{i L_T}{(L_{1}+L_{2})}\right)\right],
	\label{eq:meson}
\end{equation}
\noindent which is in agreement with the mesonic string potential in 4D Eq.~\eqref{eq:Pot_NG_LO}.

Expressing the sum in Eq.\eqref{eq:vba1} in terms of Dedekind $\eta$ functions, the potential in the $3Q$ channel can formally be written such that
\begin{equation}
	V_{3Q}=\sigma L_{\rm Y}+ \dfrac{\gamma(L_{1},L_{2},L_{3})}{L_T} \ln\left[\eta \left(\dfrac{i L_T}{L_{\rm Y}}\right)\right],
	\label{eq:total}
\end{equation}
\noindent where $\gamma(L_1,L_2,L_3)$ is a geometrical factor which depends on the quark configuration and is obtained by solving Eqs.\eqref{eq:total} and \eqref{eq:vba1}. The geometrical factor corresponding to an equilateral quark triangle, $L_{i}=L_{j} $ is explicitly $\gamma(L_1,L_2,L_3)=2$. In the limit, $T=0$ the second term turns out into the baryonic L\"uscher-like correction to the $V_{3Q}$ potential at zero temperature~\cite{Jahn2004700}.

\subsection{Mesonic String Width}
\label{sec:EffBosonic_String_Width}
The quantum delocalization of the stringlike object brings about a characteristic physical width of the corresponding flux-tube. The generic definition of the MS width of the string reads as
\begin{align}
	\label{operator}
	W^{2}(\xi;\tau) = & \quad \langle \, \mathbf{X}^{2}(\xi;\tau)\,\rangle \nonumber\\ 
	= &\quad \dfrac{\int_{\mathcal{C}}\,[D\,\mathbf{X}]\, \mathbf{X}^2 \,\mathrm{exp}(-S[\mathbf{X}])}{\int_{\mathcal{C}}[D\,\mathbf{X}] \, \mathrm{exp}(-S[\mathbf{X}])}.
\end{align}
At the center plane of the fluctuating string the above expectation value yields in 4D a logarithmic broadening versus the string length/interquark spacing $R$,  
\begin{equation}
	\label{Wid}  
	W^{2} = \frac{1}{\pi\sigma}\log\left( \frac{R}{R_{0}} \right),
\end{equation}
\noindent where $R_{0}$ is an ultraviolet (UV) cutoff scale. This is the famed prediction, made many years ago by L\"uscher, M\"unster, and Weisz \cite{Luscher:1980iy}, implying a universal logarithmic divergence  (gauge-group independent) common to  confining gauge groups. 
This term represents the leading order term of the mean-square width of the noninteracting NG string.

With the increase of the temperature, higher-order gluonic modes come into play altering the broadening pattern of the MS width versus both the string length and the temperature scale. Allais and Casselle~\cite{allais} calculated the MS width of the string delocalization at all transverse planes to the line connecting the quark pair which accordingly reads as 
\begin{equation}
	\label{sol}
	\scalemath{0.9}{	W^{2}(x) = \frac{1}{\pi\sigma}\log\left( \frac{R}{R_{0}} \right) + \frac{1}{\pi\sigma}\log\Bigg|\,\dfrac{\theta_{2}\left(-\dfrac{\pi}{R}\left(\dfrac{R}{2}-x\right),q_2\right)} {\theta_{1}^{\prime}(0,q_2)} \Bigg|,}
\end{equation}
where $\theta_i$ are the Jacobi elliptic functions
with the nome, $q_{2}=e^{\frac{i \pi}{2} \tau}$ and $x \in [0,R]$  signifies the coordinate (in length units) of the transverse planes. 

This expression converges for modular parameters near unity and includes, in addition to the logarithmic divergence term, a correction term that expresses the dependency of the width at various transverse planes on the modular parameter.
\begin{figure}[t]
	\centering		
	\includegraphics[scale=0.48]{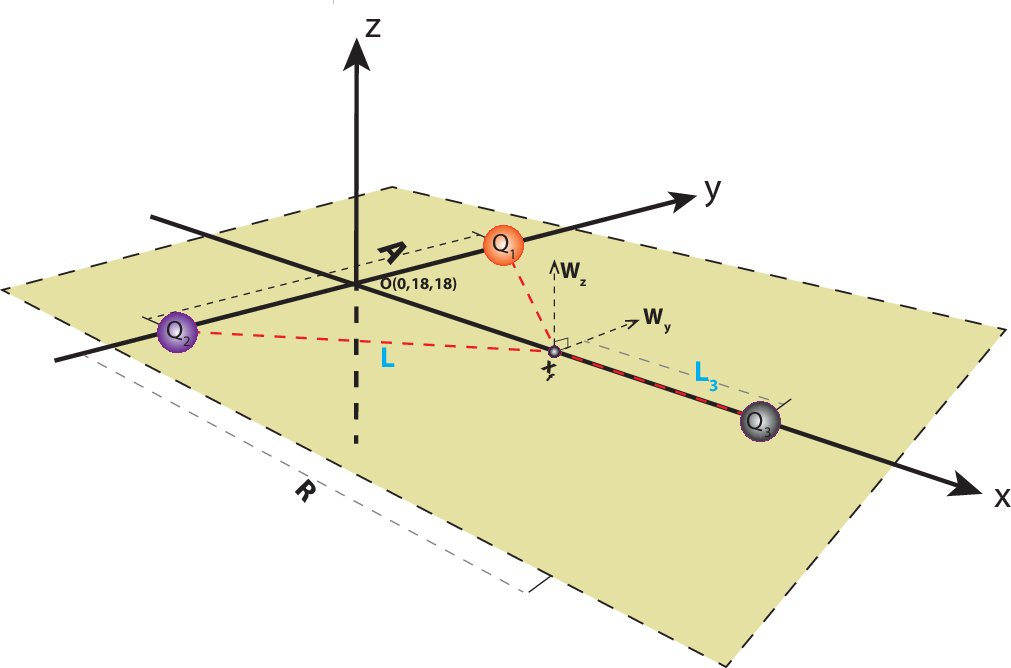}
	\caption {Schematic diagram showing the configuration of the $Y$ string relative to the quark source positions. The junction’s locus is fixed at Fermat point $x_f$. The isosceles base is denoted as $A=2d$, the Fermat point is located at $x_{F}=A/2\sqrt{3}$.}
	\label{fig:gam}	
\end{figure}

\subsection{Baryonic String Width}
The width of the string at the junction can be calculated~\cite{PhysRevD.79.025022} taking the expectation value of $\phi^2$
\begin{equation}
	\langle{\phi}^2\rangle=\frac{\int\,  [D \, \phi] \, \phi^2\,
		e^{-S}} {\int \,  [D \phi] \,e^{-S}}\:.
\end{equation}

The above second moment of the junction can be decomposed (Fig.\ref{fig:gam}) into perpendicular $z$ and parallel (in-plane) $xy$ fluctuations
\begin{equation}
	\langle{\phi}^2 \rangle=\langle{\phi_z}^{ 2}\rangle+\langle{ \phi_{xy}}^{2}\rangle.
\end{equation}
That is,
\begin{equation}
	\begin{split}
		\langle \phi^{2}_{z} \rangle = \frac{2}{L_{T}}\sum_{w > 0}\frac{1}{mw^{2}+\sigma w \sum_{i} {\rm{coth}}(wL_{i}) \psi(w,L_{i})}.
	\end{split}
	\label{eq:perpen_StrFluct}   
\end{equation}

It is more convenient for our further discussion of the in-plane fluctuations on the lattice to consider the  rotated decoupled form of $\langle{\phi_{xy}}^{2}\rangle$ given in ~\cite{Bakry:2014gea} as
\begin{equation}
	\begin{split}
		&\scalemath{0.9}{ \langle \phi^{2}_{x} \rangle = \frac{2}{L_{T}}\sum_{w > 0}\frac{1}{Q_{x,w}+Q_{y,w}-\sqrt{Q^{2}_{xy,w}+(Q_{x,w}-Q_{y,w})^{2}}} },\\
		&\scalemath{0.9}{ \langle \phi^{2}_{y} \rangle = \frac{2}{L_{T}}\sum_{w > 0}\frac{1}{Q_{x,w}+Q_{y,w}+\sqrt{Q^{2}_{xy,w}+(Q_{x,w}-Q_{y,w})^{2}}} }.
		\label{eq:inplane_StrFluct}   
	\end{split}
\end{equation}

Taking into account the convoluted fluctuations $\phi \to \int_{-\infty}^{\infty} \phi(\tau) \psi(t-\tau) d\tau $  which account for the thermal effects~\cite{Bakry:2014gea} in the series sums Eq.\eqref{eq:perpen_StrFluct} and Eq.\eqref{eq:inplane_StrFluct}. The corresponding modes $\psi(w,L_{i})$ are given by
\begin{equation}
	\begin{split}
		\scalemath{0.8}{
			\psi(w_{n},L_{i}) = \frac{-kw_{n}}{2\sigma {\rm{coth}}(w_{n}L_{i})}-\frac{2n-1}{4n{\rm{coth}}(w_{n}L_{i})}
			\left( \frac{2L_{i}\chi(\tau_{i})+1}{2L_{i}\chi(\tau_{i})-1}\right)^{2n-1} 
		}.
		\label{eq:Smoothing_term}   
	\end{split}
\end{equation}
where $k$ is an ultraviolet cutoff. The relation
between $k$ and $R_0$ Eq.\eqref{sol} is shown in  
~\cite{PhysRevD.79.025022}, where the addition of UV cutoff to log growth is justified.

The quantities $Q_x,Q_y$ and $Q_{xy}$ in the above equation are defined ~\cite{PhysRevD.79.025022,Bakry:2014gea} as
\begin{equation}
	\label{eq:cof}
	\begin{split}  
		Q_x = &\left( k w^2 +{\sigma} w \sum_i\coth(wL_i)\psi(w,L_i)\right) \\
		& + \left(\frac{\sigma} {2} w + \frac{w^3}{12\pi} \right)
		\left[
		\sum_i \eta_{i,x}^2 \coth(w L_i)\psi(w,L_i)
		\right], \\
		Q_y =& \left(k w^2 +{\sigma} w \sum_i\coth(wL_i)\psi(w,L_i)\right) \\
		& + \left( \frac{\sigma} {2} w + \frac{w^3}{12\pi}\right)
		\left[
		\sum_i \eta_{i,y}^2 \coth(w L_i)\psi(w,L_i)
		\right], \\
		Q_{xy} =& \left( \frac{\sigma} {2} w + \frac{w^3}{12\pi} \right)\left[\sum_i \eta_{i,x} \eta_{i,y} \coth(w L_i)\psi(w,L_i)\right].
	\end{split}             
\end{equation}

Although formula Eq.\eqref{eq:perpen_StrFluct} and Eq.\eqref{eq:inplane_StrFluct} in \cite{PhysRevD.79.025022} reproduces the logarithmic growth Eq.\eqref{Wid} in the mesonic limit,  the mesonic string width formula Eq.~\eqref{sol} is not attained at finite temperature. The  convoluted modes~\cite{Bakry:2014gea} are derived from the mesonic limit  such that the resulting modified width would approximate the MS width of the junction fluctuations at finite temperature.
\begin{figure}[b]
	\centering
	\subfigure[$Q_3$ position $R=14$]{\includegraphics[scale=0.37]{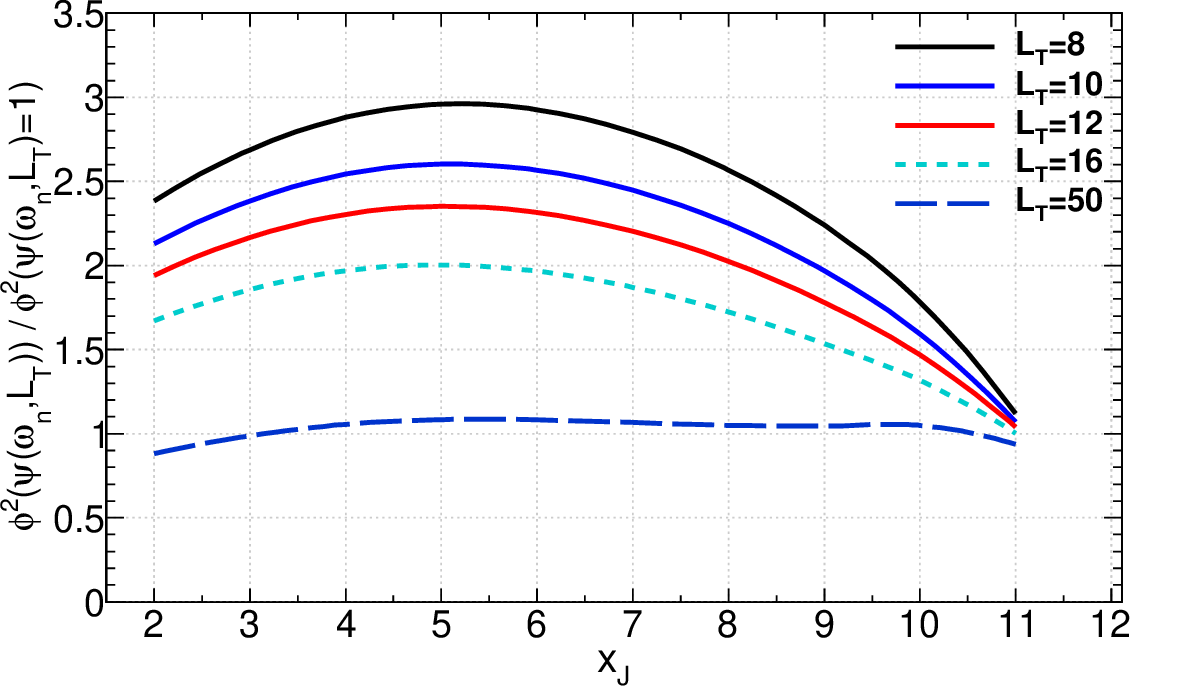}}
	\subfigure[$Q_3$ position $R=14,18$]{\includegraphics[scale=0.37]{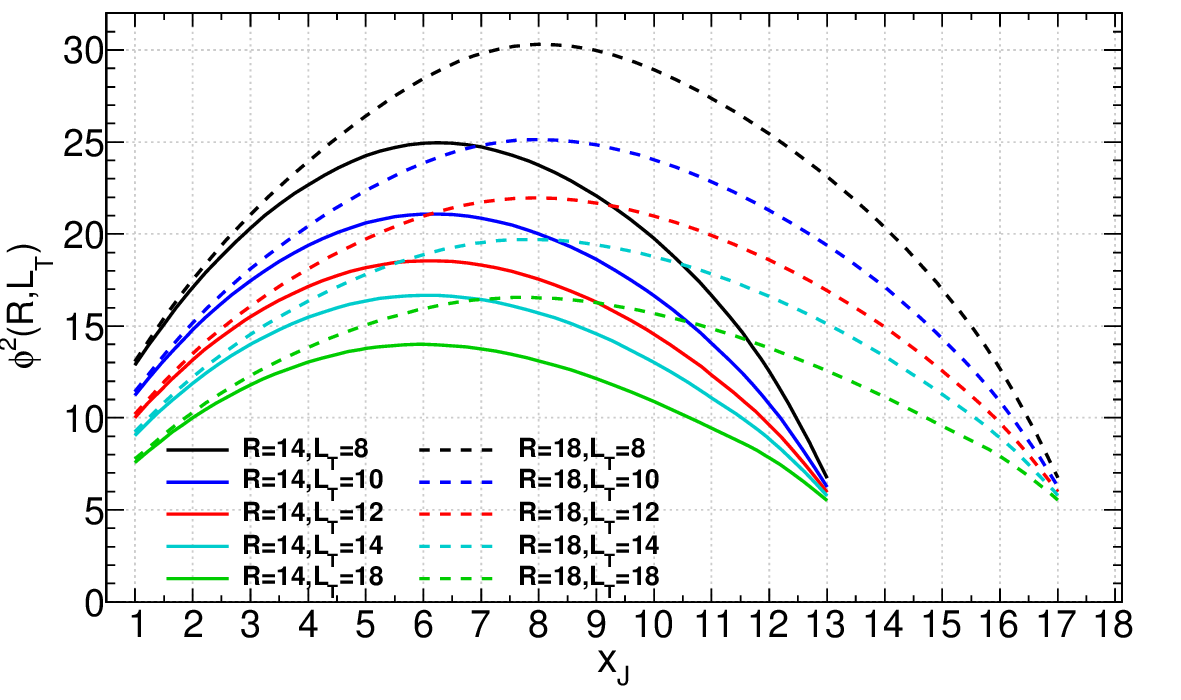}}  
	\subfigure[Junction position $x_{J}=2$]{\includegraphics[scale=0.37]{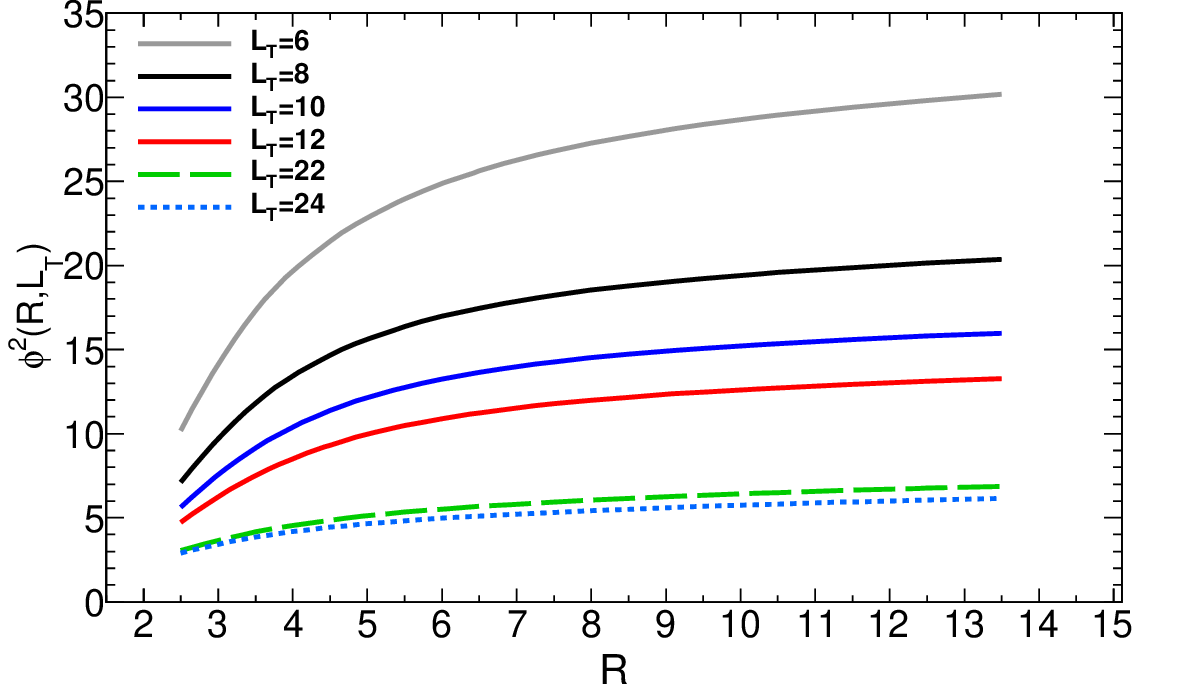}}
	\caption{(a) The ratio of the formulas of the junction width Eqs.~\eqref{eq:perpen_StrFluct} and Eq.\eqref{eq:Smoothing_term} to the unconvoluted width corresponding to $\psi_{n}(T,X_{j})=1$ \cite{PhysRevD.79.025022}. (b) The width of the Y-junction's  at each transverse plane $x_{J}$, the solid and dashed lines correspond to two $Q_3$ location (Fig.\ref{fig:gam})(c) The broadening of the junction's MS width at a selected transverse plane $x_{J}=2$ versus $Q_3$ location $R$. (Fig.\ref{fig:gam}).}
	\label{fig:epsi_nT}
\end{figure}

In Fig.\ref{fig:epsi_nT} provided are three plots each illustrates the effect of the convoluted modes on the MS width of the junction at the depicted temperature scale.

The effects of the convoluted modes are evidently disclosed in Fig.\ref{fig:epsi_nT}(a). We plot the ratio of the MS width of the junction $ \phi^{2}_{z}$, with the convoluted modes, Eq.\eqref{eq:perpen_StrFluct} to $ \phi^{2}_{z} $ taking $\psi(w_{n},L_{i}) \rightarrow 1$ corresponding to the formula derived in~\cite{PhysRevD.79.025022}. The abscissas correspond to the continuous change of the junction position $x_{J}=[0,R]$ along the $x$-axis while the quark positions are kept fixed (See Fig.\eqref{fig:gam}). The corrections are gradually decreased with the temperature and converge in the zero temperature limit $L_{T} \rightarrow \infty$ yielding a ratio that tends to unity.

Figure \ref{fig:epsi_nT}(b) illustrates the width of the junction $\phi^{2}_{z}$ Eq.\eqref{eq:perpen_StrFluct}. Same as Fig.\ref{fig:epsi_nT}(a) the quark positions are kept fixed whereas the junction position  is made to vary along $x$-axis such that $x_{J}=[0,R]$. The two subsets of curves (solid and dashed) correspond to two $3Q$ configurations only with different locations of $Q_{3}$, (Fig.\eqref{fig:gam}) $R_{1}$ and $R_{2}$. The junction width peaks at the middle of $R$, it is also where  the effects of the temperature increase on the MS width broadening are most pronounced. Interestingly, the mesonic MS-width given by Eq.\eqref{sol} exposes a similar profile to that shown in Fig.\ref{fig:epsi_nT}(b) (see Ref.\cite{Bakry:2010zt}).

The MS-width of the junction is projected to fit the baryonic flux-tube width by  selecting a transverse plane $x_{J}$ and comparing the width broadening versus the quark positions $R$ as shown in Fig.\ref{fig:epsi_nT}(c). The same procedure is applicable to the MS width fits of the meson Eq.\eqref{sol}.

Nevertheless, in regard to the systematic of the fit of the junction to the baryonic width data, it ought to emphasize that only exactly where the string's junction it is possible to identify the width of the baryonic string with the width at the junction in Eq.\eqref{eq:perpen_StrFluct} and \eqref{eq:inplane_StrFluct}. Otherwise, a formula characterizing the individual blades has to be used. That is, the fit of the width data to the width formula is valid  only at the junction location. Consequently, the junction used to calculate the width formula must be located on the plane where the width is determined (as we shall see in Sec.\ref{sec:AcDensity_onLattice}). The returned $\chi^{2}$ from the fit can then provide relevant information on the locus of the junction.

\section{Lattice measurements}
\label{sec:Operators}
\subsection{Potential operators}
The expectation value of the Wilson loop is obtained within the overlap formalism of a given gluonic wave function with the quark-antiquark state. The Polyakov loops address the free energy of the system.

The static mesonic state is constructed using  a pair of Polyakov loops such that
\begin{equation}
	\langle \mathcal{P}_{Q\bar{Q}}(\mathbf{r}_{1}, \mathbf{r}_{2})\rangle = \langle P(\mathbf{r}_{1}) P^{\dagger}(\mathbf{r}_{2})\rangle 
	\label{eq:QantiQ}   
\end{equation}
where the Polyakov loop on an Euclidean lattice of size $N^{3}_{s}\times N_{t}$. The Polyakov loop  is defined as a product of gauge field variables $U_{\mu=4}(\mathbf{r}_{i},n_{t})$:
\begin{equation}
	P(\mathbf{r}_{i}) = \frac{1}{3}\mbox{Tr} \left[ \prod^{N_{t}}_{n_{t=1}}U_{\mu=4}(\mathbf{r}_{i},n_{t}) \right],
	\label{eq:PolyakovLoop}   
\end{equation} 
The Monte Carlo evaluation of the temperature-dependent quark-antiquark potential at each $R$ is calculated through the Polyakov loop correlators in the standard manner  \cite{Luscher:2001up,Luscher:2002qv} as
\begin{align}
	\langle \mathcal{P}_{Q\bar{Q}}(0,R)\rangle 
	&= \langle P(0) P^{\dagger}(R)\rangle  \notag \\ 
	&= \int d[U] \,P(0)\,P^{\dagger}(R)\, \mathrm{exp}(-S_{w})  \notag\\
	& = e^{-\frac{1}{T}V_{Q\bar{Q}}(R,T)},
	\label{eq:QQ_correlators}
\end{align}
where $S_{w}$ is the Wilson action and $T$ is the physical temperature. 

The quark-diquark potential can be identified via a three-loop correlator such as
\begin{align}
	\langle \mathcal{P}_{3Q}\rangle &= \langle P(\mathbf{r}_{1} ) P(\mathbf{r}_{2} ) P(\mathbf{r}_{3} ) \rangle \notag  \\
	& = e^{-\frac{1}{T}V_{3Q}(\mathbf{r}_{1},\mathbf{r}_{2},\mathbf{r}_{3}) }.
	\label{eq:QQQ_potential} 
\end{align} 

The above correlators respect the center symmetry transformation \cite{Bakry:2014gea} all across the confinement phase.

\subsection{Energy-density operators}

To characterize the Euclidean action density on the lattice we utilize a plaquette operator at position $\boldsymbol\rho$ defined by 
\begin{equation}
	\mathcal{P}_{\mu\nu}(\boldsymbol\rho)= \left[U_{\mu}(\boldsymbol\rho)U_{\nu}(\boldsymbol\rho+\boldsymbol\mu)U_{\mu}^{\dagger}(\boldsymbol\rho+\boldsymbol\nu)U_{\nu}^{\dagger}(\boldsymbol\rho)\right],   
\end{equation}
with the indices $\mu$ and $\nu$ corresponding to Lorentz indices.

The Euclidean action density is given by
\begin{equation}
	S(\boldsymbol\rho) =\beta \sum_{\mu<\nu} \left( 1-\frac{1}{3} Re \, Tr\mathcal{P}_{\mu\nu}(\boldsymbol\rho) \right),  
\end{equation}
where $\beta$ is the coupling of Yang-Mills theory. The plaquette $\mathcal{P}_{\mu\nu}$ can be expanded in a power series \cite{Gattringer} in the symmetric field strength tensor $F_{\mu\nu}$ such that
\begin{equation}
	S(\boldsymbol\rho) = a^{4} \sum_{\mu<\nu} Tr F^{2}_{\mu\nu}(\boldsymbol\rho) +\mathcal{O}(a^{2})+ \mathcal{O}(a^{2}g^{2}),
\end{equation}
with $g^{2}=\dfrac{6}{\beta}$.

A dimensionless scalar field characterizing the Euclidean action-density distribution in the Polyakov vacuum, i.e., in the presence of color sources \cite{Bissey:2006bz} can be defined as
\begin{equation}
	\scalemath{0.9}{
		\mathcal{C}_{Q\bar{Q}}(\boldsymbol\rho;\mathbf{r}_{i} )= \frac{\langle\, \mathcal{P}_{2Q}(\mathbf{r}_{1},\mathbf{r}_{2})\,\rangle\, \,\langle S(\boldsymbol\rho)\, \rangle-\langle\mathcal{P}_{2Q}(\mathbf{r}_{1},\mathbf{r}_{2}) \, S(\boldsymbol\rho)\,\rangle } {\langle\, \mathcal{P}_{2Q}(\mathbf{r}_{1},\mathbf{r}_{2})\,\rangle\, \,\langle S(\boldsymbol\rho)\, \rangle}
	},
	\label{eq:mesonic_Correlator}
\end{equation}
with the vector $\boldsymbol\rho$ referring to the spatial position of the flux probe, $\mathbf{r}_{i}$ are the spatial positions of the color sources and the bracket $\langle ...\rangle$ stands for averaging over gauge configurations and lattice symmetries. Another dimensionful definition of the correlator \eqref{eq:mesonic_Correlator} yields an equivalent representation of the width (see, for example, Ref.~\cite{Okiharu:2003vt}). 

The measurements are repeated for each time slice and then averaged, 
\begin{eqnarray}
	S(\boldsymbol\rho)=\frac{1}{N_{t}} \sum_{n_{t}=1}^{N_{t}}S(\boldsymbol\rho,t).
	\label{eq:average}
\end{eqnarray}

Cross-correlations between measurements made at various distances on the same gauge configuration may be likely to happen in this arrangement when the size of the flux probe operator $S(\boldsymbol\rho)$ is large. For this reason, we refrain from using improved field operators.

For baryonic systems, a dimensionless scalar field that characterizes the gluonic field can be defined as
\begin{equation}
	\scalemath{0.9}{
		\mathcal{C}_{3Q}(\boldsymbol\rho;\mathbf{r}_{i} )= \frac{\langle\, \mathcal{P}_{3Q}(\mathbf{r}_{1},\mathbf{r}_{2},\mathbf{r}_{3})\,\rangle\, \,\langle S(\boldsymbol\rho)\, \rangle-\langle\mathcal{P}_{3Q}(\mathbf{r}_{1},\mathbf{r}_{2},\mathbf{r}_{3}) \, S(\boldsymbol\rho)\,\rangle } {\langle\, \mathcal{P}_{3Q}(\mathbf{r}_{1},\mathbf{r}_{2},\mathbf{r}_{3})\,\rangle\, \,\langle S(\boldsymbol\rho)\, \rangle}
	}.
	\label{eq:baryonic_Correlator}
\end{equation}
Due to the cluster decomposition of the operators, $\mathcal{C}$ in Eq.\eqref{eq:mesonic_Correlator}and Eq.\eqref{eq:baryonic_Correlator} should be approaching a value that is $\mathcal{C}\simeq 0$ away from the interquark space.

\subsection{Lattice parameters and Monte-Carlo updates}
The potential and action density are measured on a set of $SU(3)$ pure gauge configurations. For a coupling value of $\beta = 6.0$, the configurations are generated using the standard Wilson gauge action \cite{PhysRevD.10.2445} on two lattices with a spatial volume of $36^{3}$. The string tension of value, $\sqrt\sigma_{0} = 440$ MeV~\cite{Bali:1994de, Sommer:1993ce,Luscher:1984xn} is applied to reproduce the value of the lattice spacing of $a = 0.1$ fm. 

The action is chosen such that the Monte-Carlo simulations with the typical Wilson gauge action guarantee locality~\cite{Luscher:2001up} which reduces cross correlation among adjacent locations.

The two lattices  temporal extents are $N_{t} = 8$ and $N_{t} = 10$. These temporal lengths correspond to temperatures $T/T_{c} = 0.9$ and $T/T_{c} = 0.8$, respectively. The latter temperature lies near the end of  the plateau of QCD phase diagram such that of the quark-gluon condensate~\cite{QCDPhase} or the string tension~\cite{PhysRevD.85.077501}. 

The $SU(3)$ gluonic gauge configurations have been generated employing a pseudo-heat bath algorithm ~\cite{Fabricius:1984wp, Kennedy:1985nu} updating of the corresponding three $SU(2)$ subgroup elements \cite{Marinari:1982}. Each update step consists of one heat bath and five microcanonical reflections.

The autocorrelation time is reduced as a result of mixing the heat-bath and overrelaxation/microcanonical steps~\cite{QCD-TARO:1992iba}. That is, the use of microcanonical reflections would help to produce less correlated configurations through Monte-Carlo time.
\begin{figure}[t]
	\centering
	\includegraphics[scale=0.38]{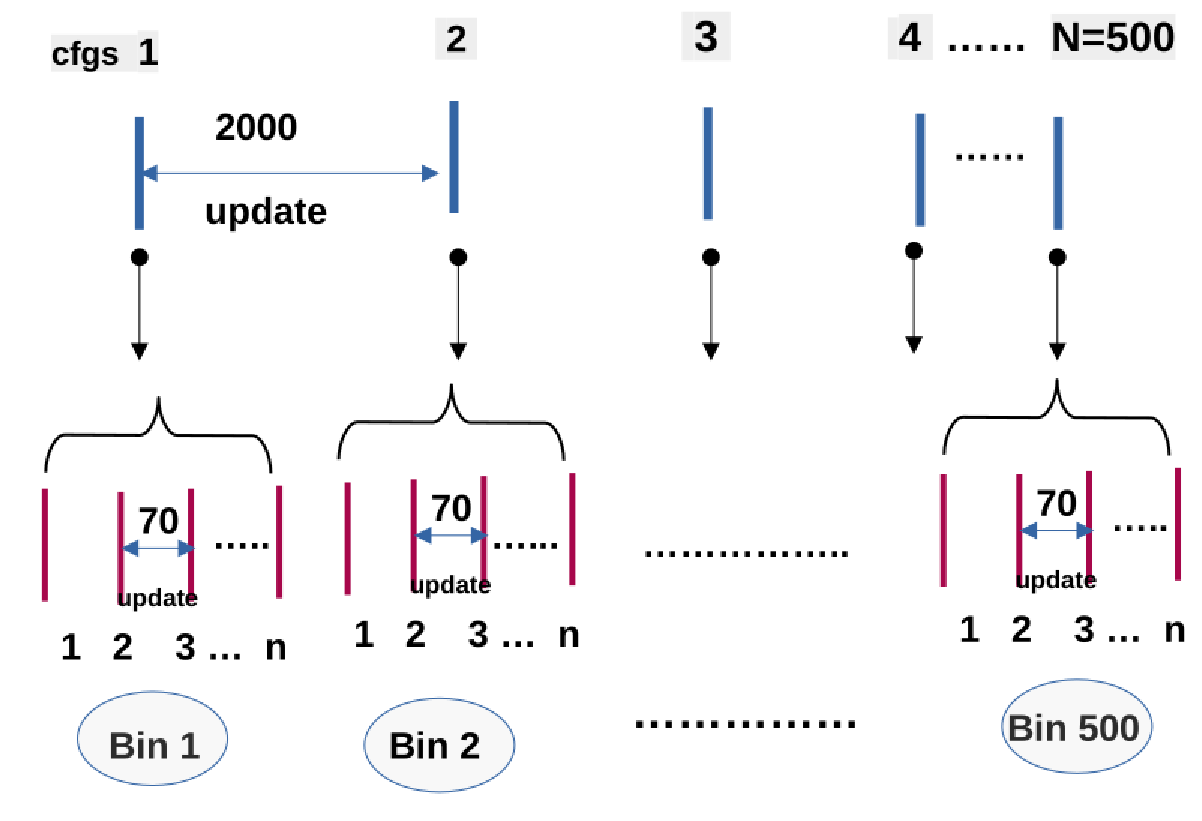}
	\caption{Schematic diagram illustrating the hierarchical updating of gauge configurations and measurements. Vertical line segments denote configuration and horizontal arrows denote skipped updates }
	\label{fig:Updating_Scheme}
\end{figure}

After the thermalization of the gauge configurations, a set of $N=500$ configuration is generated. The number of Monte-Carlo updates between the configurations is 2000 updating sweeps.

Each configuration of the set $N=500$ is updated 70 sweeps then a measurement is taken. This process is repeated $n$ times, then the resultant $n$ measurements are averaged and binned together into a single bin entry. The jackknifed standard deviation is thus calculated with $N=500$ bins. The illustration of the described scheme is shown on Fig.\ref{fig:Updating_Scheme}.

The measurements correspond to evaluating Polyakov lines correlators Eqs.\eqref{eq:QantiQ} and \eqref{eq:QQQ_potential} for the meson and baryon respectively, in addition to evaluating the corresponding action-density correlations of both systems Eqs.\eqref{eq:mesonic_Correlator} and \eqref{eq:baryonic_Correlator}, respectively. 

The correlators of $Q\bar{Q}$ system are evaluated on bins inclusive of $n=6$ updates, on the other hand, the baryonic correlators $(QQ)Q$ are evaluated on bins containing $n=20$ updated configurations.

The measurements at color source separation $R \ge 12a$ are disregarded for a careful investigation due to possible correlations from the opposite side of the lattice generated by the periodic boundary of the 3D torus.

To reduce the noise from the signal, we make use of translational invariance  by computing the correlation on every node of the lattice, averaging the results over the volume of the 3D torus, in addition, to averaging the action measurements taken at each time slice in Eq.\eqref{eq:average}.

The volume of lattices employed in this investigation  are chosen reasonably large in order to gain high statistics in a gauge-independent manner and also minimize the mirror impacts and correlations across the boundaries ~\cite{Bali:1994de,Bissey:2005sk}. These effects will be investigated below.
\begin{figure}[t]
	\centering
	\includegraphics[scale=0.45]{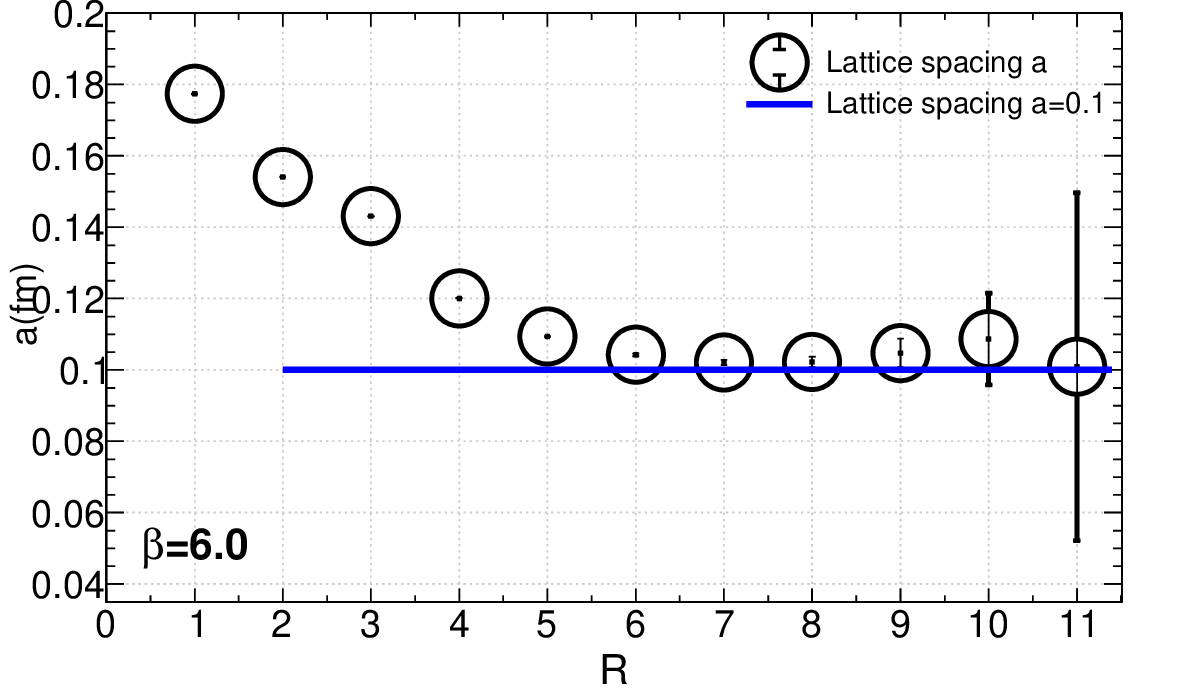}
	\caption{The lattice spacing versus separation between two Polyakov loops measured in accord with Eq.\eqref{LS} on an ensemble of 100 gauge configuration.}
	\label{fig:Lattice_spacing}
\end{figure}

We examine the local string tension and translate it to the lattice spacing $a$ in Fermi units, 
\begin{align}
	& \alpha(\Lambda) \, e^{- \sigma R L_T}  = \langle P(0) P^{\dagger}(R) \rangle, \\			
	& \sigma a^2   = - \frac{1}{L_{T}}\log \left[ \frac{\langle P(0) P^{\dagger}(R+1)\rangle}{\langle P(0) P^{\dagger}(R) \rangle}    \right].
\end{align} 
The lattice spacing in Fermi units is then
\begin{equation}
	a = \frac{0.1973 \rm{GeV\, fm} }{\sqrt{0.440} \rm{GeV}}  \left[ -\frac{1}{ L_{T}}\log \left( \frac{\langle P(0) P^{\dagger}(R+1)\rangle}{\langle P(0) P^{\dagger}(R)\rangle }\right)  \right]^{\frac{1}{2}},
	\label{LS}
\end{equation}
on each gauge configuration. To enhance the statistics in a gauge-independent way, the aforementioned expectation values are computed as the average across all lattice points.
\begin{figure}[t]
	\centering
	\includegraphics[scale=0.35]{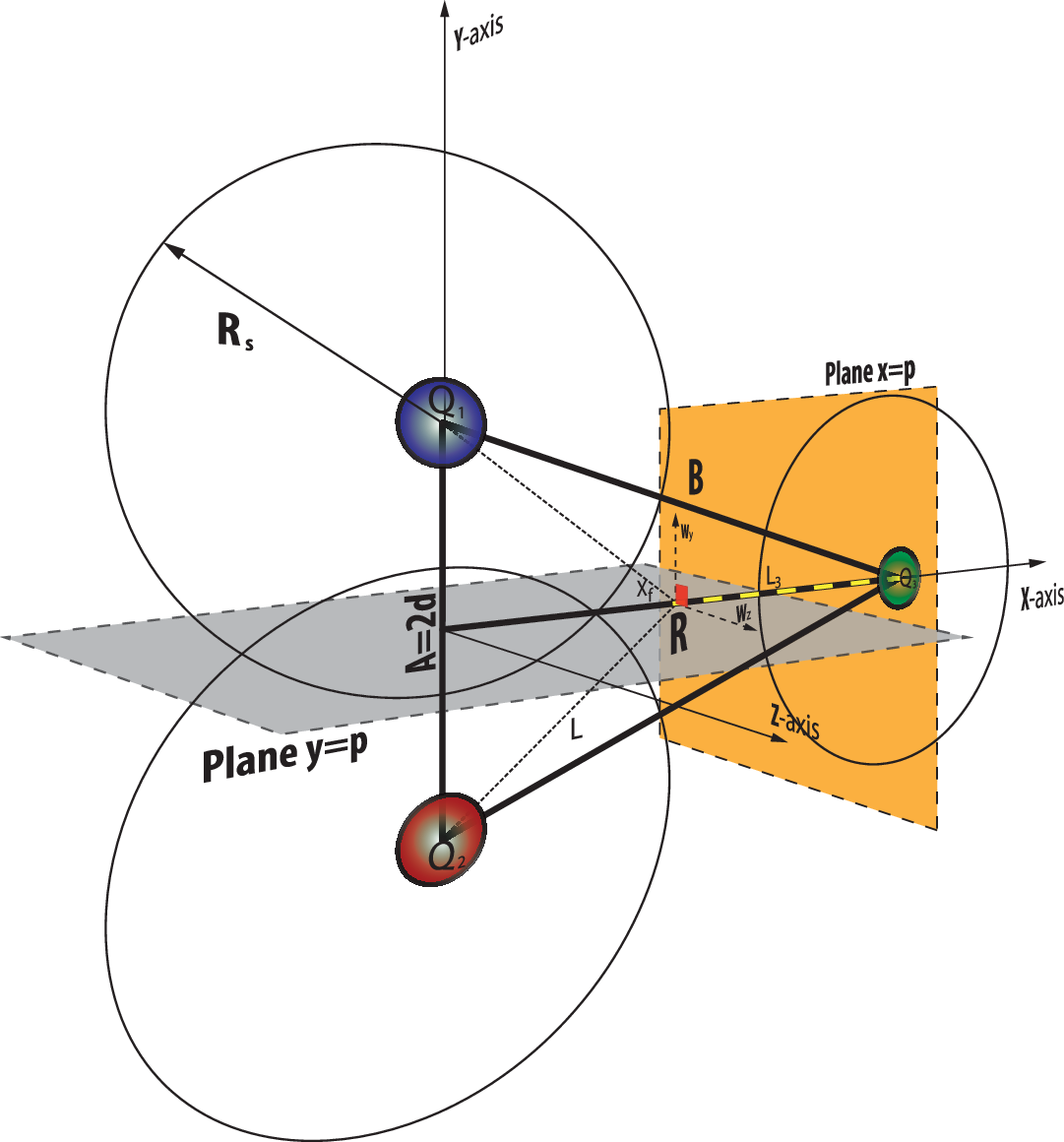}
	\caption{
		The geometry of the three well-separated quarks on the nutshell. The large spheres represent the motion of the diffused field of a characteristic smearing radius of $R_{s}$ centered at the quarks (small spheres).
	}
	\label{fig:QQQ_config}
\end{figure}
\begin{figure*}[!hpt]
	\centering		
	\subfigure[]{\includegraphics[scale=0.42]{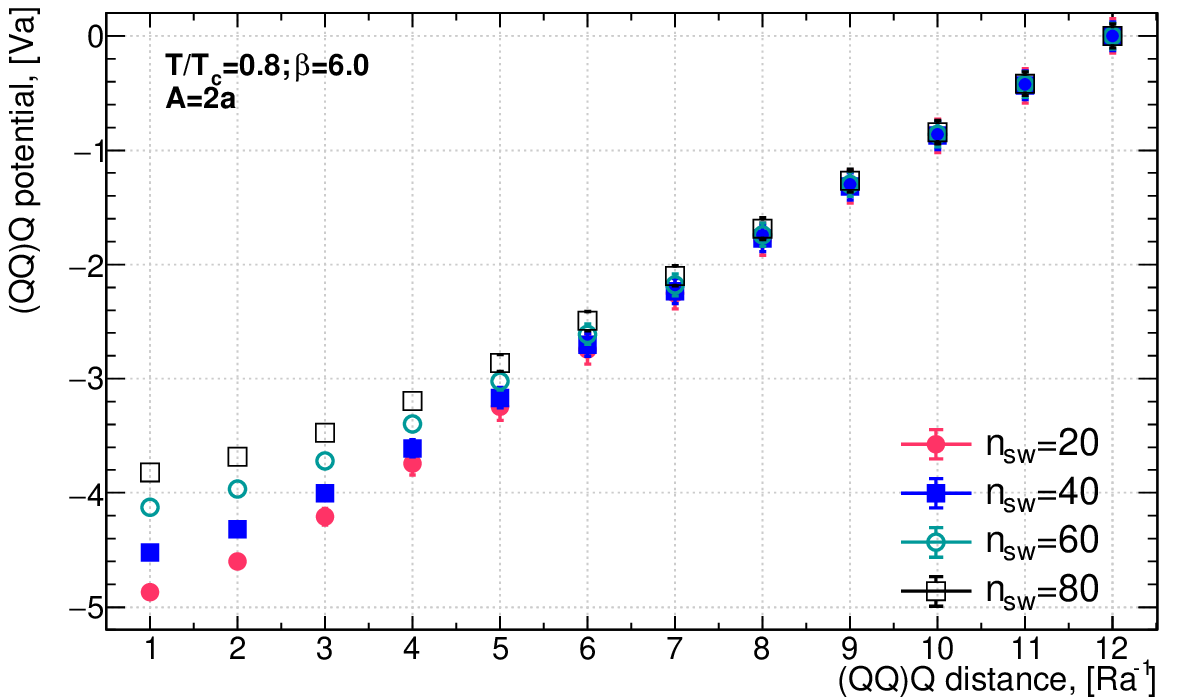}}
	\subfigure[]{\includegraphics[scale=0.42]{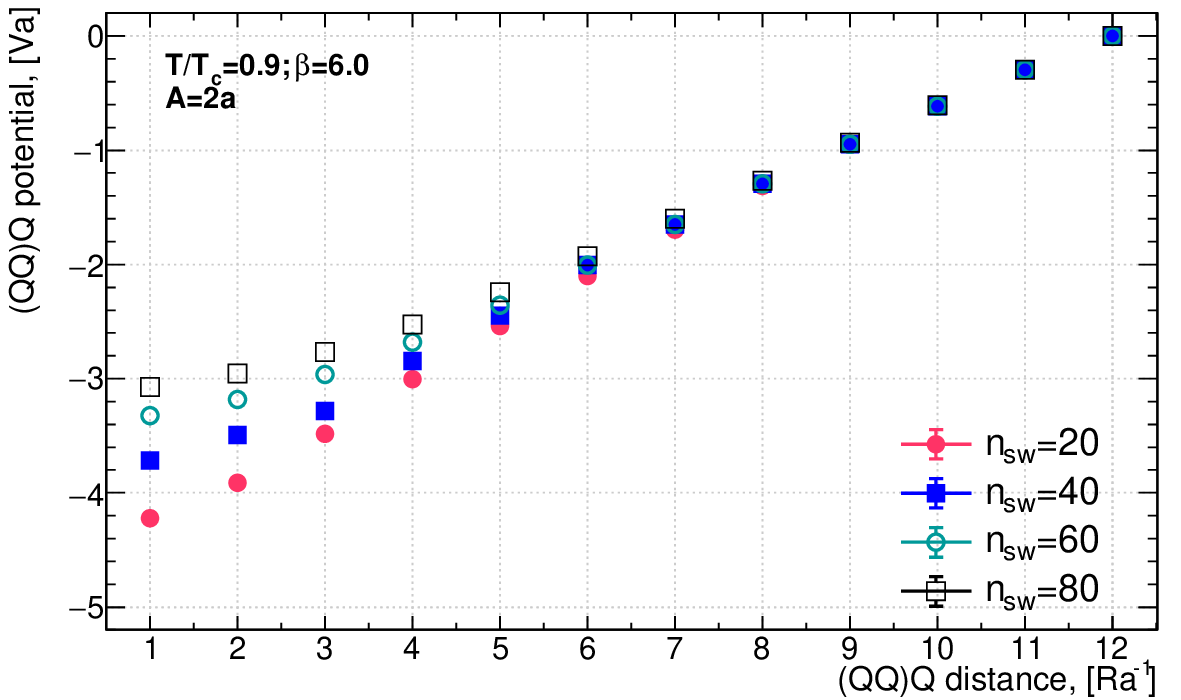}} 
	\caption{
		The $(QQ)Q$ potential of a planar  arrangement (Fig.\ref{fig:QQQ_config}) with base diameter $A=2a$ versus the $(QQ)-Q$ distance $R$ measured at the depicted cooling levels (a) temperature $T/T_{c}=0.8$, (b) temperature $T/T_{c}=0.9$.
	}
	\label{fig:3Qlevels}	
\end{figure*}

The lattice spacing in Eq.\eqref{LS} is an observable which depends on measurements of operators that are placed at two adjacent spatial positions on the lattice. That is, the expectation value of the difference between the logarithm of the two correlators is estimated. The statistical measurement of this observable is likely to disclose whether the two correlators fluctuate comparatively. The average over an ensemble that is correlated may bring about overestimated error bars. This would lead to incorrect measurement of lattice spacing at large $R$.

In Fig.\ref{fig:Lattice_spacing} the physical lattice spacing, $a$, measured through formula Eq.~\eqref{LS} with $L_T=10a$ is plotted versus the $Q\bar{Q}$ separation $R$. The error bars are calculated on an ensemble of 100 configurations \cite{Ph}(from first level as shown in Fig.\ref{fig:Updating_Scheme}). The line corresponds to the lattice spacing $a=0.1$ fm, considering the results at $R>5a$ the lattice spacing can be extracted with good accuracy as $a=0.098(2)$ which reproduces the spacing obtained, for example, in Ref.~\cite{Luscher:2002qv} at the same $\beta$.

\subsection{Cooling method}
\label{sec:UV filtering}
Our measurements of the action density distribution across the lattice are preceded by an ultraviolet (UV) filtering procedure. To achieve a decent signal-to-noise ratio in the aforementioned correlations, the UV-filtering of the gauge configurations suppresses the short-range quantum fluctuations of the vacuum.

At large source separations, it was demonstrated that the effective string physics in the heavy meson are independent of the UV fluctuations ~\cite{Bakry:2011zz}. With careful choice of the number of cooling sweeps, it can be shown that the lattice data compare with the predictions of the free string model at the intermediate and large source separation distance at high temperatures~\cite{Bakry:2010sp}.

As an alternative to ~\cite{Thurner1999, Feurstein:1997rz}, who employed the Cabbibo-Marinari cooling~\cite{Marinari:1982}, we have chosen to cool the gauge field using a stout-link algorithm ~\cite{Morningstar:2003gk}. Filtering techniques of this type are categorized within the set of so-called analytic link-blocking methods.

In the present analysis, we adopt a smearing parameter of value $\rho=0.06$ in the standard stout-link cooling~\cite{Morningstar:2003gk}. To lessen cross-correlations between adjacent lattice measurements we renounce using improved versions of stout-link algorithms which employ large-sized operators.

The link-fuzzing approach is comparable to the Brownian motion of a dispersed sharp field ~\cite{Takahashi:2002bw}. The diffused field is Gaussian distributed around a sphere centered at point $r$ evolving in the smearing time $\tau$ as described in~\cite{Takahashi:2002bw}. Figure \ref{fig:QQQ_config} schematically represent this $(QQ)Q$ system along with the accompanying blurring spheres.
\begin{table}[!t]
	\begin{tabular}{c|cccccc}
		Number of sweeps $n_{\rm{sw}}$    &20   &40   &60   &80\\
		\hline
		Characteristic radius $R_{s}$ &0.27 &0.38 &0.47 &0.54 
	\end{tabular}
	\caption{
		The characteristic smearing radius $R_{s}$ at each smearing level $n_{\rm{sw}}$.
	}	
	\label{tab:ChRadius}	
\end{table}

The Gaussian diffuseness model of the smearing procedure establishes a characteristic radius that can scale the effects of each smearing level. Table~\ref{tab:ChRadius} collects the characteristic radius of the Brownian motion at each selected number of sweeps. That is, one can find a mapping between the smearing radii and the distances at which smearing effects can be neglected. This distance scale could be established by a careful examination of a lattice observable such as the confining potential among $3Q$ system.

The confining potential of the $(QQ)Q$ system Eq.\eqref{eq:QQQ_potential} is tested by taking measurements over four ensembles of cooled gauge configurations. These correspond to configurations with varying cooling levels   $n_{\rm{sw}}=20, 40, 60$, up to $80$ sweeps. 

The potential is plotted versus the $QQ-Q$ distance $R$ in Fig.\ref{fig:3Qlevels}. 
The potential is normalized with respect to a fixed point such as $R=12a$. 

The two quarks $Q_1$ and $Q_2$ that make up the diquark system at the basis are interspersed with overlapping patches as a result of the link-fuzzing procedure (see Fig.\ref{fig:QQQ_config}) which we found that it has negligible impact on the potential measurement.

Further inspection of the potential of the $(QQ)Q$ at the two considered temperatures, in Fig.\ref{fig:3Qlevels}(a,b), reveals that the numerical outcomes measured after 40 cooling sweeps can be identified for $(QQ)-Q$ distance $R \geq 4a$  (almost identical within the uncertainty of measurements). Similarly, $n_{\rm{sw}} \leq 60$  cooling sweeps at $T/T_{c}=0.8$ leave the $(QQ)Q$ potential intact on distance $R \geq 5a$.

The above observational outcomes can be related to the cooling radii listed in Table~\ref{tab:ChRadius}. At both temperatures, the potential measured at $(QQ)-Q$ distance $R \geq R_{s}$ receives a minimal impact of cooling at the number of sweeps $n_{\rm{sw}}$.

In the subsequent analyses, we evaluated measurements on two sets of ensembles corresponding to two levels of cooling sweeps. Because the signal-to-noise ratio decreases with temperature, we have had to use two different smearing levels for each temperature scale in the evaluation of the action-density correlation functions. That is,  $n_{\rm{sw}} = 40$ sweeps at $T /T_c = 0.9$  and $n_{\rm{sw}} = 60$ sweeps at $T/T_c = 0.8$.

Nevertheless, the correlation functions relevant to the potential operators, which are less noisy, have been analyzed considering gauge links subject to $n_{\rm{sw}}= 40$ sweeps at both temperatures all over the present analysis, unless otherwise stated.

Many lattice measurements are affected by renormalization effects~\cite{Peter} via lattice spacing, which creates a natural cutoff for the underlying QFT. Cutoff and renormalization effects are entangled such that an energy scale inverse to the lattice spacing is set; hence, any change in the scale also affects the cutoff. A shift in the cutoff causes the width to be measured with a constant overall offset~\cite{Bakry:2017fii}.

This deduction is supported by the observation of an identical impact on the width of the profile, Eqs.\eqref{eq:mesonic_Correlator} and \eqref{eq:baryonic_Correlator}, when a different number of cooling sweeps are applied, owing to increasing lattice spacing or modifying the UV cutoff scale. In the configuration under scrutiny, we investigate the behavior of the IR quantum broadening of the width, which are not impacted~\cite{Bakry:2010sp} by the identical global shifts.

\section{Diquark-quark (QQ)Q potential}
\label{sec:QQQPotential} 
To determine the heavy quarks potential of the $Q\bar{Q}$ and that for $(QQ)Q$ systems, we evaluate and analyze the correlators Eq.\eqref{eq:QQ_correlators} corresponding to the meson and Eq.\eqref{eq:QQQ_potential} of the baryon. 
The baryonic arrangement corresponds to isosceles triangles with bases $A$ and diquark-quark distance $(QQ)-Q$ distance $R$, as shown in the schematic Fig.\ref{fig:gam}. 
\begin{figure*}[!hptb]
	\centering
	\subfigure[]{\includegraphics[scale=0.4]{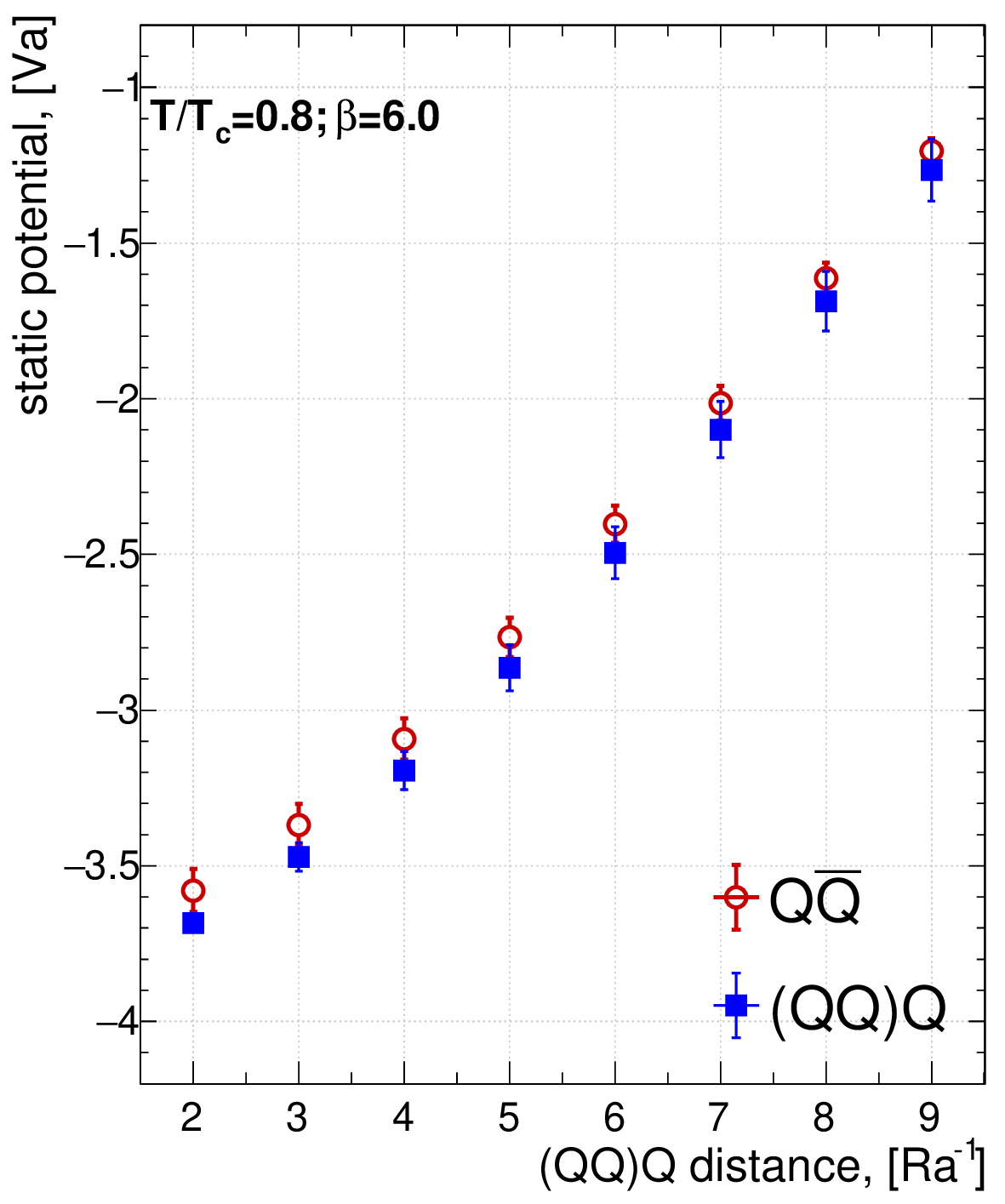}\label{fig:effpotsT08}}
	\subfigure[]{\includegraphics[scale=0.4]{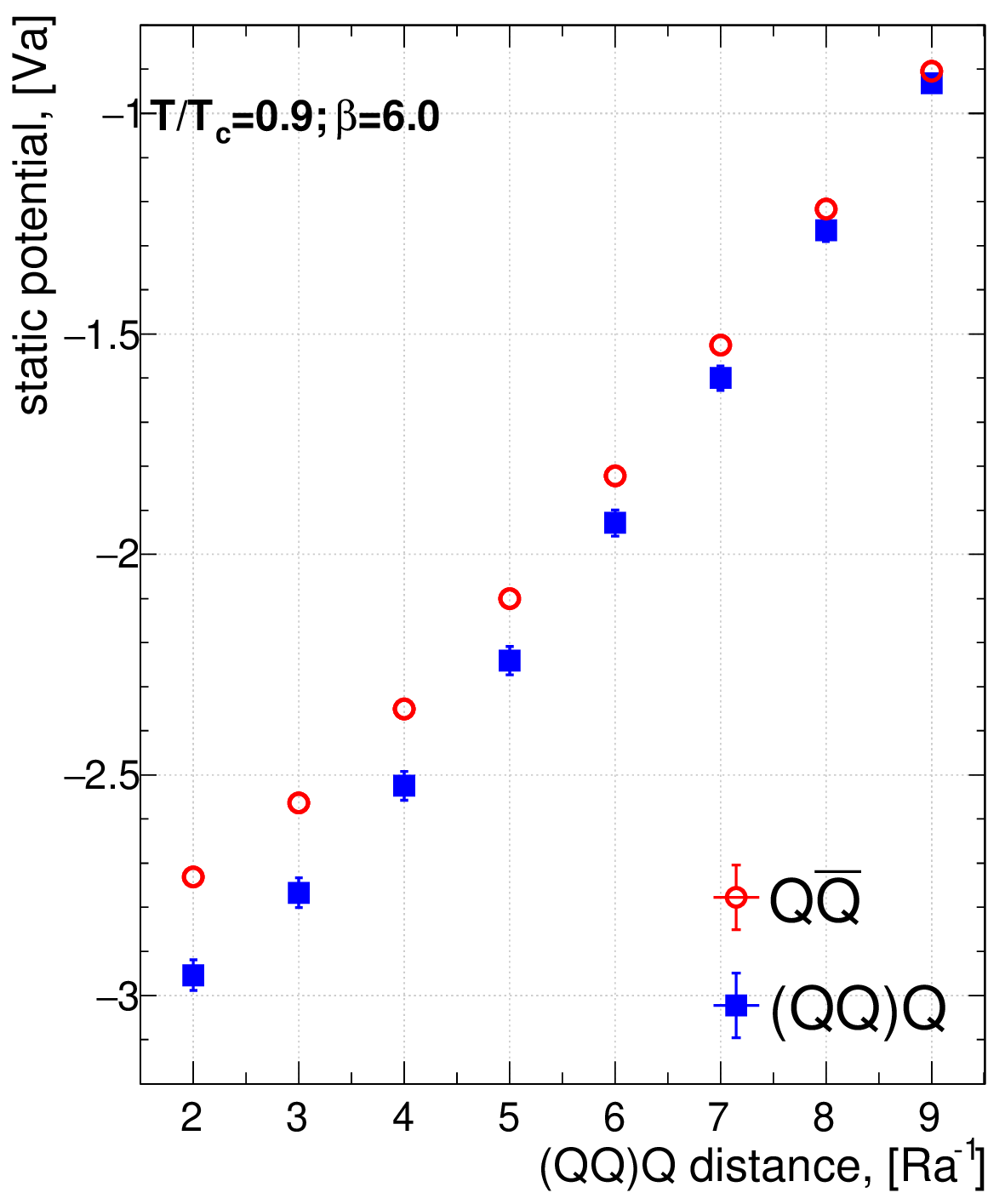}\label{fig:effpotsT09}}
	\caption{
		(a) Compares the static potential of quark-antiquark $Q\bar{Q}$ and diquark-quark $(QQ)Q$ configuration of base length $A=2a$ at $T/T_{c}=0.8$ and $\beta = 6.0$. (b) Compares of the static potential of quark-antiquark $Q\bar{Q}$ and diquark-quark $(QQ)Q$ configuration of base length $A=2a$ at $T/T_{c}=0.9$ and $\beta = 6.0$.
	}
	\label{fig:effpots}
\end{figure*}

The potential data of the static meson $Q\bar{Q}$ and the diquark-baryon $(QQ)Q$ are plotted in Fig.\ref{fig:effpots} at the two close temperatures $T/T_{c} = 0.8$ in (a) and $T/T_{c} = 0.9$ in (b). The comparison in Fig.\ref{fig:effpots} depicts the $(QQ)Q$  potential as the quark $Q_3$ is pulled a distance $R$ apart from the diquark $(Q_2,Q_3)$  at the base. The diquark  diameter is set to $A=2a$. 

It is interesting to find that while comparing the $Q\bar{Q}$ and $(QQ)Q$ systems with gauge connections subjected to the same number of cooling sweeps, the potential of the two systems conformally change with respect to each other. 
That is, the data behaves similarly and preserve the potential differences at a given distance $R$. To clearly contrast the quark systems, only in Fig.\ref{fig:effpots} the rendered potential is evaluated on gauge ensemble cooled $n_{\rm{sw}}=80$ sweeps.

At the temperature, $T/T_c=0.8$, the $(QQ)Q$ system in Fig.\ref{fig:effpotsT08} exposes an identical potential to the mesonic string. 
At this temperature, thermal factors only account for around $10\%$ of the decrease in string tension ~\cite{Kac}. This outcome coincides with the result in Ref.\cite{Koma:2017hcm} which displays $(QQ)Q$ identical to the meson configuration at a much lower temperature using Polyakov loops of length $N_t=20$ time slices and lattice at the same coupling $\beta=6.0$.

The potential corresponding to each system, however, differs noticeably close to the deconfinement point $T/T_c=0.9$. The remarkable findings displayed in Fig.\ref{fig:effpotsT09}  is that the $(QQ)Q$, at either short or intermediate distance scales $R<10a$, does not manifest the similarity in the confining potential with the $Q\bar{Q}$ system.

Thus, we discovered a splitting of the identical confining force in the bosonic and fermionic arrangements that do occur with the approach of the temperature scale the deconfinement point from below. At this point, one is naturally inclined to question both the mesonlike and baryonlike aspects of each system. 

The effective bosonic string model is a well-suited framework to further explore and assess the lattice data. Within this paradigm, for example, one can pose the question: If the meson-like gluonic field of the diquark is excited, would a crossover into the Y-junction behavior take place? 

In the baryon, the analysis of the lattice data would suggest two types of parametrization depending on the interquark distances, i.e., the $\Delta$ and Y-type potentials.~\cite{Takahashi:2000te, Takahashi:2002bw, Alexandrou:2001ip, Alexandrou:2003ip}. The $\Delta$-potential describes a sum of two-body forces and is proportional to the perimeter of the $3Q$ triangle with a string tension half that of the corresponding $Q\bar{Q}$ system. The $\Delta$-potential is given by
\begin{equation}
	V_{\rm 3Q}(\mathbf{r}_{1},\mathbf{r}_{2},\mathbf{r}_{3})=\dfrac{-A_{Q\bar{Q}}}{2} \sum_{i<j}\frac1{|\mathbf{r}_{i}-\mathbf{r}_{j}|}+\frac{\sigma_0}{2} \sum_{i<j} |\mathbf{r}_{i}-\mathbf{r}_{j}|+\mu_c,
	\label{eq:delta_ansatz}
\end{equation}
with $A_{Q\bar{Q}}$ signifying the strength of the one-gluon-exchange (OGE) Coulomb term derived from perturbative QCD (see Ref.~\cite{Takahashi:2000te,Takahashi:2002bw,Bazavov98}).
\begin{figure}[!hpt]
	\centering		
	\includegraphics[scale=0.3]{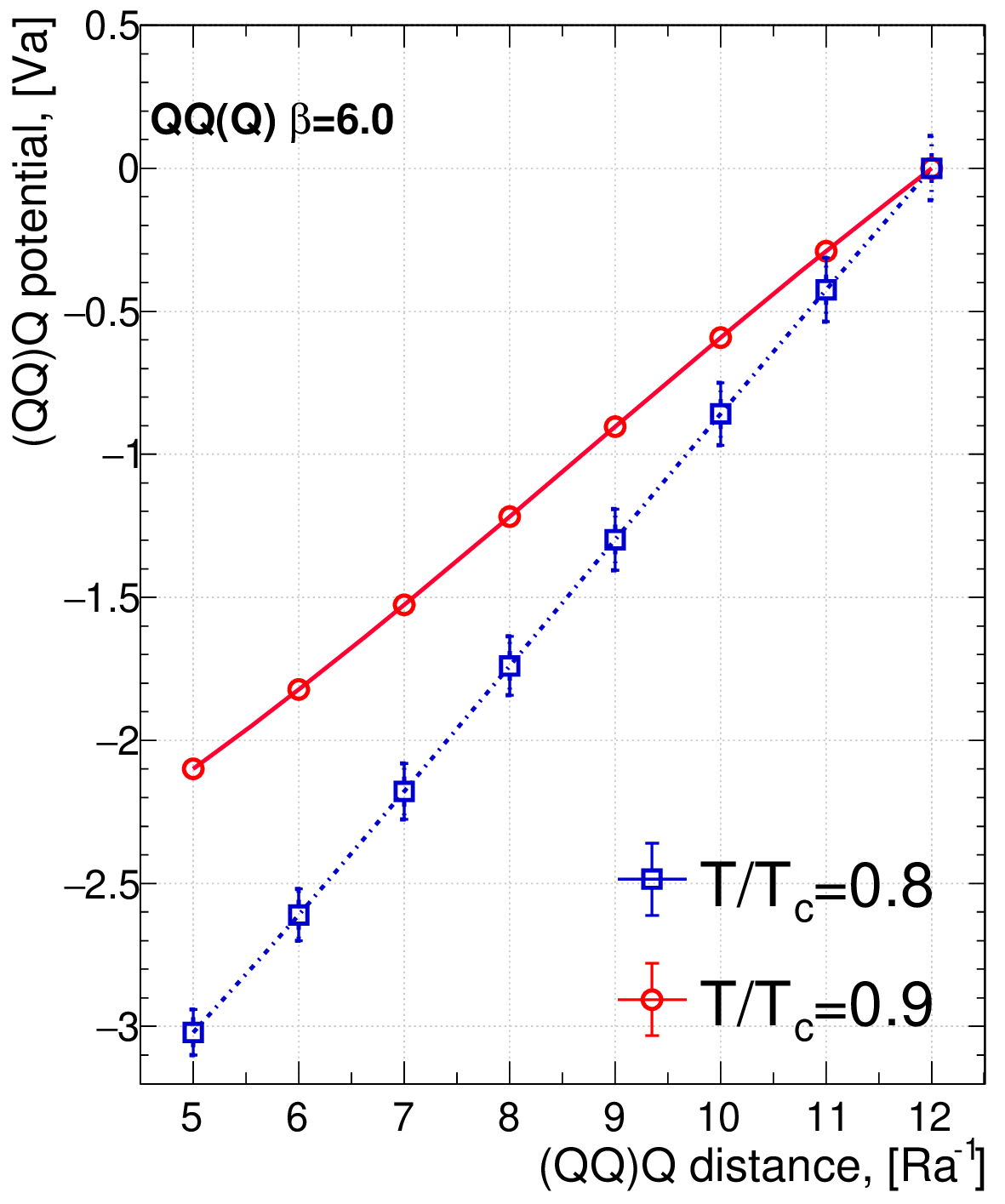}        
	\caption{The $3Q$ potential of planar isosceles configurations of a base length $A = 8a$. The two lines correspond to the best fits to baryonic Y-string Eq.\eqref{eq:vba1} at two temperature scales $T/T_c=0.8$ and $T/T_c=0.9$ over fit  range $R \in [5a, 12a]$.}
	\label{fig:YandDT0908}
\end{figure} 

We systematically examine each model on a selected $3Q$ configuration with a significant separation between the two quarks $Q_1$ and $Q_2$. Considering the data corresponding to the potential of an isosceles $3Q$ quark configuration with base length $A=6a$  at the highest temperature $T/T_{c}=0.9$.
\begin{table}[t]
	\begin{center}
		\begin{tabular}{c|cc|cc|cc}
			\hline
			\hline
			Fit Range           
			&\multicolumn{2}{c}{$R\in [4a-12a]$}
			&\multicolumn{2}{c}{$R \in [5a-12a]$}
			&\multicolumn{2}{c}{$R \in [6a-12a]$}\\
			\hline
			Fit Parameters   
			&$\chi^{2}$ &$\sigma_{0}a^{2}$
			&$\chi^{2}$ &$\sigma_{0}a^{2}$ 
			&$\chi^{2}$ &$\sigma_{0}a^{2}$\\
			\hline
			$\Delta$-ansatz  &22.4 &0.352(3)   &13.0  &0.341(5)  &6.58 &0.329(5)\\
			Y-string model   &5.68 &0.319(6)   &5.17  &0.317(6)  &3.39 &0.311(2)\\
			\hline
			\hline
		\end{tabular}
		\caption{The returned $\chi^{2}(x)$ for fits of the lattice data to $3Q$ isosceles of width $A=6a$  at $T/T_{c}=0.9\,$. The fits compare Eq.\eqref{eq:delta_ansatz} for the $\Delta$-ansatz, Eq.\eqref{eq:vba1} for the Y-string model.}
		\label{tab:T}	
	\end{center}
\end{table}
\begin{figure*}[!hpt]
	\centering
	\subfigure[]{\includegraphics[scale=0.35] {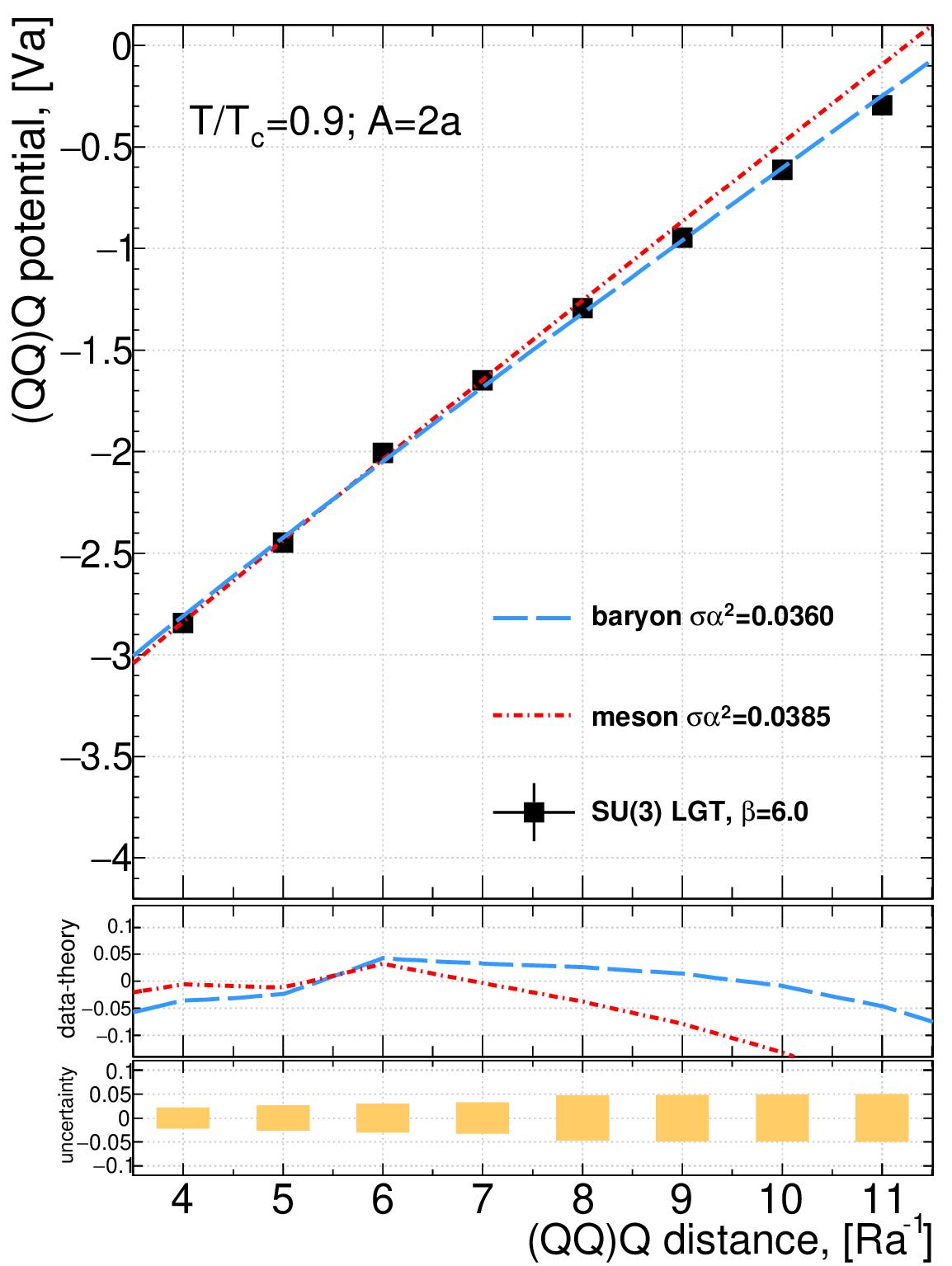}}
	\subfigure[]{\includegraphics[scale=0.35] {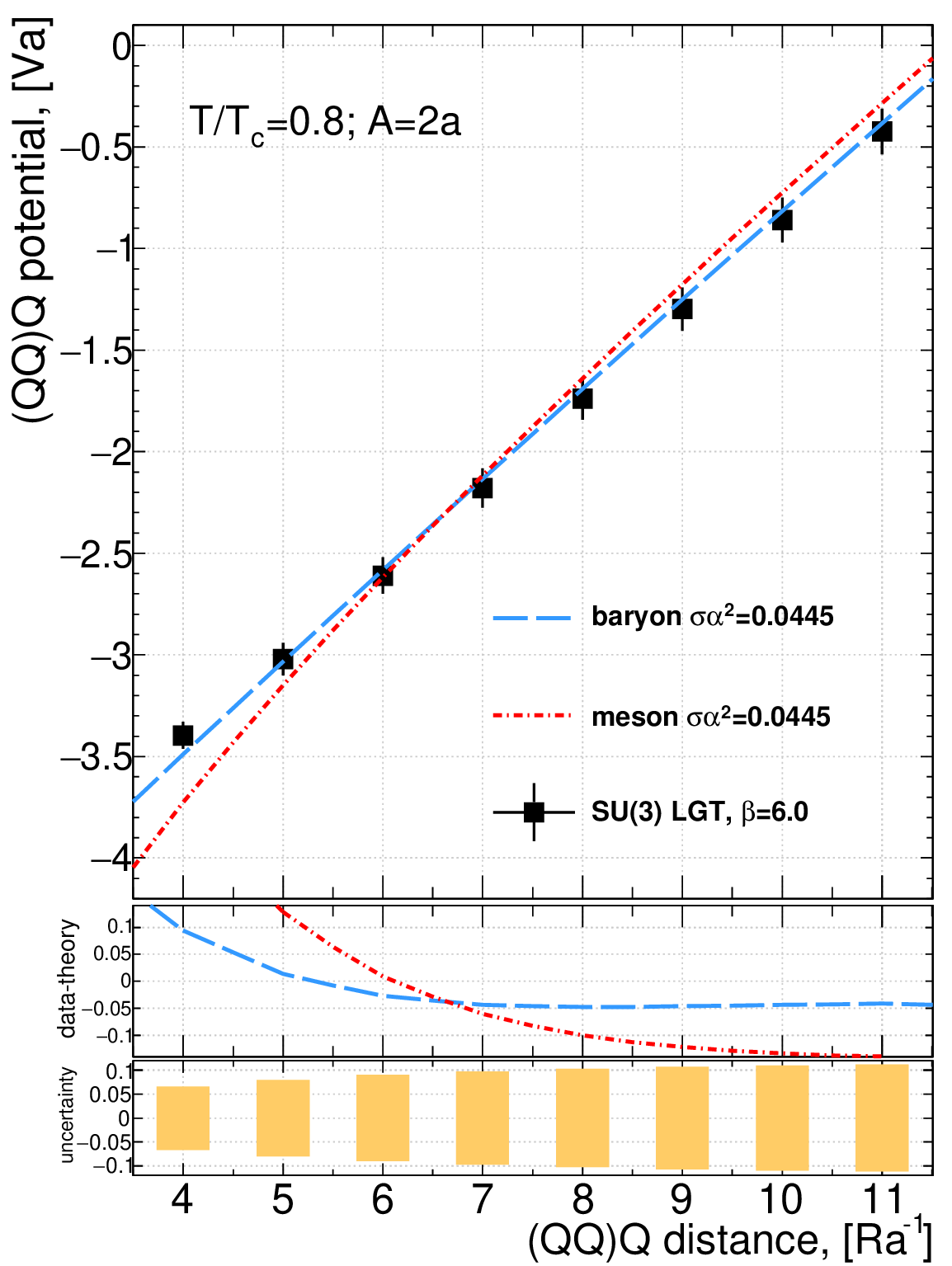}}
	\caption{(a) The $(QQ)Q$ potential at base width $A=2a$. The dashed and dotted lines compares the fit of Eq.\eqref{eq:total} of the baryonic Y-string to Eq.\eqref{eq:Pot_NG_LO} of the mesonic string potential. (b) The same as (a); however, the temperature scale is set to $T/T_{c}=0.8$.
		Bottom pads show the data-theory residuals, and the lattice data uncertainties respectively.
	} 
	\label{fig:p2}
\end{figure*}

Before proceeding with fitting the baryonic Y-string potential to the lattice data, the Y-string’s length could be minimized, i.e., setting the node's position at the Fermat point of the $3Q$ triangular configuration.
Using elementary variational calculus, the position of the Fermat point of the planar isosceles arrangement may be simply proven to reside at the point $x_f=\dfrac{A}{2 \sqrt{3}}$(see Fig.\ref{fig:gam}).

At the outset, the limits where the long string approximation is valid ought to be elucidated. The match of the free mesonic string with the lattice data suggests a minimal distance corresponding to $R=0.5$ fm~\cite{Bakry:2020flt}. In the same context the isosceles $3Q$ geometry with $A=2a,4a$ assumes a minimal total Y-string length $L_{Y}=R+\frac{\sqrt{3}A}{2}$ which offsets the limit to $R=0.4$ fm at the smallest base $A=2a$.

In Table~\ref{tab:T} we collected the returned values of $\chi^{2}$ from the resultant fits of the $3Q$ potential to the Y-string model formula Eq.\eqref{eq:vba1}  and $\Delta$-model Eq.\eqref{eq:delta_ansatz}. The string tension has been taken as a fit parameter together with ultraviolet (UV)-cutoff $\mu_c$.

Inspection of $\chi^{2}$ all over the fit intervals subsets from $R \in [4a, 12a]$, reveals that the Y-string model appears to provide the best match compared to the $\Delta$-model. A  physical realization of this observation is that the 
confining force in the baryon appears to be consistent
with a three-body force proportional to the total length of the Y-string and its subleading terms owing to the junction fluctuations. 

The good fit displayed in Fig.\ref{fig:YandDT0908}, represents the $3Q$ potential of an isosceles triangle shape with a wider base length $A=8a$. The lines are the best fits of the Y-string formula Eq.\eqref{eq:vba1} to the $3Q$ potential data, at the two considered temperature scales.
\begin{table}[!t]
	\begin{center}	
		\begin{tabular}{c|cc|cc|cc}
			\hline
			\hline
			Fit range 
			&\multicolumn{2}{c|}{$[4a-12a]$}
			&\multicolumn{2}{c|}{$[5a-12a]$}
			&\multicolumn{2}{c}{$[7a-12a]$}\\
			Fit parameters&$\chi^2$ &$\sigma_{0} a^{2}$&$\chi^{2}$&$\sigma_{0} a^{2}$&$\chi^2$ &$\sigma_{0} a^{2}$  \\
			\hline
			\tiny{\bf{Meson String}} &   &    &     &          &      &       \\
			$A=2a$     &14.4 &0.0361(4) &8.62 & 0.0353(6) & 0.86 &0.033(1)\\ 
			$A=4a$     &3.24 &0.0302(6) &2.85 & 0.0299(7) & 0.96 &0.029(1) \\
			\hline
			\hline
		\end{tabular}	
		\caption{The string tension and $\chi^{2}$ form the fit of $(QQ)Q$ potential to the meson string Eq.\eqref{eq:Pot_NG_LO} at $T/T_{c}=0.9$.}
		\label{T9M}
	\end{center}
\end{table}

\begin{table}[!t]
	\begin{center}	
		\begin{tabular}{c|cc|cc|cc}
			\hline
			\hline
			Fit range 
			&\multicolumn{2}{c|}{$[4a-12a]$}
			&\multicolumn{2}{c|}{$[5a-12a]$}
			&\multicolumn{2}{c}{$[7a-12a]$}\\
			Fit parameters&$\chi^2$ &$\sigma_{0} a^{2}$&$\chi^{2}$&$\sigma_{0} a^{2}$&$\chi^{2}$&$\sigma_{0} a^{2}$  \\
			\hline
			\tiny{\bf{Meson String}}           &   &    &   &&\\
			$A=2a$      &3.04   &0.045(1)  &0.39 &0.045(1) &0.2  &0.044(3)  \\ 
			$A=4a$      &4.21   &0.040(5)  &7.68 &0.044(5) &0.6  &0.043(3) \\
			\hline
			\hline
		\end{tabular}	
		\caption{The string tension and $\chi^{2}$ form the fit of $(QQ)Q$ potential to the mesonic string Eq.\eqref{eq:Pot_NG_LO} at $T/T_{c}=0.8$.}
		\label{T8M}
	\end{center}
\end{table}

\begin{table}[!t]
	\begin{center}	
		\begin{tabular}{c|cc|cc|cc}
			\hline
			\hline
			Fit range 
			&\multicolumn{2}{c|}{$[4a-12a]$}
			&\multicolumn{2}{c|}{$[5a-12a]$}
			&\multicolumn{2}{c}{$[7a-12a]$}\\
			Fit parameters&$\chi^2$ &$\sigma_{0} a^{2}$&$\chi^{2}$&$\sigma_{0} a^{2}$&$\chi^2$ &$\sigma_{0} a^{2}$\\
			\hline
			\tiny{\bf{Baryon String}} &  &  &  &  &  &\\
			$A=2a$ &11.1  &0.358(4) &7.22  &0.351(6)  &0.73 &0.34(1)\\ 
			$A=4a$ &2.02  &0.295(6) &2.01  &0.295(7)  &0.80 &0.28(1)\\
			\hline
			\hline  
		\end{tabular}	
		\caption{The string tension and $\chi^{2}$ from the fit of $(QQ)Q$ potential to the baryonic Y-string  Eq.\eqref{eq:vba1} at $T/T_{c}=0.9$.}
		\label{tab:T09B}
	\end{center}
\end{table}

The Y-string model, thereof, comes out as the most suitable framework to delve into the characteristics of the three body forces of a given $3Q$ configuration. In what follows we oppose the Y-string model and its mesonic counterpart while dissecting the lattice data of  the $(QQ)Q$ system's potential.

On the account that we are interested in spotting the mesonic string signatures of the $(QQ)Q$ system, a fit of the mesonic string potential Eq.\eqref{eq:Pot_NG_LO} to the baryonic $(QQ)Q$ potential data has to be looked over. The outcome of the fit, $\chi^2$ and string tension $\sigma_{0} a^{2}$ values, are collected in Tables.~\ref{T9M} and \ref{T8M} corresponding to temperatures $T/T_{c}=0.9$ and $T/T_{c}=0.8$, respectively.

It is essential to mention beforehand that the fit of mesonic string potential to $Q\bar{Q}$ potential Eq.\eqref{eq:Pot_NG_LO} at $T/T_{c}=0.9$~\cite{Bakry:2020flt} returns a string tension of a value amount to $\sigma_{0} a^{2}=0.036$, this is smaller than  $\sigma_{0} a^{2}=0.044$ measured at $T=0$~\cite{Koma:2017hcm}. However, the fit of the mesonic string at the lower temperature $T/T_{c}=0.8$ reproduces the correct string tension $\sigma_{0} a^{2}=0.044$ as in Eq.\eqref{eq:Tension_NG_LO}. The free bosonic string model yields $\sigma_{0} a^{2}=0.044$, at the higher temperature $T/T_{c}=0.9$, only if other effects beyond the free string theory ~\cite{Bakry:2020flt} are included, such as self-interaction~\cite{PhysRevD.27.2944, Billo:2012da}, rigidity~\cite{Kleinert, Vis} and boundary effects~\cite{Billo:2012da,Bakry:2020flt}.

The values in Table.~\ref{T9M} indicates that the free string model do parametrize well the $(QQ)Q$ potential data at base length, $A=2a$ and $4a$, returning small $\chi^{2}$ on the considered distance scale. Despite this, it is important to highlight how the fit of the free string model to $Q\bar{Q}$ potential returns very large $\chi^{2}$ on most fit intervals~\cite{Bakry:2020flt} with increased string tension. Specifically, for the fit range $R\in[5a,12a]$, $\sigma_0 a^{2}=0.0385$.

On the other hand, the results in Table~\ref{T8M} of the mesonic string fits to the baryonic data at $T/T_c=0.8$ replicate the fits to $Q\bar{Q}$ potential data ~\cite{Bakry:2020flt}, since the data at this temperature coincide (Fig.\ref{fig:effpotsT08}). The model matches well the data and returns correct $\sigma_0 a^{2}=0.044$ for $(QQ)Q$ at both base diameters $A=2a,4a$.

The string tension and $\chi^2$ retrieved from the fit of the baryonic string potential Eq.\eqref{eq:vba1} to the $(QQ)Q$ potential data, at the temperature $T/T_c=0.9$ and $T/T_c=0.8$, are collected in Table~\ref{tab:T09B} and Table~\ref{tab:T08B}, respectively.

The data compare with the baryonic string model at both temperatures. The fit parameters collected in Table~\ref{tab:T09B} show that, for $(QQ)Q$ configuration with base diameters $A=2a$, the Y-string retrieves the smallest $\chi^{2}$ on the interval $R\in [7a, 12a]$ with $\sigma_{0}=0.033(1)$. The fit to the baryonic arrangement with $A=4a$ is less tight on all intervals returning smaller string tension $\sigma_{0}=0.0302(6)$ on $R\in [4a, 12a]$.

\begin{table}[!t]
	\begin{center}	
		\begin{tabular}{c|cc|cc|cc}
			\hline
			\hline
			Fit range 
			&\multicolumn{2}{c|}{$[4a-12a]$}
			&\multicolumn{2}{c|}{$[5a-12a]$}
			&\multicolumn{2}{c}{$[7a-12a]$}\\
			Fit parameters&$\chi^2$ &$\sigma_{0} a^{2}$&$\chi^{2}$&$\sigma_{0} a^{2}$&$\chi^{2}$&$\sigma_{0} a^{2}$\\
			\hline
			\tiny{\bf{Baryonic String}}  &  &  &  & &\\
			$A=2a$   &0.10  &0.45(1)  &0.05  &0.45(1) &0.02 &0.44(2)\\ 
			$A=4a$   &0.15  &0.43(1)  &0.08  &0.44(1) &0.05 &0.43(2)\\
			\hline
			\hline
		\end{tabular}	
		\caption{The string tension and $\chi^{2}$ from the fit of $(QQ)Q$ potential to the baryonic Y-string Eq.\eqref{eq:vba1} at $T/T_{c}=0.8$.}
		\label{tab:T08B}
	\end{center}
\end{table}

That the $\chi^{2}_{\rm{d.o.f}}$ is tiny
on this long fit interval, it is possible to
retry the fit and set the string tension to a value such as $\sigma_{0}a^2=0.036$, obtained from the mesonic fits of the  $Q\bar{Q}$ potential  at long source separation $R\in [9a, 12a]$(least $\chi^{2}$). The resultant fit appears to match the lattice data $\chi^{2}_{\rm{d.o.f}}=1.29$ and $\chi^{2}_{\rm{d.o.f}}=1.13$ on the fit interval $R\in [7a, 12a]$ and 
$R \in [8a, 12a]$, respectively.

Regarding the same temperature scale $T/T_c=0.9$, Fig.\ref{fig:p2}(a) opposes the mesonic string Eq.\eqref{eq:Pot_NG_LO}, at fixed string tension $\sigma_0 a^{2}=0.0385$ which is reproduced~\cite{Bakry:2020flt} from the fits of $Q\bar{Q}$ data on the fit interval $R\in[5a,12a]$, with the baryonic string Eq.\eqref{eq:vba1} adopting the string tension  value $\sigma_0 a^{2}=0.036$.

The goal of this comparison is to point out that the mesonic string potential brought about by fitting the $Q\bar{Q}$ data on the entire range  $R\in[5a,12a]$ deviates significantly from the baryonic potential data $(QQ)Q$. The baryonic string, nevertheless, compares with the $(QQ)Q$ lattice data on the whole interval $R\in[5a,12a]$. The plot attests to the baryonic nature of the string, which in agreement with a $3Q$ potential proportional to the length of the Y-string with a string tension the same as that of the $Q\bar{Q}$.

Furthermore, the string tension retrieved from the fits of the baryonic Y-string model at $T/T_{c}=0.8$ matches, within uncertainties, the correct value $\sigma_{0}=0.044$ at both base width $A=2a,4a$. It's an interesting result, in its own right, since the findings support the validity of the free baryonic Y-string model to interpret the lattice $3Q$ data ~\cite{Jahn2004700,deForcrand:2005vv} for the presented geometry, at low temperatures.

At the lower temperature scale $T/T_c=0.8$, both fits to the $(QQ)Q$ potential data for the mesonic string Eq.\eqref{eq:Pot_NG_LO} and that of the baryonic Y-string are plotted in Fig.\ref{fig:p2}(b).

The graphic shows the agreement between the two models and a good fit to the potential data for the $(QQ)Q$ system at the same string tension as the $Q\bar{Q}$ system, $\sigma_0 a^2=0.044$.

\section{Diquark-quark (QQ)Q gluonic profile}   
\label{sec:AcDensity_onLattice}      
\subsection{Vacuum's Action-density}

Further characteristics of the confining force can be explored by analyzing the profiles of the flux tubes. We examine the action density of the vacuum in the presence of quarks through the correlation functions $\mathcal{C}_{3Q}(\boldsymbol\rho)$ and $\mathcal{C}_{Q\bar{Q}}(\boldsymbol\rho)$ of Eq.\eqref{eq:mesonic_Correlator} and Eq.\eqref{eq:baryonic_Correlator}. The correlations correspond to $Q\bar{Q}$ and the planar $(QQ)Q$ systems.

As demonstrated in Fig.\ref{fig:gam}, the vector $\boldsymbol\rho = (x, y, 18)$ describes either the plane of the color sources or its perpendicular plane $\boldsymbol\rho = (x,18, z)$. The intersection of the orthogonal lines is the point $\mathbf{O}=(0,18,18)$, that is, the coordinate $x=0$ coincides with the diquark center or the antiquark position in the case of $Q\bar{Q}$ system. The location of the quark $Q_3$ is at the position $\boldsymbol\rho=(R, 18, 18)$  such that $R$ is varied from $R=0$ to $R=na$ steps for each selected base diameter $A$ between the quarks $Q_2$ and $Q_3$ (constituting the diquark) residing on the y=axis.

\begin{figure}[!hptb]
	\centering
	\includegraphics[scale=0.28]{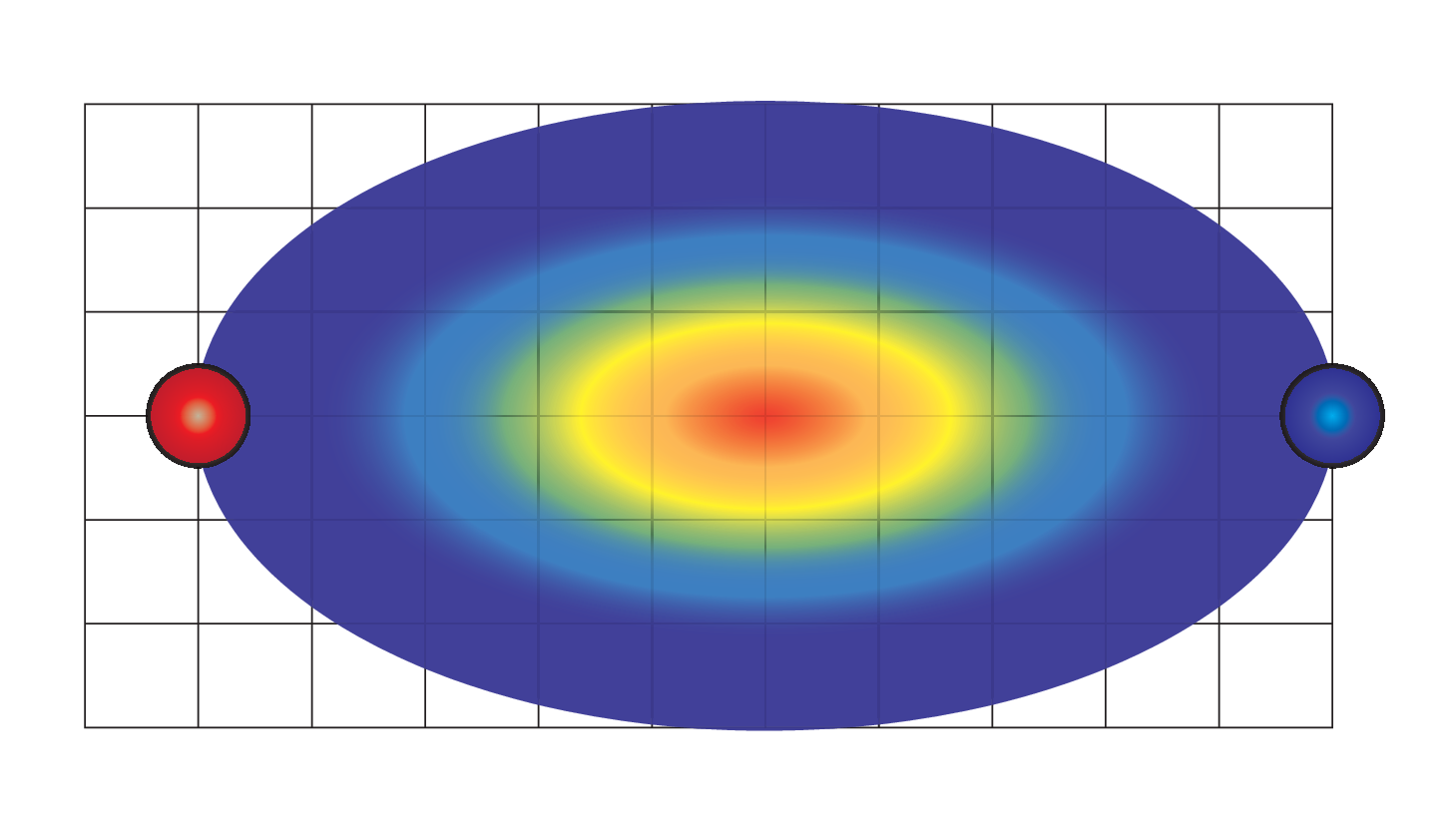}
	\caption{
		The action density Eq.\eqref{eq:mesonic_Correlator} of meson $Q\bar{Q}$ at $T/T_{c}=0.8$ and color source separation $R=10a$.} 
	\label{fig:ActionDens}
\end{figure}

\begin{figure}[!hpt]
	\centering
	\subfigure[Baryon $(QQ)Q$ with base $A=2a$ ]{\includegraphics[scale=0.30]{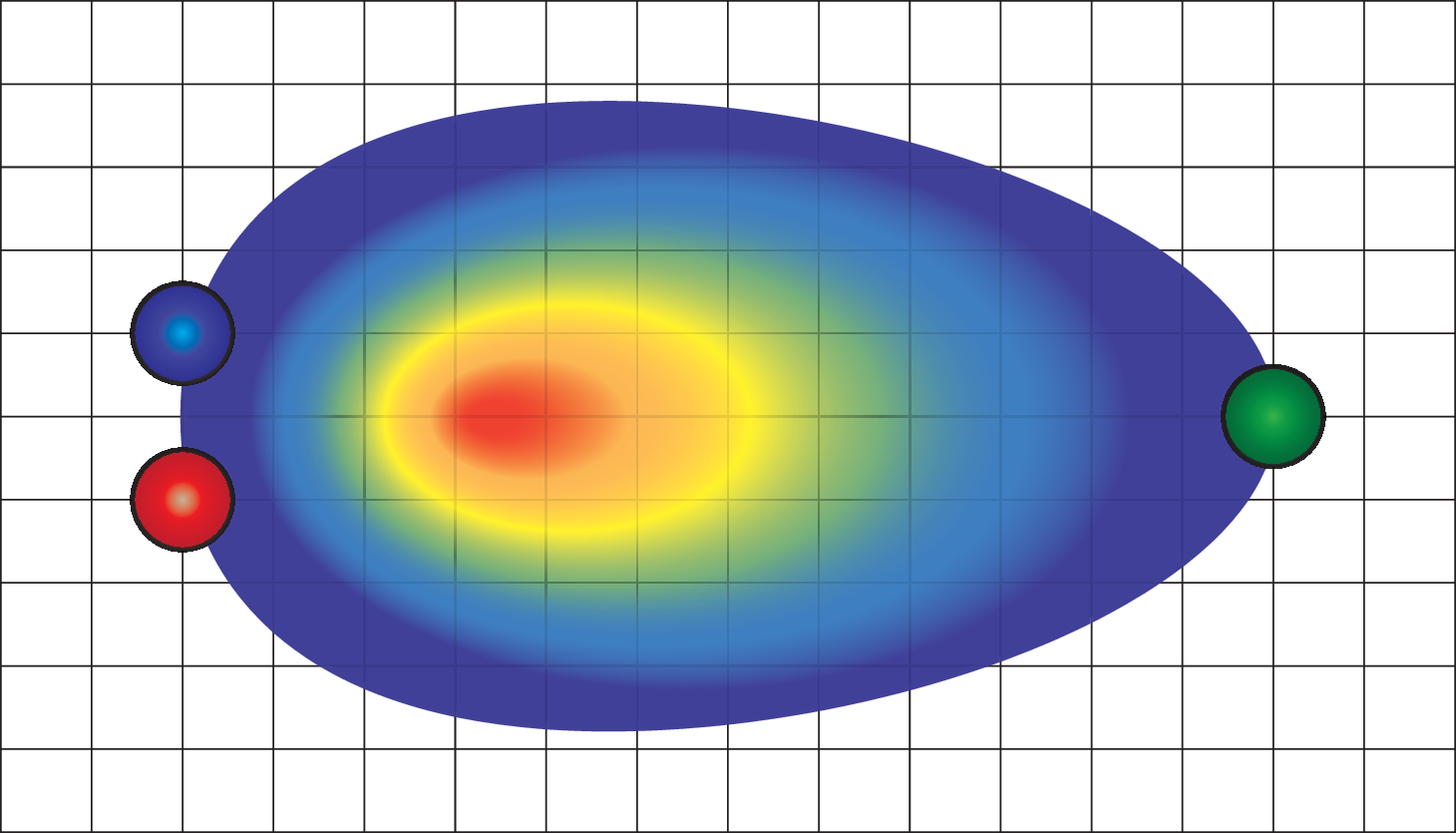}}
	\subfigure[Baryon $3Q$ with base $A=6a$ ]{\includegraphics[scale=0.30]{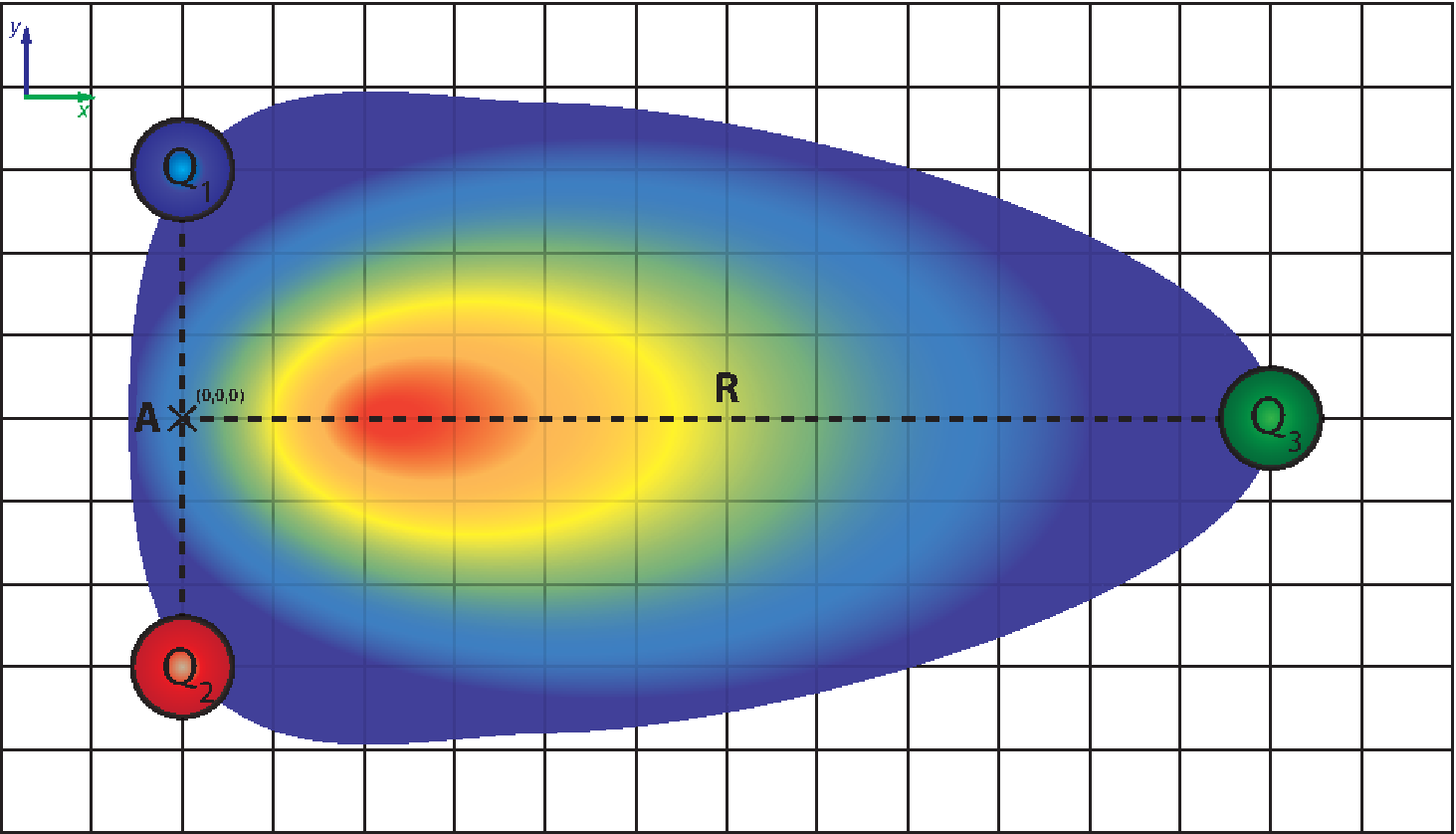}}
	\caption{The gluonic action density Eq.~\eqref{eq:baryonic_Correlator} in the baryon with base diameters $A=2a$ and $A=6a$ and $QQ-Q$ distance $R=12a$  at $T/T_{c}=0.8$}
	\label{fig:T-tube10}
\end{figure}

Figure.\ref{fig:ActionDens} and Fig.\ref{fig:T-tube10} display two snapshots of the expulsion of vacuum fluctuations at $T/T_{c}=0.8$ and $R=10a$. The density profiles correspond to the formation of flux tubes owing to the presence of the static color charges $Q\bar{Q}$ and $3Q$ in the vacuum. The flux density of the baryon in Fig.\ref{fig:T-tube10} is exposed in the plane-$xy$ at base diameters $A=2a$ and $A=6a$. The density map shows the formation of $(QQ)Q$ flux tube, with the decrease of base diameter $A$, which identifies with $Q\bar{Q}$ system. 

\begin{figure}[!t]
	\centering
	\includegraphics[scale=0.38]{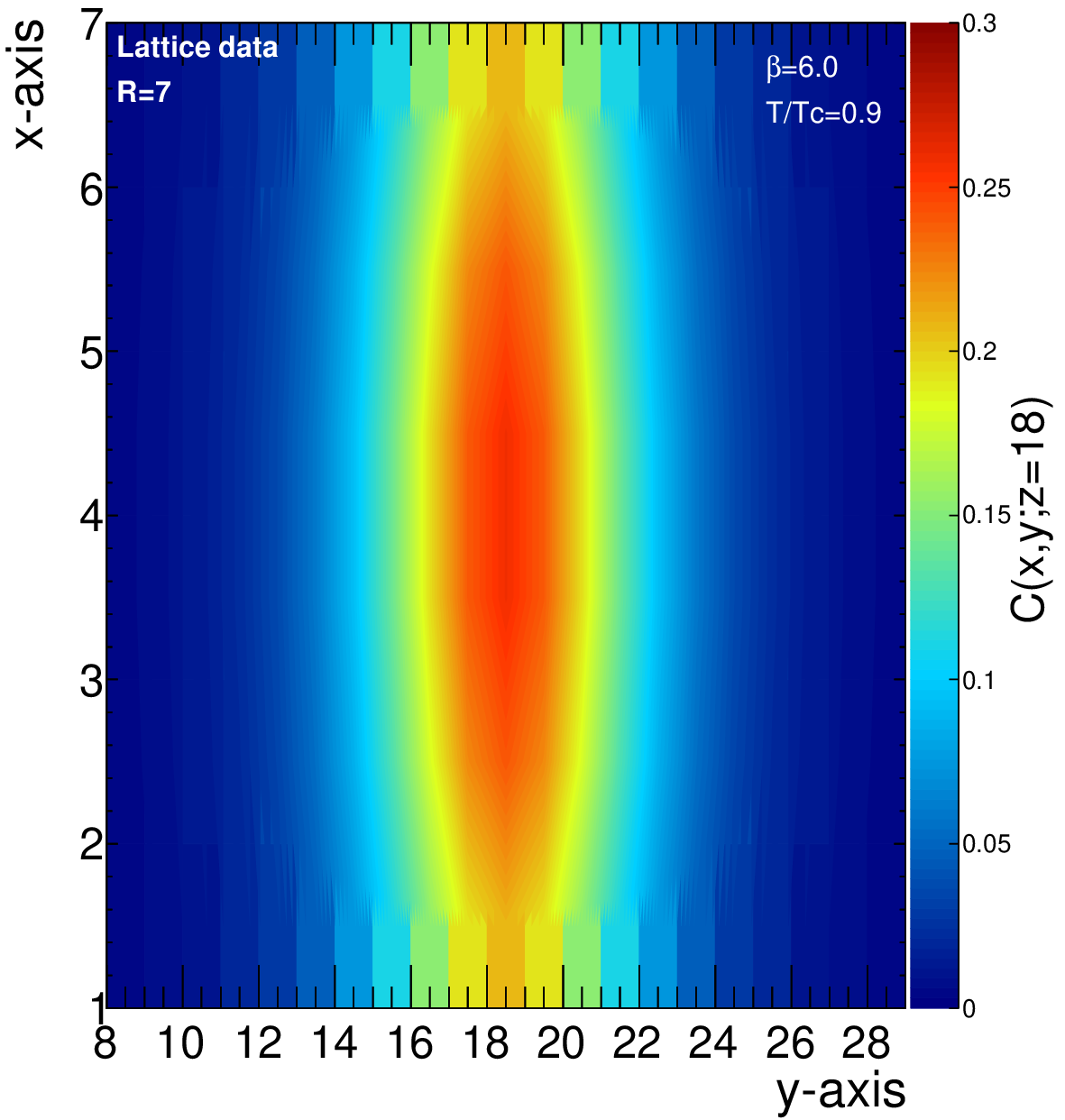}
	\caption{
		The in-plane action density Eq.\eqref{eq:mesonic_Correlator} for meson $Q\bar{Q}$ at $T/T_{c}=0.9$ and color source separation $R=7a$.
	}
	\label{fig:abcT091}
\end{figure}

Similar patterns of the action density are represented; however, at the higher temperature $T/T_c=0.9$ in Fig.\ref{fig:abcT091} corresponding to a meson $Q\bar{Q}$ with color source separation $R=7a$. The planar map exposes the action density of $(QQ)Q$ diquark-quark with the quark $Q_{3}$ residing at $R=7a$ and base length corresponding to $A=2a$ as shown in Fig.\ref{fig:abcT092}(a) and $A=4a$ in Fig.\ref{fig:abcT092}(b).

In Appendix.~A, 2D density maps of the action are shown off at $R= 4a, 5a, 6a, 9a$, and $R=12a$. Figures \ref{fig:2DFluxTubes_TTc09} and \ref{fig:2DFluxTubes_TTc08} assimilate the $(QQ)Q$ system at the temperature scales $T/T_c=0.9$ and $T/T_c=0.8$, respectively. Each row corresponds to the action density in the $xy$ plane at base length $A=2a$  and $A=4a$. On the other hand, Figs.\ref{QQT09} and \ref{QQT08} amount to the density map of $Q\bar{Q}$ system at the corresponding $R$ and temperature scales.

The action density plots are rendered using {\texttt{ROOT}} package~\cite{link}. We've implemented the  (CONT) flag for the 2D interpolation. The option corresponds to drawing plots using surface colors that distinguish the contours with the so-called Delaunay triangles~\cite{link}. 
That is, the unique triangulation $DT(S)$ of a set of points $S$ in the Euclidean plane such that no point in $S$ is inside the circumcircle of any triangle in $DT(S)$ ~\cite{link}.

The action density distribution within the $Q\bar{Q}$ system is not uniformly distributed. The planar distribution  peaks at the center of the planar distribution $\boldsymbol\rho=(\frac{R}{2},18,18)$. 
The $(QQ)Q$ system in Fig.\ref{fig:T-tube10}(a) discloses also a nonuniform correlation function; $\mathcal{C}_{3Q}(\boldsymbol\rho (x,y,18))$, that peaks, however, at a point displaced roughly one lattice spacing from the center $\boldsymbol\rho=(\frac{R}{2}-a,18,18)$ toward the diquark~\cite{Bakry:2011zz, Bakry:2011vnk}.
\begin{figure}[!hpt]
	\centering
	\subfigure[]{\includegraphics[scale=0.38]{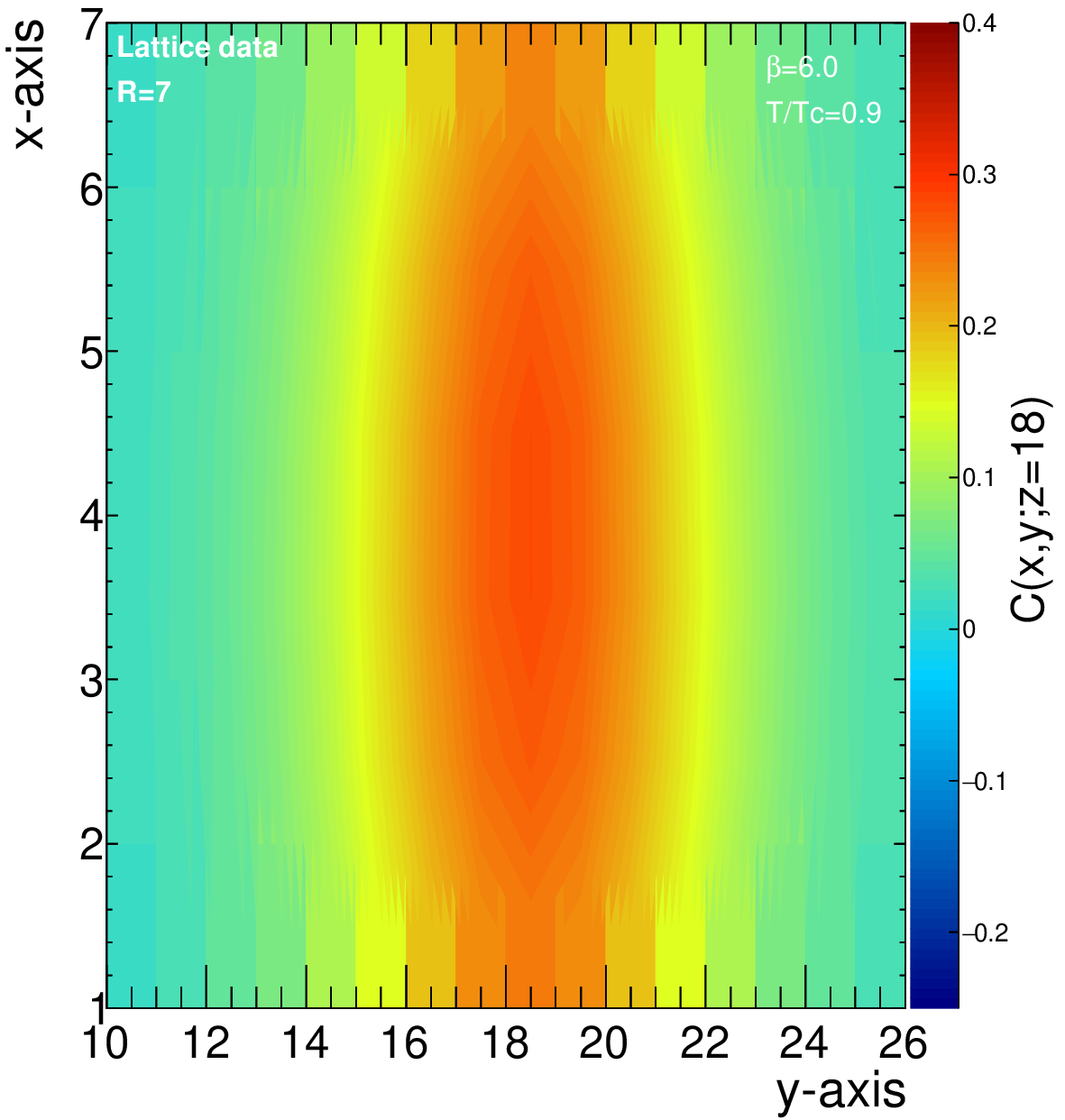}}
	\subfigure[]{\includegraphics[scale=0.38]{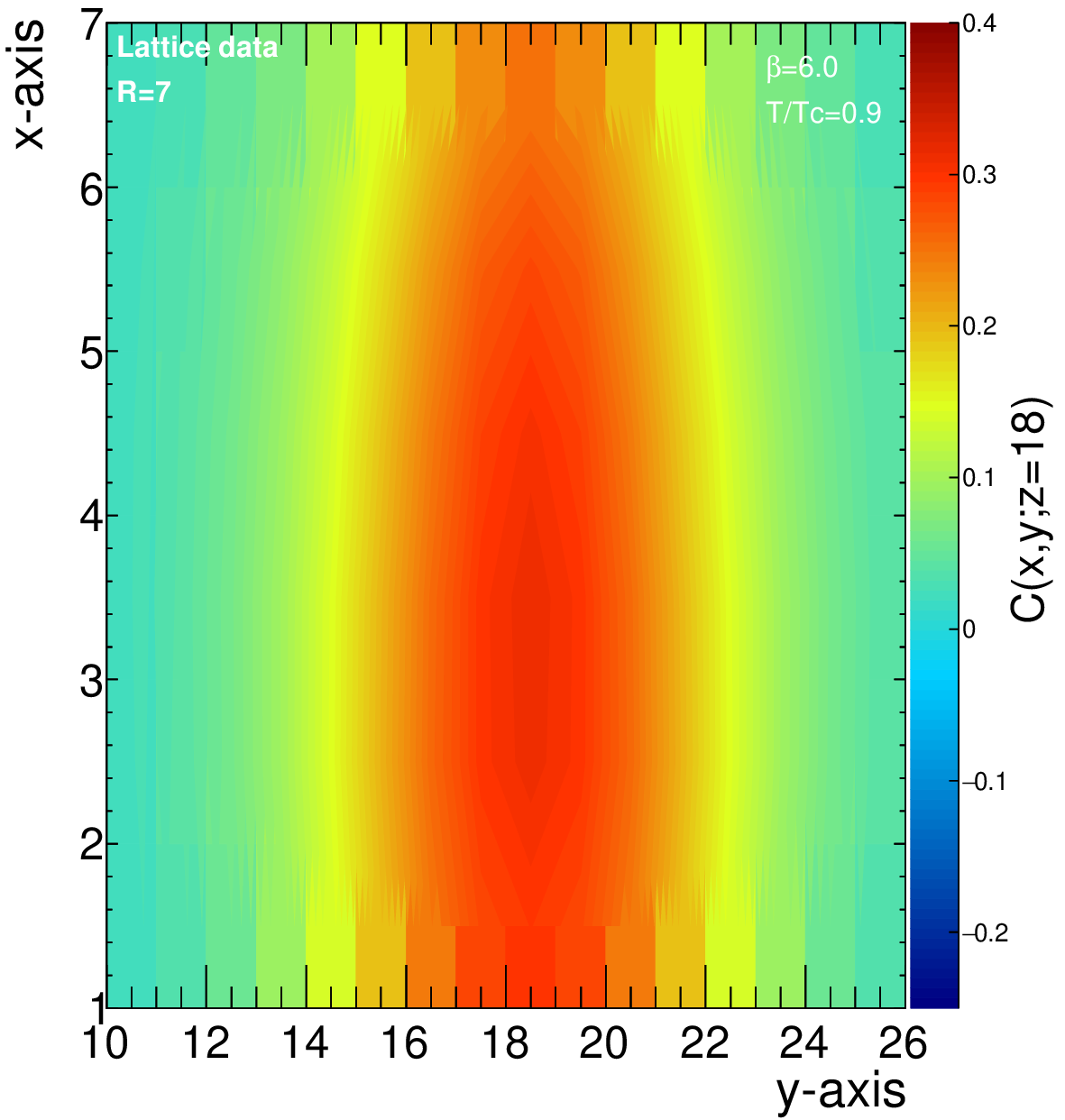}}   
	\caption{The in-plane action density Eq.\eqref{eq:mesonic_Correlator} for baryon $(QQ)Q$ at $T/T_{c}=0.9$. 
		The baryonic geometry corresponds to a triangle with $(QQ)-Q$ distance  $R=7a$ and base length $A=2a$ in (a) and $A=4a$ in (b).
	}
	\label{fig:abcT092}
\end{figure}
At both temperature scales, the longitudinal profiles of $(QQ)Q$ and $Q\bar{Q}$ flux tubes along the line  $\boldsymbol\rho = (x, 18, 18)$ unveils vacuum expulsion that is stronger near the diquark than it is near the quark  (Fig.\ref{fig:2DFluxTubes_TTc09} and Fig~\ref{fig:2DFluxTubes_TTc08}). The two flux tubes, however, exhibit almost similar profiles close to the quark.

The noteworthy finding is that, despite having reported the same $Q\bar{Q}$ potential as the $(QQ)Q$ system in the preceding section at $T/T_c=0.8$. The vacuum expulsion map does not absolutely align with each other as discussed above. 
\begin{figure*}[!hpt]
	\centering
	\subfigure[$\parallel$-plane, $A=2a$]{\includegraphics[scale=0.34]{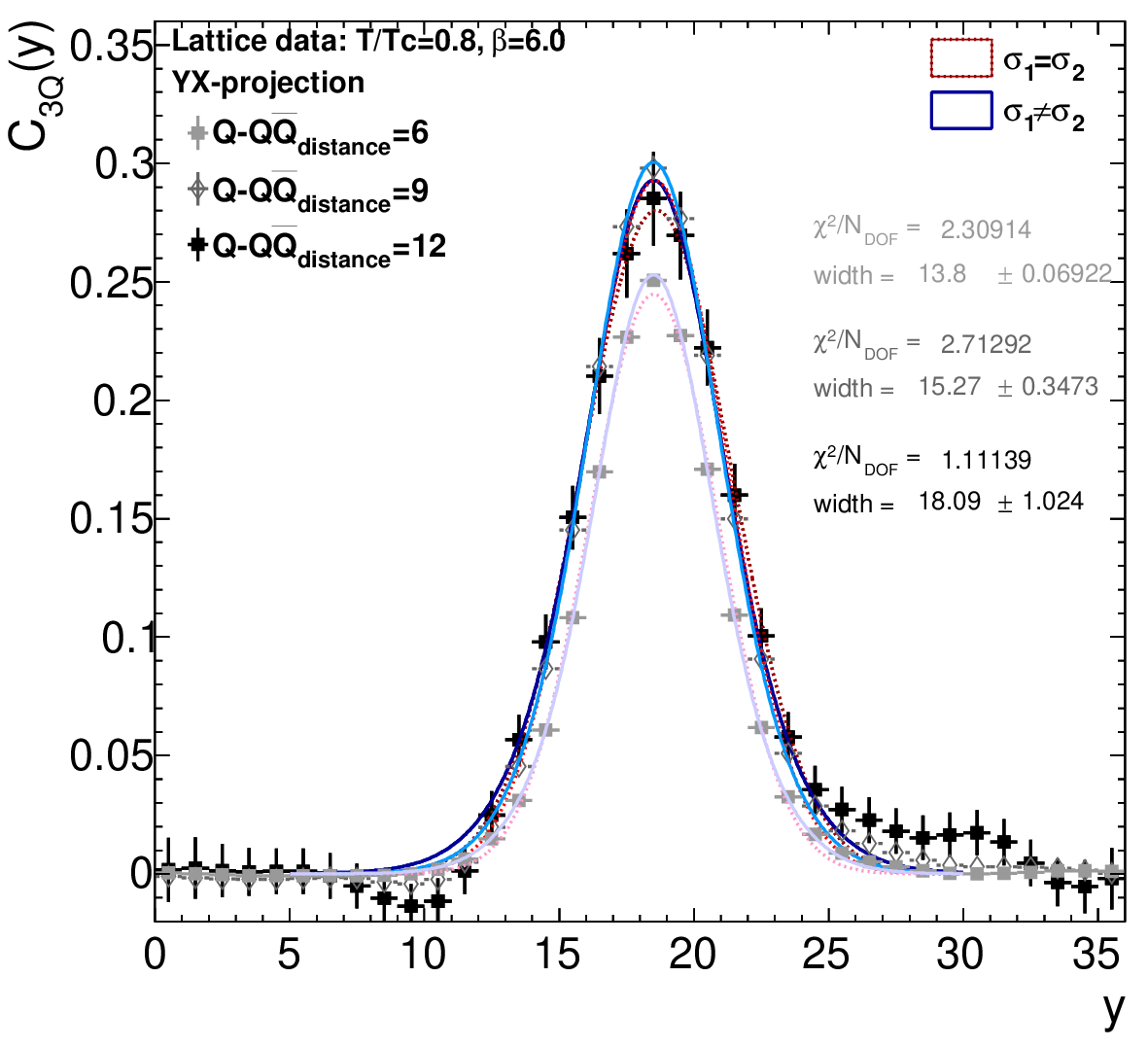} }
	\subfigure[$\parallel$-plane, $A=4a$]{\includegraphics[scale=0.34]{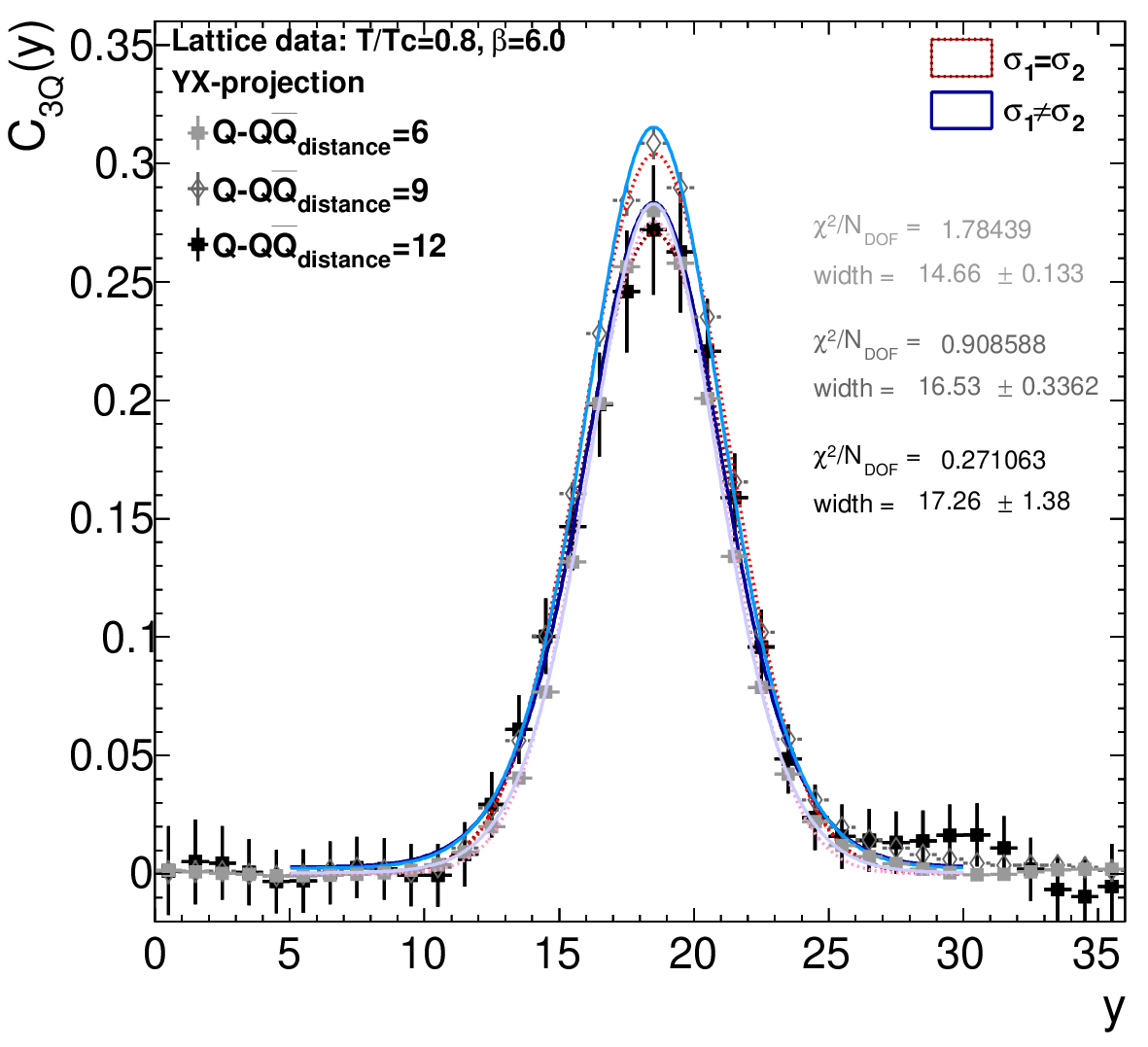} }
	
	\subfigure[$\parallel$-plane, $A=2a$]{\includegraphics[scale=0.34]{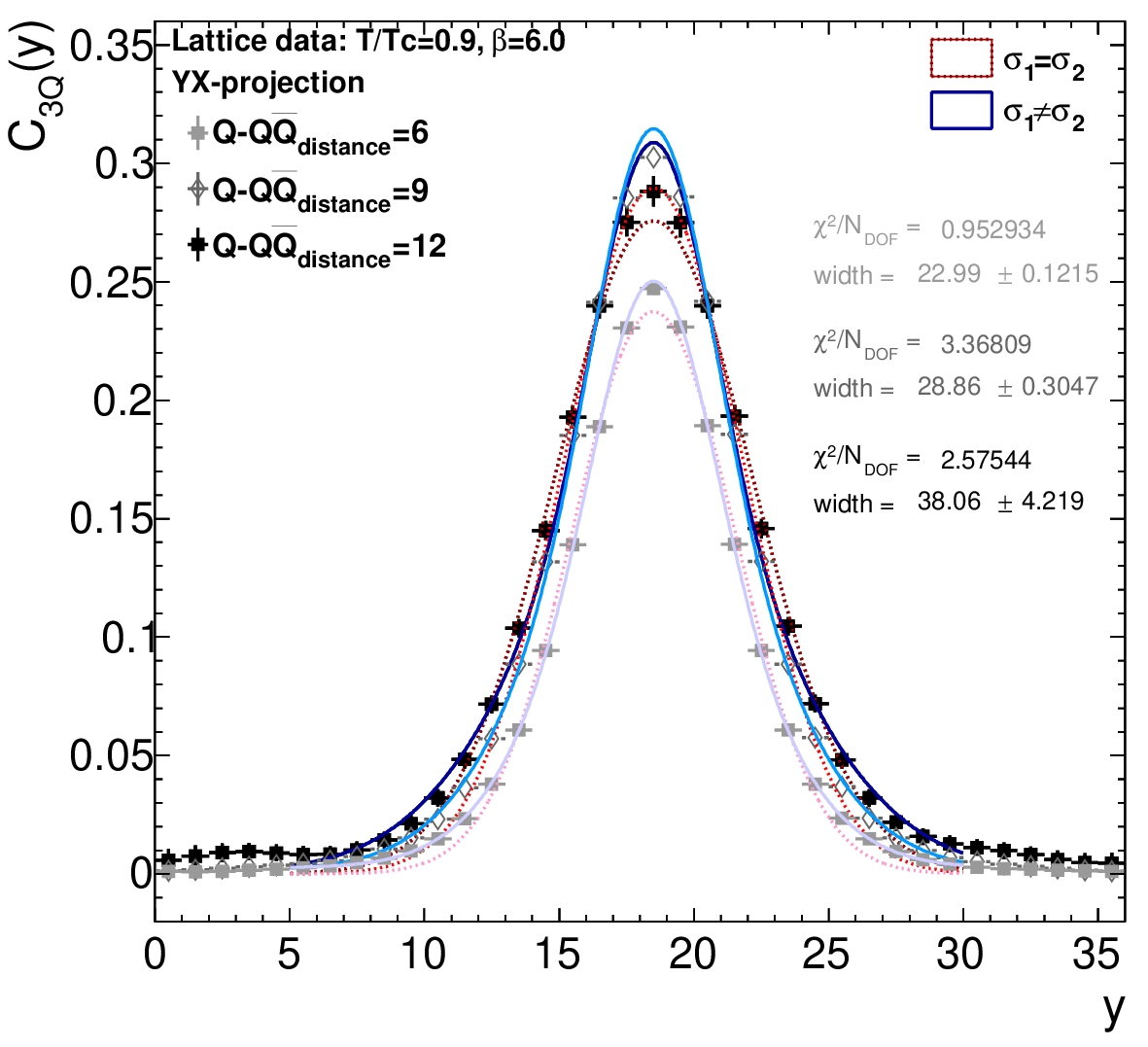} }
	\subfigure[$\parallel$-plane, $A=4a$]{\includegraphics[scale=0.34]{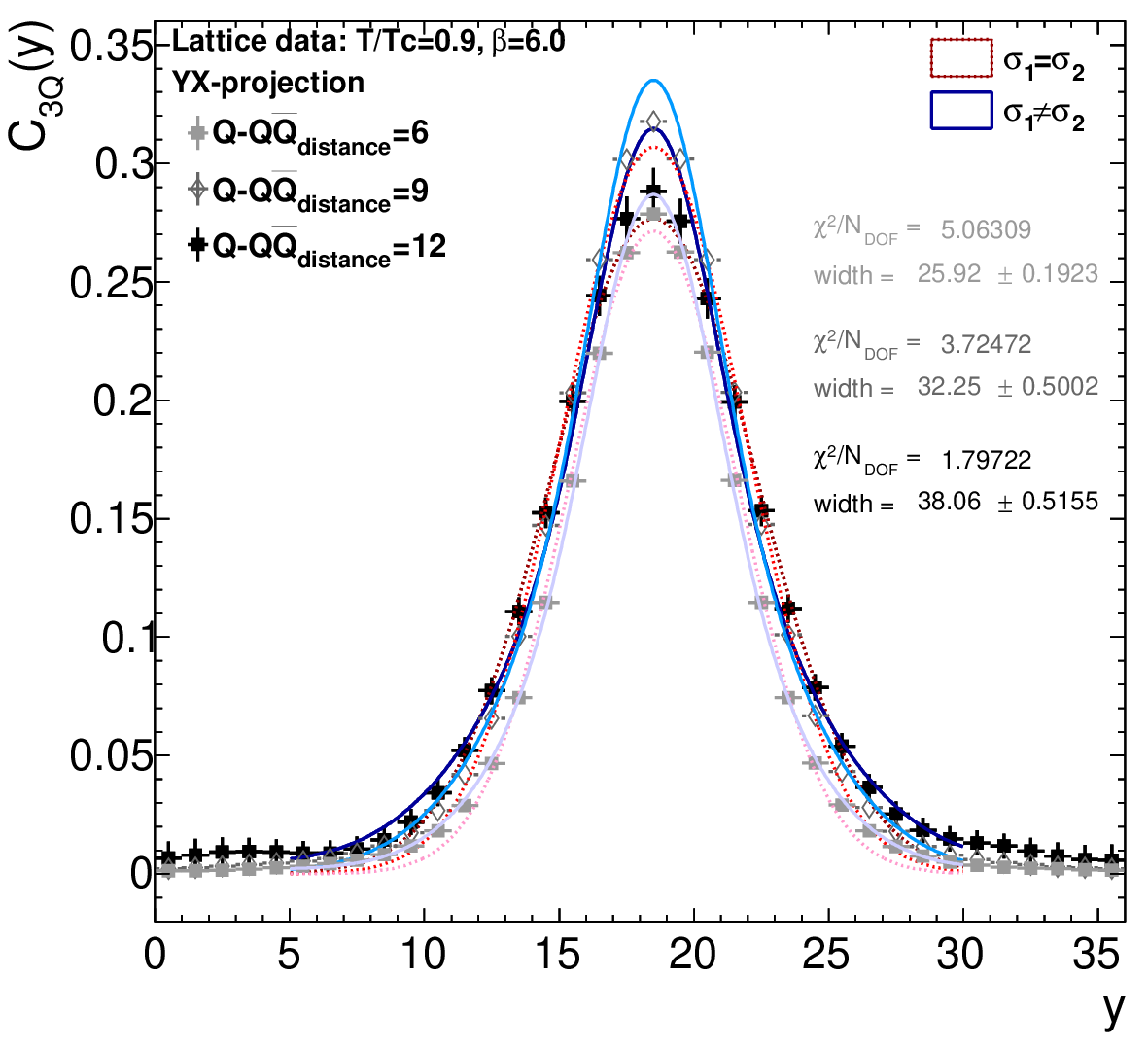} }
	\caption{(a,b) The $(QQ)Q$ action density $\mathcal{C}_{3Q}(R/2,y,18)$ Eq.\eqref{eq:baryonic_Correlator} (denoted as $\parallel$-plane), QQ-Q distance $R=6a, 9a, 12a$, base diameter $A=2a$(a) and $A=4a$(b), at $T/T_{c} = 0.8$. The red dotted lines correspond to fits using the standard Gaussian, $\sigma_1=\sigma_2$ Eq.\eqref{eq:conGE}, the blue solid lines correspond to unconstrained form, $\sigma_1\neq\sigma_2$. (c,d) Same as (a,b) except that  $T/T_{c}=0.9$.}	\label{fig:Baryon_AcDens_20swT08}
\end{figure*}

\begin{figure*}[]
	\subfigure[]{\includegraphics[scale=0.33]{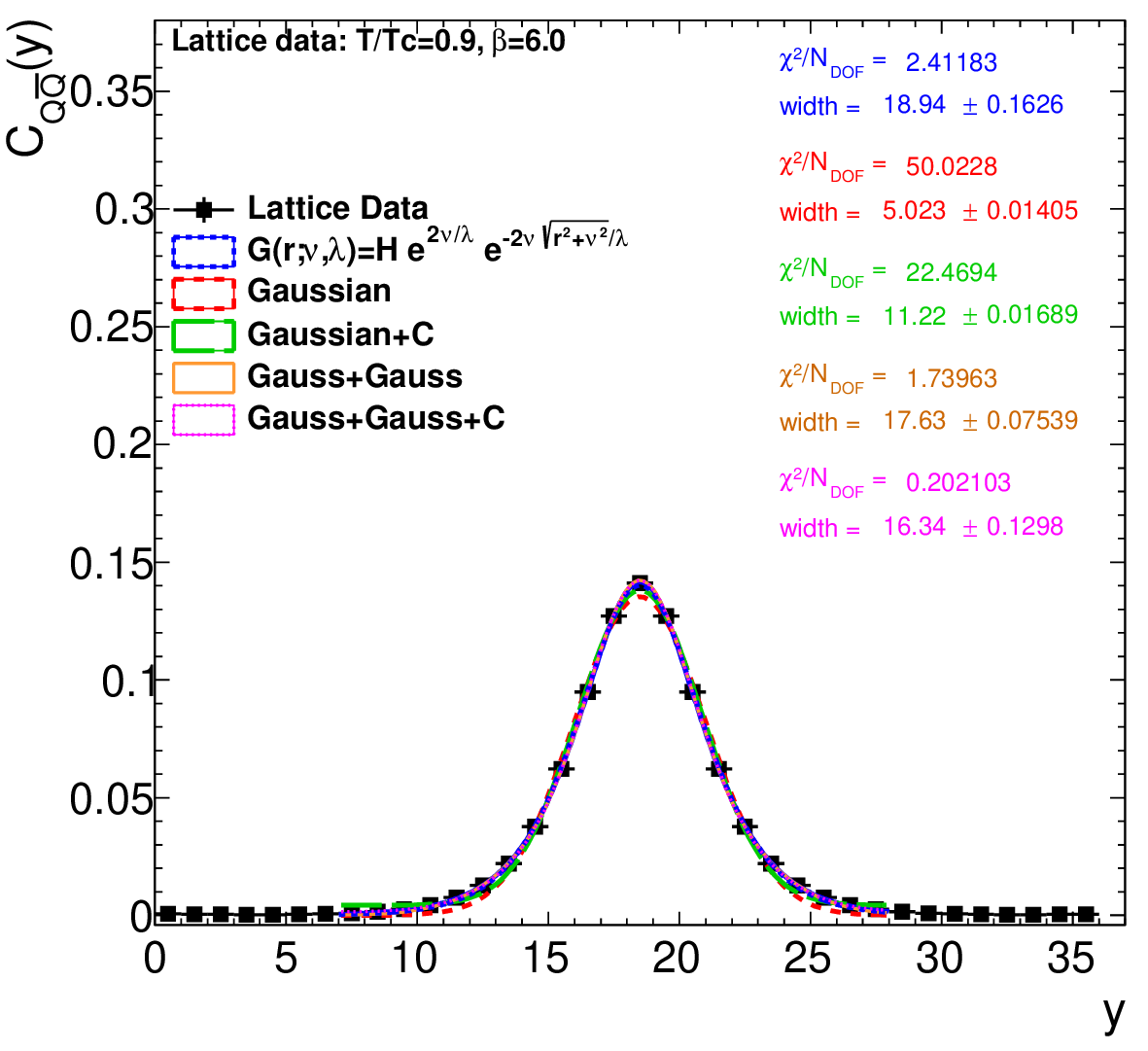}}
	\subfigure[]{\includegraphics[scale=0.33]{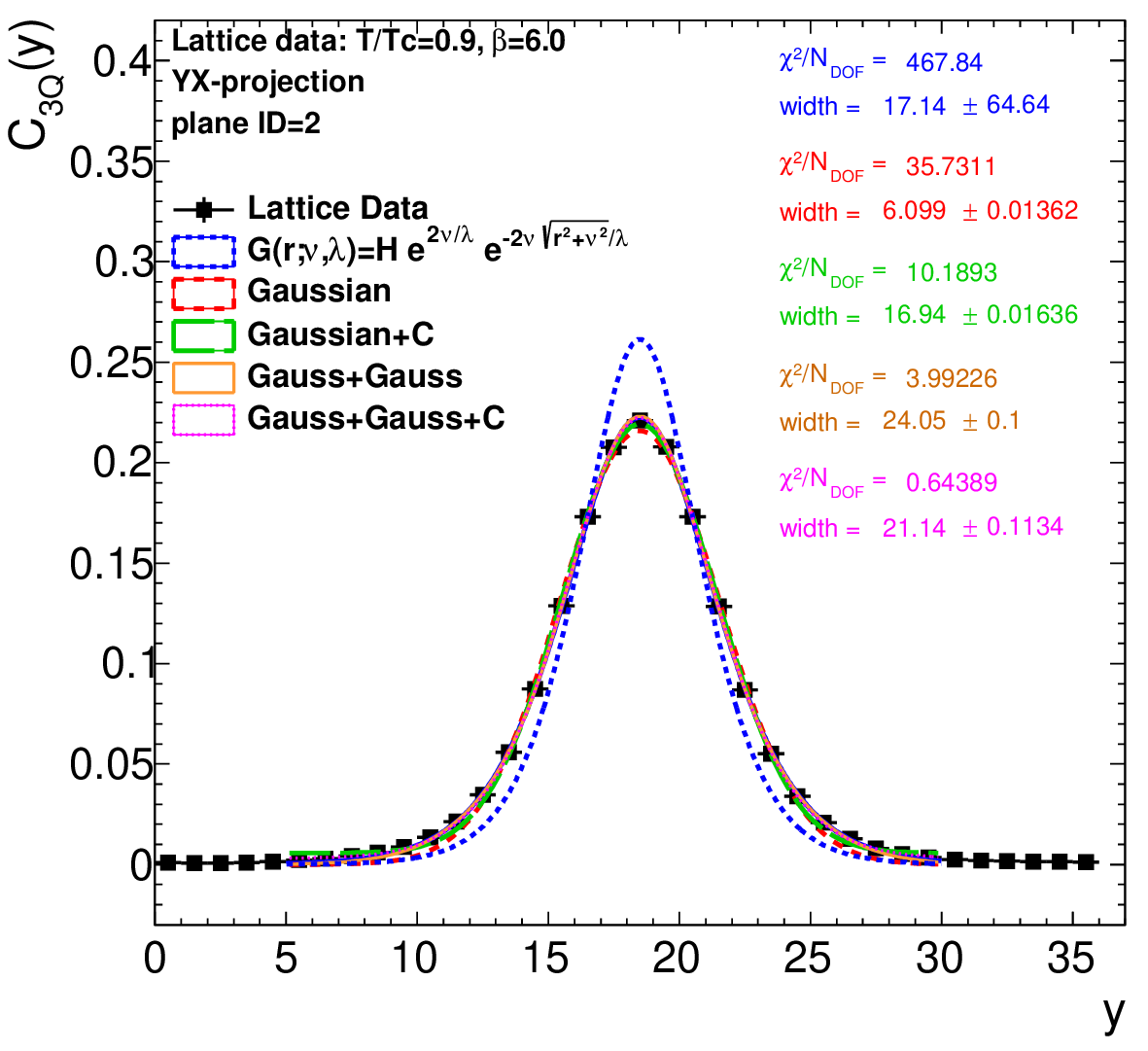}}
	\caption{
		The resultant fits using five different fit functions and returned $\chi^2_{\rm{d.o.f}}$ and width measurements of a selected mesonic and baryonic configuration. 
		(a) The $Q\bar{Q}$ action density $\mathcal{C}_{Q\bar{Q}}(R/2,y,18)$ Eq.\eqref{eq:mesonic_Correlator}, at the middle plan, at $T/T_{c} = 0.9$.
		(b) The $(QQ)Q$ action density $\mathcal{C}_{3Q}(x_i=2,y,18)$ Eq.\eqref{eq:baryonic_Correlator}, (QQ)-Q distance $R=4a$, plane $x_i=2$, $A=4a$, at $T/T_{c} = 0.9$.
	}		
	\label{fig:Meson_Baryon_AcDens}	
\end{figure*}
\begin{figure}[!ht]
	\centering
	\includegraphics[scale=0.56]{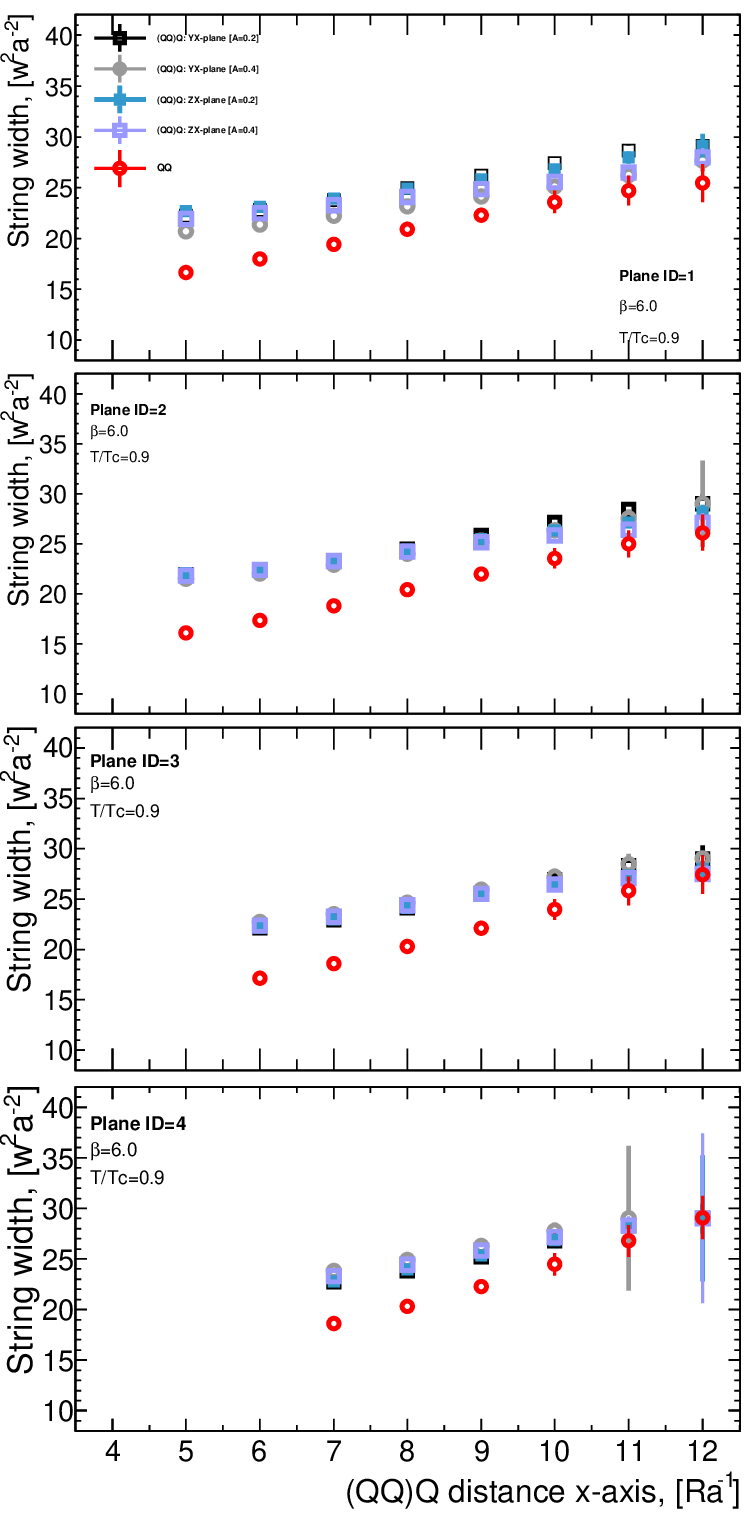}
	\caption{The in-plane and perpendicular-plane MS width at $T/T_{c}=0.9$ for $(QQ)Q$ and $Q\bar{Q}$ systems at planes $x=1,2,3$ and $x=4$, diquark base diameters $A=2a,4a$.}
	\label{fig:planesT09}
\end{figure}
\begin{figure*}[!hbt]
	\centering
	\includegraphics[scale=0.52]{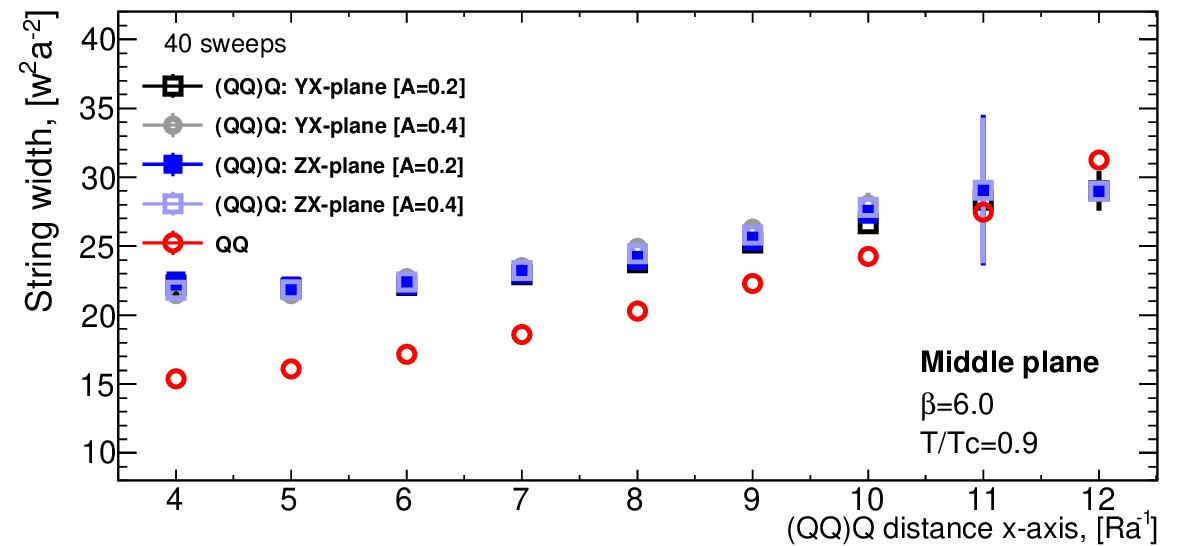}    
	\caption{Compares the in-plane and perpendicular MS width at $T/T_{c}=0.9$ for $(QQ)Q$ and $Q\bar{Q}$ systems at the middle plane  for diquark base  diameters $A=2a,4a$.}
	\label{fig:MidplaneT09}
\end{figure*}
\begin{figure}[!hpt]
	\centering
	\subfigure[]{\includegraphics[scale=0.30]{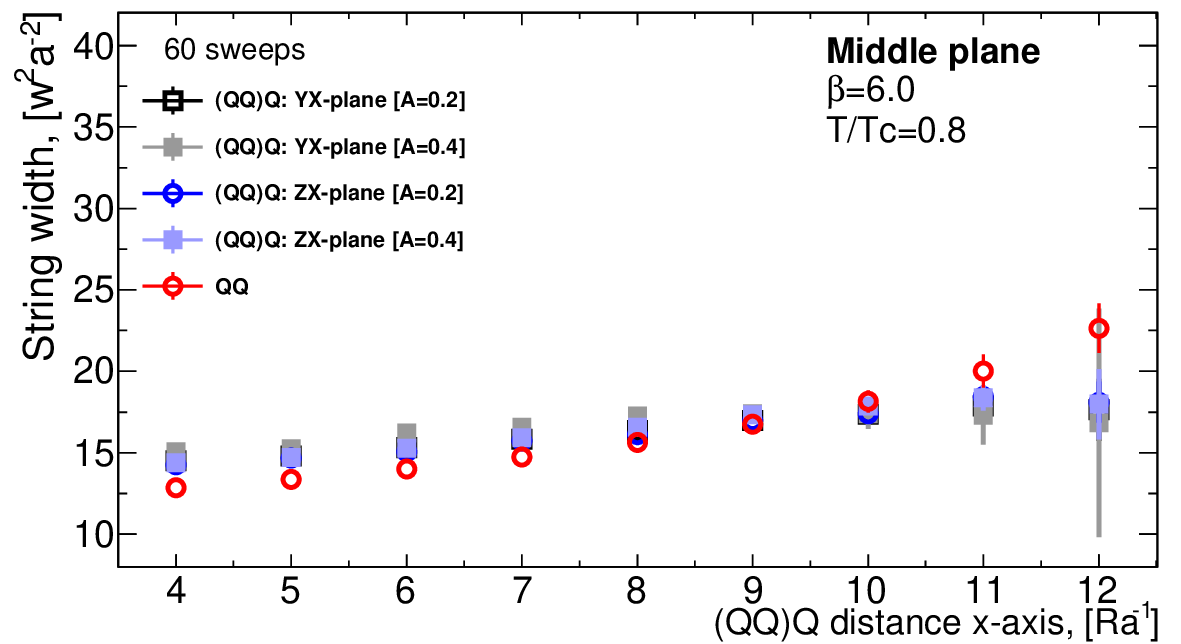}}   
	\subfigure[]{\includegraphics[scale=0.35]{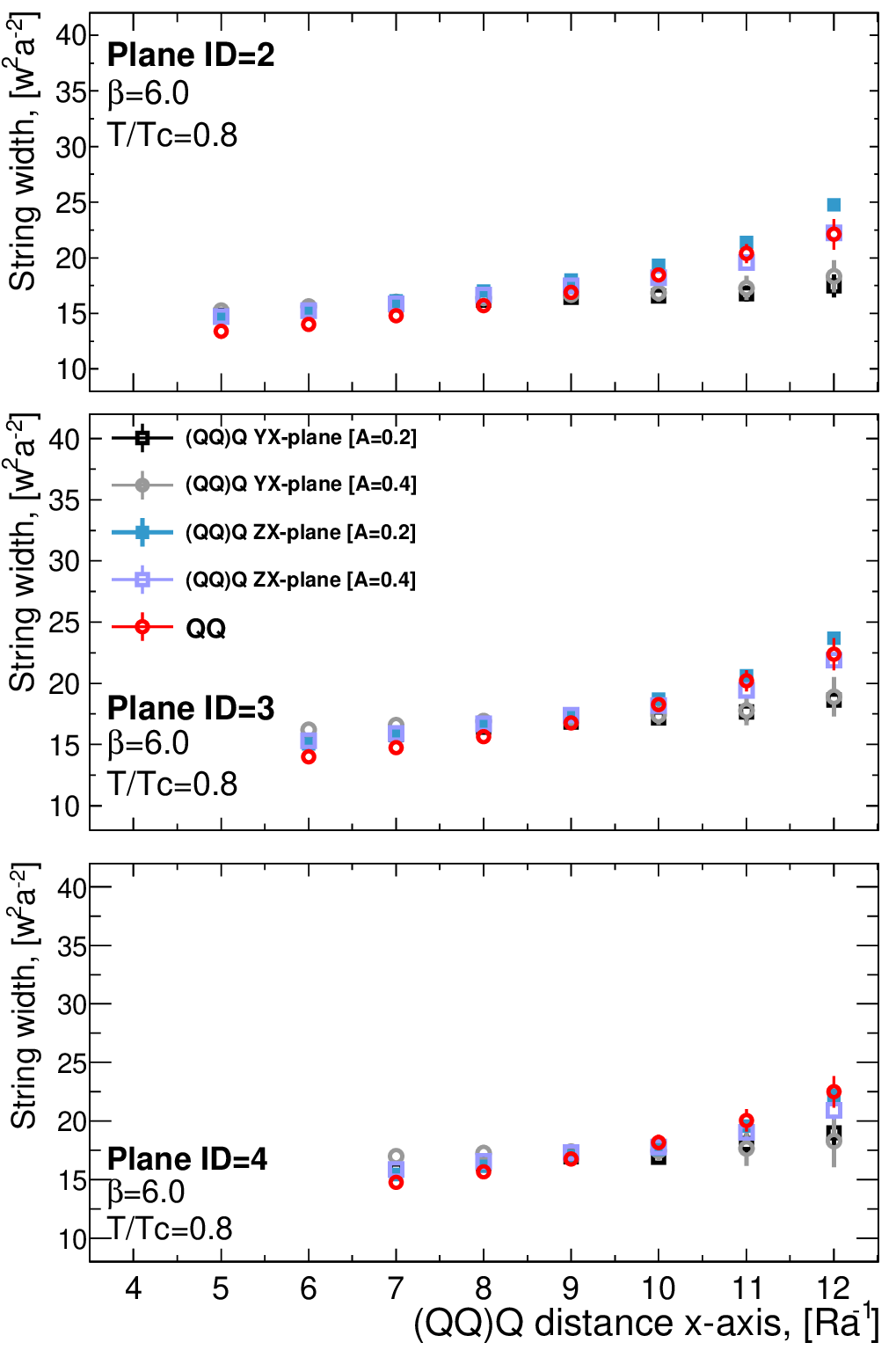}}     
	\caption{(a) The in-plane and the perpendicular MS width of action density at the middle plane for $Q\bar{Q}$ and $(QQ)Q$ of base diameters $A=2a,4a$ at $T/T_c=0.8$. (b) Same as (a), however, the plots depicts the width at the planes $x=2,3,4$.
	} 
	\label{fig:MidplaneT08}
\end{figure}

Apart from the above discussed qualitative aspects, taking measurements specifying the characteristics of the action density is mandatory to fully dissect the manifested profile. The amplitude and second moment of the density distribution come out as the most significant features that are possible to estimate with fits to an appropriate functional shape.

Introducing cylindrical coordinate system $\boldsymbol\rho(x,r^2=(z-18)^2+(y-18)^2,\theta)$. At each $x_i$ we proceed to fit the action density values along the line $\boldsymbol\rho(x_i, r^2=(y-18)^2,\theta=0)$ to a double-Gaussian function of amplitude $A$ and fit parameters ($\sigma_{1},$ $\sigma_{2}$), such as  
\begin{equation}
	G(r;\theta=0,x_i)= A (e^{-r^2/\sigma_1^2}+e^{-r^2/\sigma_2^2}).
	\label{eq:conGE}
\end{equation}
Similarly, we fit $G(r;\theta=\frac{\pi}{2},x_i)$ along the orthogonal line  $\boldsymbol\rho(x_i,r^2= (z-18)^2, \theta=\frac{\pi}{2})$. The fit interval is taken such that $y,z\in[7,28]$, respectively.

The fits to the action density distribution of the $(QQ)Q$ Eq.\eqref{eq:baryonic_Correlator} at the center of the tube $\boldsymbol\rho(R/2,r^2=(y-18)^2,\theta=0)$ are shown for $QQ-Q$ separation $R=6a, 9a$ and $R=12a$ in Figs.\ref{fig:Baryon_AcDens_20swT08} at both temperatures $T/T_{c} = 0.8$ and $T/T_{c} = 0.9$.

The retrieved $\chi^{2}$ from the fits of the $Q\bar{Q}$ and $(QQ)Q$ systems are collected in Tables~\ref{tab:AAAw2a2T08}, \ref{tab:AAAw2a2T09} and \ref{tab2:AAAw2a2T08}, the values indicate suitable fits at most planes.

The in-plane $\boldsymbol\rho(x_i,r,\theta=0)$ and the orthogonal $\boldsymbol\rho(x_i, r, \theta=\frac{\pi}{2})$ MS width can be extracted from Eq.\eqref{eq:conGE} such that
\begin{equation} 
	\begin{split}
		W_{\parallel}^{2}(x_i)=2&\quad \frac{\int \, dr\,r^{2}\,G(r;\theta=0,x_i)} {\int \,dr \, \: G(r;\theta=0,x_i) },\\
		W_{\perp}^{2}(x_i) =2&\quad \frac{\int \, dr\,r^{2}\,G(r;\theta=\frac{\pi}{2},x_i)} {\int \,dr
			\, \: G(r;\theta=\frac{\pi}{2},x_i)) },
		\label{eq:widthG}
	\end{split}
\end{equation}
respectively. The multiplicative factor of 2 in the above equation is to keep intuitive contact with the cylindrical-integrated definition of the meson string width~\cite{Bakry:2020ebo} which reproduces width values of subtle differences. The formulas of the Y-string model are scaled accordingly when we come to compare the data with the models as described below.

In addition to the favorable parametrization behavior over the specified distance scale, the form Eq.\eqref{eq:conGE} is selected on the grounds that the returned values of the fit parameters $\sigma_1 $ and $ \sigma_2$ signifying the width, 
\begin{equation}
	W_{(\parallel,\perp)}^{2}(x_i)=\sigma_{1}^{2}+\sigma_{2}^{2}-\sigma_1-\sigma_2,
\end{equation}
divulge~\cite{Bakry:2020flt} a significant splitting, that is, $\sigma_1 \gg \sigma_2$.

It is conceivable that the flux tube is made up of a solid, vortex like core whose fluctuations are described by bosonic string. 
For instance, a good agreement was found ~\cite{Caselle:2016mqu,Cea:2012qw,Cea:2014uja} between the action density profile with an exponential decline as opposed to the Gaussian profile anticipated by the bosonic string model. Nevertheless, it has been shown~\cite{PhysRevD.88.054504} that the profile may be analyzed by a heuristic convolution of the bosonic string (Gaussian) profile and vortex (exponential) profile, which suggests that the flux tube shares traits with both.

This is in consonance and would explain the splitting of the profile Eq.\eqref{eq:conGE} into a wide and a narrow Gaussian distribution  appropriate for quantum oscillations and also accommodating for the exponential decline of the vortex flux-tube at the center.

We have tested the fit function with the exponential decay such as that in ~\cite{PhysRevD.88.054504,Bicudo:2017uyy}, as shown in Fig.\ref{fig:Meson_Baryon_AcDens}. The fits using two Gaussian Eq.\eqref{eq:conGE} and exponential form \cite{PhysRevD.88.054504,Bicudo:2017uyy} reveal a correspondence between the two functions, which may suggest an approximate mathematical equivalence on a given	subset of the parameter space. The measurements of the width are almost equal within the uncertainties. Equating the second moment of the two forms, we find
\begin{equation}
\frac{\sigma_1^4+\sigma_2^4}{\sigma_1^2+\sigma_2^2}=
\dfrac{3}{2}\lambda^{2}+2\dfrac{2 \lambda \nu^2}{\lambda+2\nu},
\end{equation}
which defines the physically meaningful parameter $\lambda$, accounting for the penetration length, in terms of the two parameters of the Gaussian form Eq.\eqref{eq:conGE}.

However, the resultant fits of the exponential form to  the baryonic or mesonic action densities reveal either inadequacy or large error bars of the measured parameters owing to redundancy (see Fig.\ref{fig:Meson_Baryon_AcDens}). This increasing size of uncertainties appears in the width measurements in \cite{PhysRevD.88.054504,Bicudo:2017uyy} as well for color source separation $R>0.75$ fm.
	As shown in Fig.\ref{fig:Meson_Baryon_AcDens}, some baryonic configuration 
	such as that with diquark base length $A=4a$
	quite poor fits of the action density are returned 
	when adopting the exponential form. In mesonic configurations, the error bars relevant to the width as measured by the exponential  function are too large that it is not possible to clearly single out/disentangle the broadening of the quantum string  of the leading and next to the leading Nambu-Goto string solutions~\cite{Gliozzi:2010zt,Gliozzi:2010zv}, when the temperature is close to deconfinement point~\cite{Bakry:2020ebo}, at the represented level of statistics.

Tables~\ref{tab:AAAw2a2T09} and \ref{tab2:AAAw2a2T08} include the width measured at the first four planes of both the $(QQ)Q$ and $Q\bar{Q}$ systems. The width measurements in Table \ref{tab:AAAw2a2T09}, at $T/T_{c}=0.9$, depict equal MS width components of the orthogonal $W^{2}_{\perp}$ and in-plane action density $W^{2}_{\parallel}$.

On the account of the fact that the flux tube of the $(QQ)Q$ exposes cylindrical symmetry we represented the symmetrized width, $W^{2}=(W^{2}_{\parallel}+W^{2}_{\perp})/2$, in Table~\ref{tab:AAAw2a2T08} relevant to the tube's middle plane $x=\frac{R}{2}$ at both temperatures $T/T_{c}=0.8$ and $T/T_{c}=0.9$. Table~\ref{tab2:AAAw2a2T08} similarly contains the averaged width at $x_i=1,2,3,4$ and temperature $T/T_{c}=0.8$. The averaging of the width at this lower temperature scale improves the signal-to-noise ratio.

Inspection of the MS width discloses that, at $T/T_{c}=0.8$, the flux tubes of the $Q\bar{Q}$ and the $(QQ)Q$ systems are almost identical within uncertainties of the measurements. However, at the higher temperature $T/T_{c}=0.9$ the measurements of the MS width reproduce identical values only for long enough flux-tubes $R\geq 10 a$.
\begin{table}[!ht]
	\centering
	\begin{tabular}{cc|c|c|c|c|c|c}
		\hline
		\hline
		\multicolumn{2}{c|}{}   
		&\multicolumn{3}{c|}{$T/T_{c}=0.8$}
		&\multicolumn{3}{c}{$T/T_{c}=0.9$}\\
		\hline
		\multicolumn{2}{c|}{R-Config}   
		&\multicolumn{1}{c|}{$A$}
		&\multicolumn{1}{c|}{$w^{2}a^{-2}$}
		&\multicolumn{1}{c|}{$\chi^2_{\rm{dof}}$}
		&\multicolumn{1}{c|}{$A$}
		&\multicolumn{1}{c|}{$w^{2}a^{-2}$}
		&\multicolumn{1}{c}{$\chi^2_{\rm{dof}}$}\\
		\hline
		\multirow{2}{*}{\begin{turn}{90} \scriptsize{$4a$} \end{turn}}
		&$QQ$    &0.0812(2)&13.0(1)&0.87 &0.0712(2)&15.7(1)&1.74\\
		&$(QQ)Q$ &0.0935(2)&13.3(1)&1.60&0.0863(2)&22.0(1)&7.16\\
		\hline
		\multirow{2}{*}{\begin{turn}{90} \scriptsize{$5a$} \end{turn}}
		&$QQ$    &0.1066(2)&13.4(1)&0.51 &0.0895(3)&16.4(2)    &1.46\\        
		&$(QQ)Q$ &0.1183(2)&13.7(1)&1.43&0.1062(2)&21.7(2)&5.85\\ 
		\hline
		\multirow{2}{*}{\begin{turn}{90} \scriptsize{$6a$} \end{turn}}
		&$QQ$ &0.1332(3)&14.0(1)&0.23 &0.1063(4)&17.4(2)    &1.08\\
		&$(QQ)Q$ &0.1413(4)&14.2(2)&1.06 &0.1244(3)&22.1(2)&4.51\\
		\hline
		\multirow{2}{*}{\begin{turn}{90} \scriptsize{$7a$} \end{turn}}
		&$QQ$ &0.1530(5)& 14.7(2)&0.07  &0.1149(5)&18.9(2)    &0.75\\
		&$(QQ)Q$ &0.1604(5)&14.8(2)&1.09 &0.1375(4)&25.5(2)&2.98\\
		\hline
		\multirow{2}{*}{\begin{turn}{90} \scriptsize{$8a$} \end{turn}}
		&$QQ$     &0.1721(9)& 15.6(3)    &0.02 &0.1209(8)&20.5(3)    &0.58\\
		&$(QQ)Q$  &0.1764(8)& 15.3(3)&0.75 &0.1475(6)&24.2(3)&1.92\\
		\hline
		\multirow{2}{*}{\begin{turn}{90} \scriptsize{$9a$} \end{turn}}
		&$QQ$     &0.183(2)& 16.7(4)&0.005 &0.120(1)&22.3(5)    &0.47\\
		&$(QQ)Q$  &0.187(2)&16.0(5)&0.88 &0.1523(8)&25.6(4)&1.09\\
		\hline           
		\multirow{2}{*}{\begin{turn}{90} \scriptsize{$10a$} \end{turn}}
		&$QQ$      &0.193(2)&18.1(7)&0.003  &0.117(1)&24.2(7)    &0.40\\
		&$(QQ)Q$   &0.195(1)&17.1(6)&0.86 &0.154(1)&27.5(5)&0.64\\
		\hline
		\multirow{2}{*}{\begin{turn}{90} \scriptsize{$11a$} \end{turn}}
		&$QQ$      &0.197(4)&20.0(1.0)&0.002 &0.111(2)&26.3(1.0)&0.31\\
		&$(QQ)Q$   &0.200(3)&18.7(9)&1.18  &0.151(1)&29.6(7)&0.35\\
		\hline           
		\multirow{2}{*}{\begin{turn}{90} \scriptsize{$12a$} \end{turn}}
		&$QQ$      &0.199(5)&22.6(1.5)&0.003 &0.103(2)&28.8(1.5)&0.24\\
		&$(QQ)Q$   &0.200(4)&20(1.4)&1.50 &0.145(2)&32.2(1.0)&0.33\\
		\hline
		\hline                
	\end{tabular}
	\caption{The symmetrized MS width $W^2=(W_{\perp}^2(R/2)+W^2_{\parallel}(R/2))/2$ and amplitude of the action-density at the middle plane $R/2$ from fits to a double-Gaussian ansatz Eq.\eqref{eq:conGE}, at temperatures $T/T_{c}=0.8$ and $T/T_{c}=0.9$.}	
	\label{tab:AAAw2a2T08}	
\end{table}

\begin{table}[!hpt]
	\caption{The returned values of the $\chi^{2}(x)$ from fits of the width of in-plane action density $W_{\parallel}^{2}(x)$ of base $A=2a$ and $A=4a$ at $T/T_{c}=0.9$ to Y-string model Eq.\eqref{eq:perpen_StrFluct} at two values of the string tension.	}
		\begin{tabular}{c|cccc|cccc}
			\hline
			\hline
			\multicolumn{1}{c|}{$\bf{A=2a}$}&\multicolumn{4}{c|}{$\bf{\sigma=0.036}$}&\multicolumn{4}{c}{$\bf{\sigma=0.044}$}\\\hline			         	
			$\chi^{2}_{\rm{d.o.f}}(x_i)$/FR\footnote{FR - denotes the fit range. We considered the following ranges 5a-12a, 6a-12a, 7a-12a, 8a-12a.}&\tiny{5a-12a}&\tiny{6a-12a}&\tiny{7a-12a}&\tiny{8a-12a}&\tiny{5a-12a}&\tiny{6a-12a}&\tiny{7a-12a}&\tiny{8a-12a}  \\
			\hline
			$\chi^{2}_{\rm{d.o.f}}(1)$  &1.26&0.92 &1.08&0.68 &2.11&1.83&1.04&0.66 \\
			$\chi^{2}_{\rm{d.o.f}}(2)$  &1.73&0.30&0.68&0.26 &0.57&0.15 &0.06&0.03\\
			$\chi^{2}_{\rm{d.o.f}}(3)$  &7.27&0.50&0.75&0.24 &4.14&0.17&0.14&0.10\\
			$\chi^{2}_{\rm{d.o.f}}(4)$  &28.4&1.63&0.07&0.06&18.6&0.83&0.76&0.62\\
			\hline
			\multicolumn{1}{c|}{$\bf{A=4a}$}&\multicolumn{4}{c|}{$\bf{\sigma=0.036}$}&\multicolumn{4}{c}{$\bf{\sigma=0.044}$}\\\hline			         				
			$\chi^{2}_{\rm{d.o.f}}(1)$ &3.04 &0.27&0.22&0.12&2.89&0.37&0.13&0.10\\
			$\chi^{2}_{\rm{d.o.f}}(2)$ &0.49 &0.05&0.04&0.05&0.13&0.03&0.65&0.03\\
			$\chi^{2}_{\rm{d.o.f}}(3)$ &1.16 &0.11&0.15&0.07&0.69&0.53&0.42&0.15\\
			$\chi^{2}_{\rm{d.o.f}}(4)$ &11.3 &0.63&0.67&0.13&6.74&0.64 &0.14&0.01\\
			\hline
			\hline		         			
		\end{tabular}
		
	\label{tab:T09P}
\end{table}
\begin{table}[!hpt]
	\caption{Same as Table~\ref{tab:T09P}; however, the values of the $\chi^{2}$ are returned from the fits of formula Eq.\eqref{eq:perpen_StrFluct} to the perpendicular width of the action density $W_{\perp}^{2}(x)$.}
		\begin{tabular}{c|cccc|cccc}
			\hline
			\hline
			\multicolumn{1}{c|}{$\bf{A=2a}$}&\multicolumn{4}{c|}{$\bf{\sigma=0.036}$}&\multicolumn{4}{c}{$\bf{\sigma=0.044}$}\\\hline			         	
			$\chi^{2}_{\rm{d.o.f}}(x_i)$/FR&\tiny{5a-12a}&\tiny{6a-12a}&\tiny{7a-12a}&\tiny{8a-12a}&\tiny{5a-12a}&\tiny{6a-12a}&\tiny{7a-12a}&\tiny{8a-12a} \\
			\hline
			$\chi^{2}_{\rm{d.o.f}}(1)$  &0.43 &0.50&0.56&0.20&6.83&1.18&0.26&0.10\\
			$\chi^{2}_{\rm{d.o.f}}(2)$  &1.81 &0.26&0.33&0.18&14.2&3.78&0.80&0.16\\
			$\chi^{2}_{\rm{d.o.f}}(3)$  &29.2 &9.06&2.12&0.38&28.1&8.59&1.97&0.35\\
			$\chi^{2}_{\rm{d.o.f}}(4)$  &42.3 &18.1&5.57&1.24&41.6&17.3&5.19&1.14\\
			\hline
			\multicolumn{1}{c|}{$\bf{A=4a}$}&\multicolumn{4}{c|}{$\bf{\sigma=0.036}$}&\multicolumn{4}{c}{$\bf{\sigma=0.044}$}\\\hline			         			
			$\chi^{2}_{\rm{d.o.f}}(1)$ &3.38 &1.03 &0.35 &0.16 &3.22&0.97&0.33&0.15\\
			$\chi^{2}_{\rm{d.o.f}}(2)$ &6.49 &1.80 &0.41 &0.01 &6.09&1.66&0.38&0.10\\
			$\chi^{2}_{\rm{d.o.f}}(3)$ &8.48 &2.87 &0.73 &0.15 &8.48&2.63&0.64&0.12\\
			$\chi^{2}_{\rm{d.o.f}}(4)$ &13.9 &4.45 &1.07 &0.20 &13.5&4.17&0.95&0.16\\
			\hline
			\hline		         			
		\end{tabular}
		
	\label{tab:T09N}	
\end{table}
\begin{table}[!hpt]
	\caption{
		The returned $\chi^{2}(x)$ from fits of the baryonic in-plane width $W_{\parallel}^{2}(x)$, $(QQ)Q$ of base length $A=2a$ and $A=4a$ at $T/T_{c}=0.9$, to the mesonic string  Eq.\eqref{sol} at planes $x$.
	}
	\centering
	\begin{tabular}{c| cc cc| cc cc }
		\hline
		\hline
		\multicolumn{1}{c|}{$\bf{A=2a}$}
		&\multicolumn{4}{c|}{$\bf{\sigma_0=0.036}$}
		&\multicolumn{4}{c}{$\bf{\sigma_0=0.044}$}\\
		\hline
		$\chi^{2}_{\rm{d.o.f}}(x_i)$/FR &\tiny{5a-12a}&\tiny{6a-12a}&\tiny{7a-12a}&\tiny{8a-12a}&\tiny{5a-12a}&\tiny{6a-12a}&\tiny{7a-12a}&\tiny{8a-12a} \\
		\hline
		$\chi^{2}_{\rm{d.o.f}}(1)$  &13.9  & 16.2 & 14.87& 10.9 & 24.7 &22.8  &19.1  &16.1 \\ 
		$\chi^{2}_{\rm{d.o.f}}(2)$  &8.68 & 10.4 & 8.89 & 6.88 &31.1 &30.0  &23.2  &16.7 \\
		$\chi^{2}_{\rm{d.o.f}}(3)$  &28.4 & 2.74&2.56  &2.68 &38.7 &39.0  &30.4  &21.0 \\
		$\chi^{2}_{\rm{d.o.f}}(4)$  & 258  &27.54 & 2.04 & 0.54 &43.1 &48.7  &41.3  &29.9 \\
		\hline
		\multicolumn{1}{c|}{$\bf{A=4a}$}
		&\multicolumn{4}{c|}{$\bf{\sigma_0=0.036}$}
		&\multicolumn{4}{c}{$\bf{\sigma_0=0.044}$}\\
		\hline
		$\chi^{2}_{\rm{d.o.f}}(1)$  &8.02   & 6.67    & 7.63     &5.38     &11.4  &14.2 &7.3 &6.55 \\
		$\chi^{2}_{\rm{d.o.f}}(2)$  &3.56    & 4.17    &3.66      &2.78     &13.3  &11.9 &9.0 &6.75 \\
		$\chi^{2}_{\rm{d.o.f}}(3)$  &9.35    & 1.15    & 1.22     & 1.30 &17.7  &16.8 &12.6 &9.3 \\
		$\chi^{2}_{\rm{d.o.f}}(4)$  &95.9    & 10.71     &0.911  &0.42     &21.1  &23.2 &19.9 &15.1 \\
		\hline
		\hline
	\end{tabular}
	\label{tab:T09M}	
\end{table}

The amplitude profiles in Table.~\ref{tab:AAAw2a2T09} demonstrate, at $T/T_{c}=0.9$, the similarity near the quark for each system. Close to the diquark, nevertheless, the magnitude of amplitudes signifies vacuum expulsions are greater than that in the proximity of the quark. The width differences $\Delta W^{2}=W^{2}(x_{i})-W^{2}(\frac{R}{2})$ along the flux tube  are very small and within the uncertainties of the fit. This is not alike with the curved MS width profile along the $Q\bar{Q}$ transverse planes shown in ~\cite{Bakry:2020ebo, Bakry:2010zt}.

The values of the MS width of the flux-tube at the first four lattice slices from the diquark system are plotted in Fig.\ref{fig:planesT09}. The width growth is presented at the middle planes $x=\frac{R}{2}$ in Fig.\ref{fig:MidplaneT09}. The MS width of the in-plane $W^{2}_{\parallel}$ and perpendicular component $W^{2}_{\perp}$ of the $Q(QQ)$ system display a cylindrical symmetry; even so, the string MS width profile is not identical to that of the $Q\bar{Q}$ system. The coincidence with the mesonic string does not manifest either at small or intermediate separation regions $R<10a$.  

At temperature scale $T/T_{c}=0.8$, we find the MS width of the energy profiles of $(QQ)Q$ to be very similar considering the middle plane as depicted in Fig.\ref{fig:MidplaneT08}. The same assertion in regard to the broadening profile at planes other than the middle holds as well.  The action density exhibits cylindrical symmetry even for $3Q$ arrangement with a larger base length $A=4a$.

\subsection{Broadening of effective strings}
The objective of this section is to lay out the characteristics of the broadening profile of the $(QQ)Q$ flux tube. 
This is another instance on the feasibility of bosonic string models to dissect the growth behavior versus color source separation. 

The lattice data of the MS width is opposed with the string model  Eq.\eqref{eq:perpen_StrFluct}, Eq.\eqref{eq:inplane_StrFluct}, and Eq.\eqref{sol} for both mesonic and baryonic strings, respectively. The fit of the MS width data is discussed considering two values of the string tension, that is, the standard value returned from the lattice simulations at zero temperature $\sigma_{0} a^{2}=0.044$~\cite{Koma:2017hcm} and also returned from the fits of the $Q\bar{Q}$ system at $T/T_{c}=0.8$~\cite{Bakry:2020flt}, and the other value $\sigma_{0} a^{2}=0.036$ obtained from the fits of $Q\bar{Q}$ at $T/T_{c}=0.9$~\cite{Bakry:2020flt}.

As the third quark $Q_3$ is being dragged distance $R$ apart from the triangle's base, we look over the broadening at the first four subsequent planes from the diquark $x=1,2,3,4$. This should unveil whether there are fairly substantial signatures of the baryonic junction on the broadening of MS of the action density.

\begin{figure}[!hptb]
	\centering
	\subfigure[mesonic string]{\includegraphics[scale=0.20]{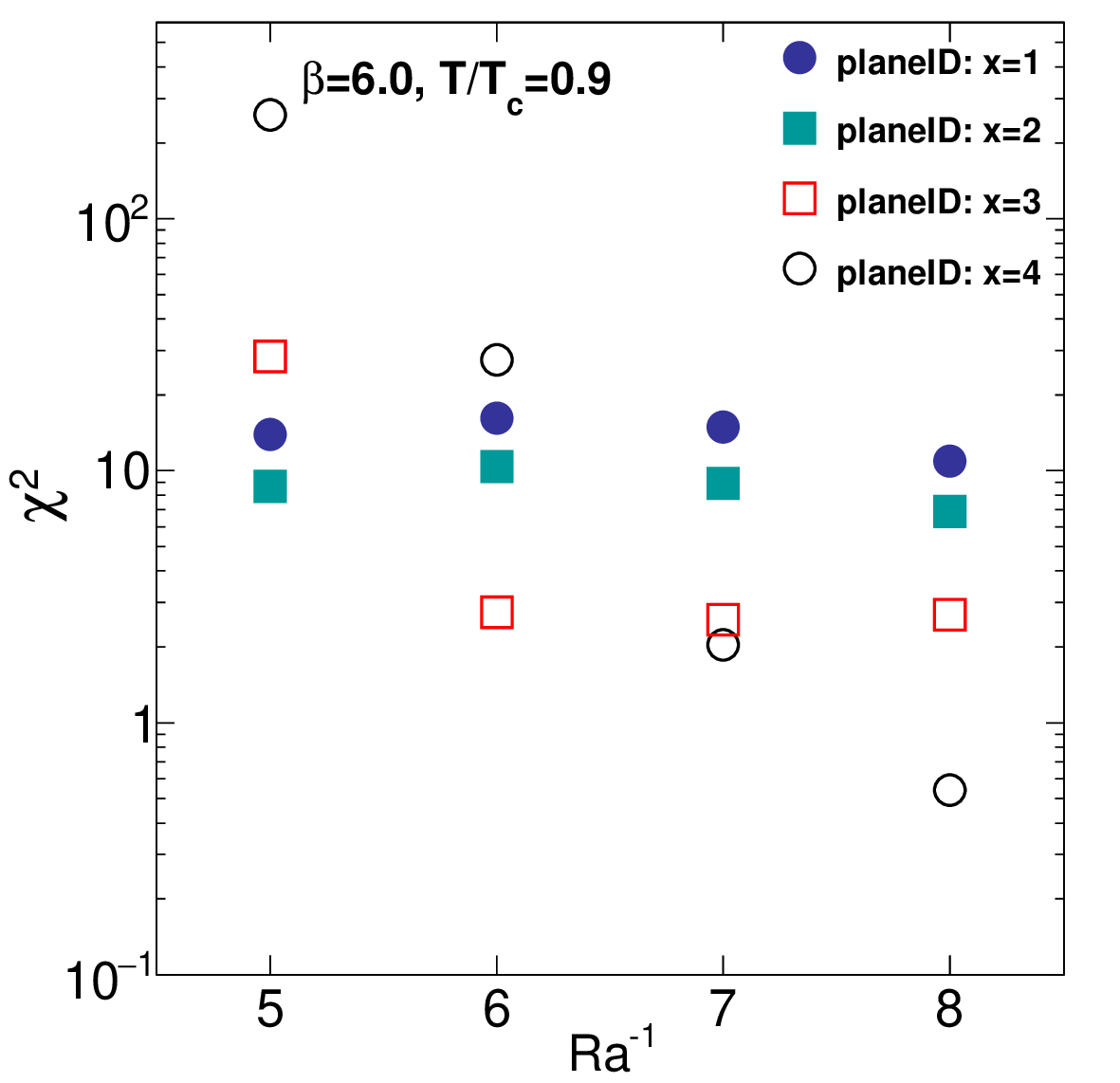}}	
	\subfigure[mesonic string]{\includegraphics[scale=0.20]{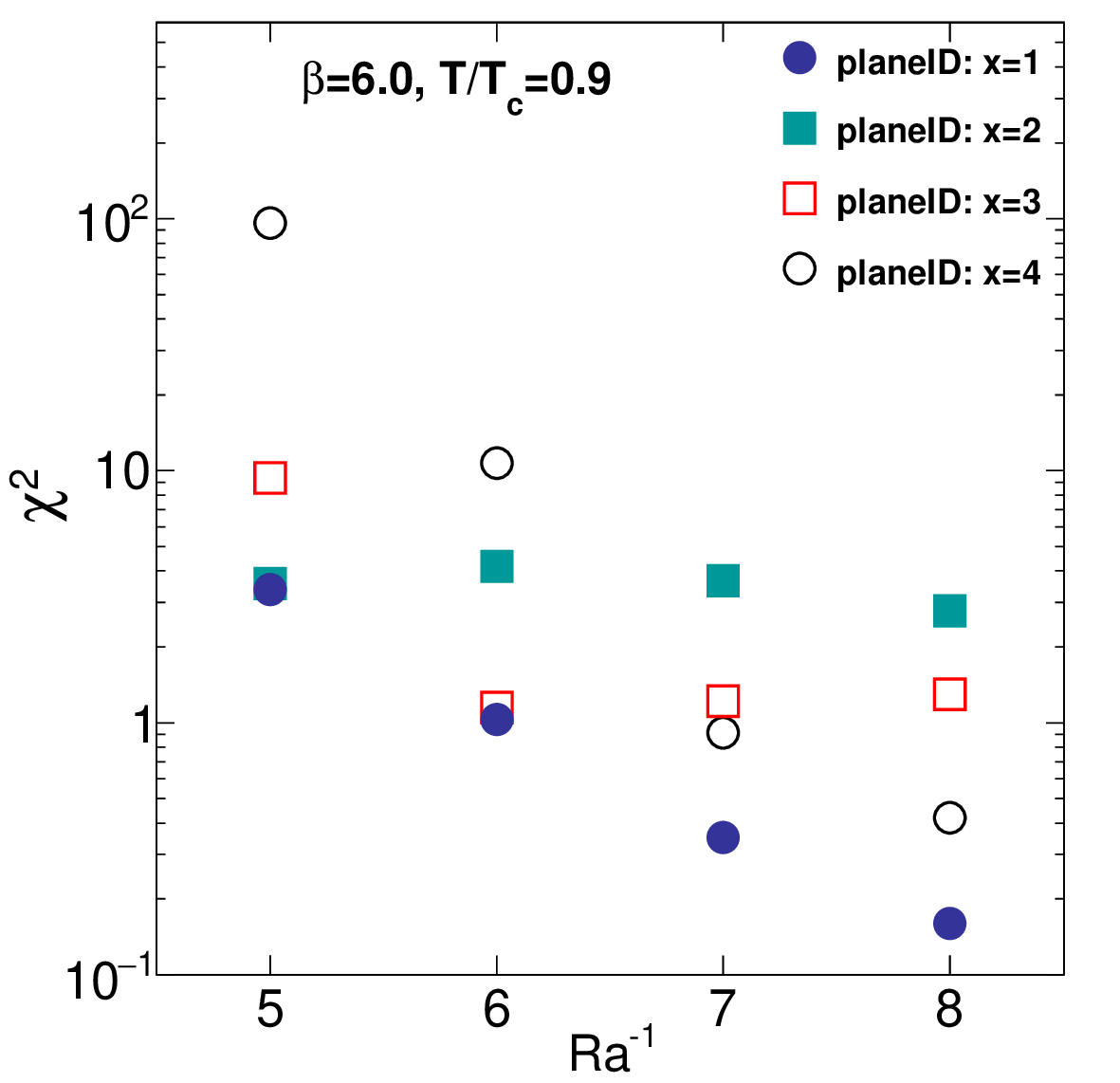}}
	\subfigure[baryonic string]{\includegraphics[scale=0.20]{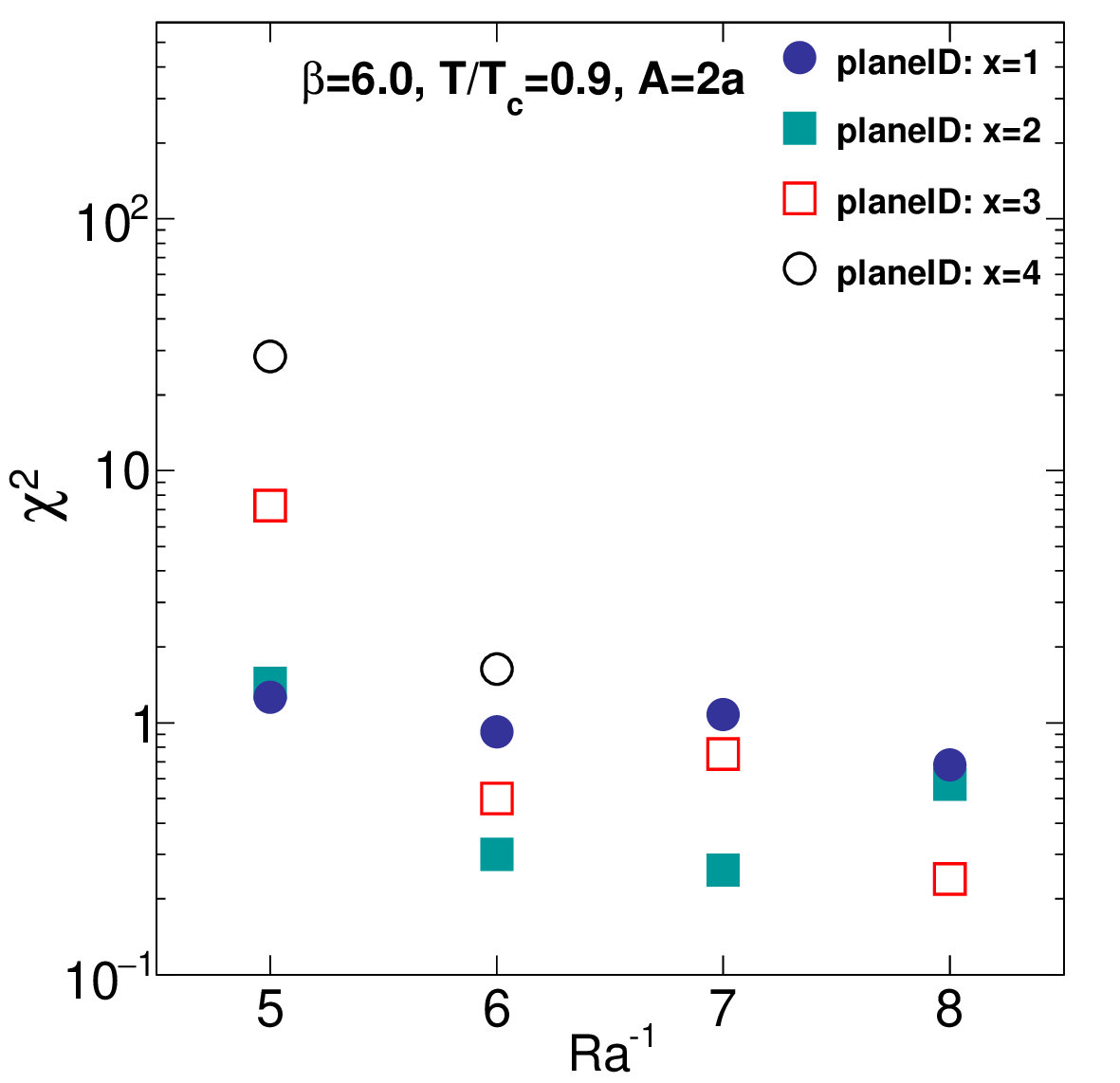}}	
	\subfigure[baryonic string]{\includegraphics[scale=0.20]{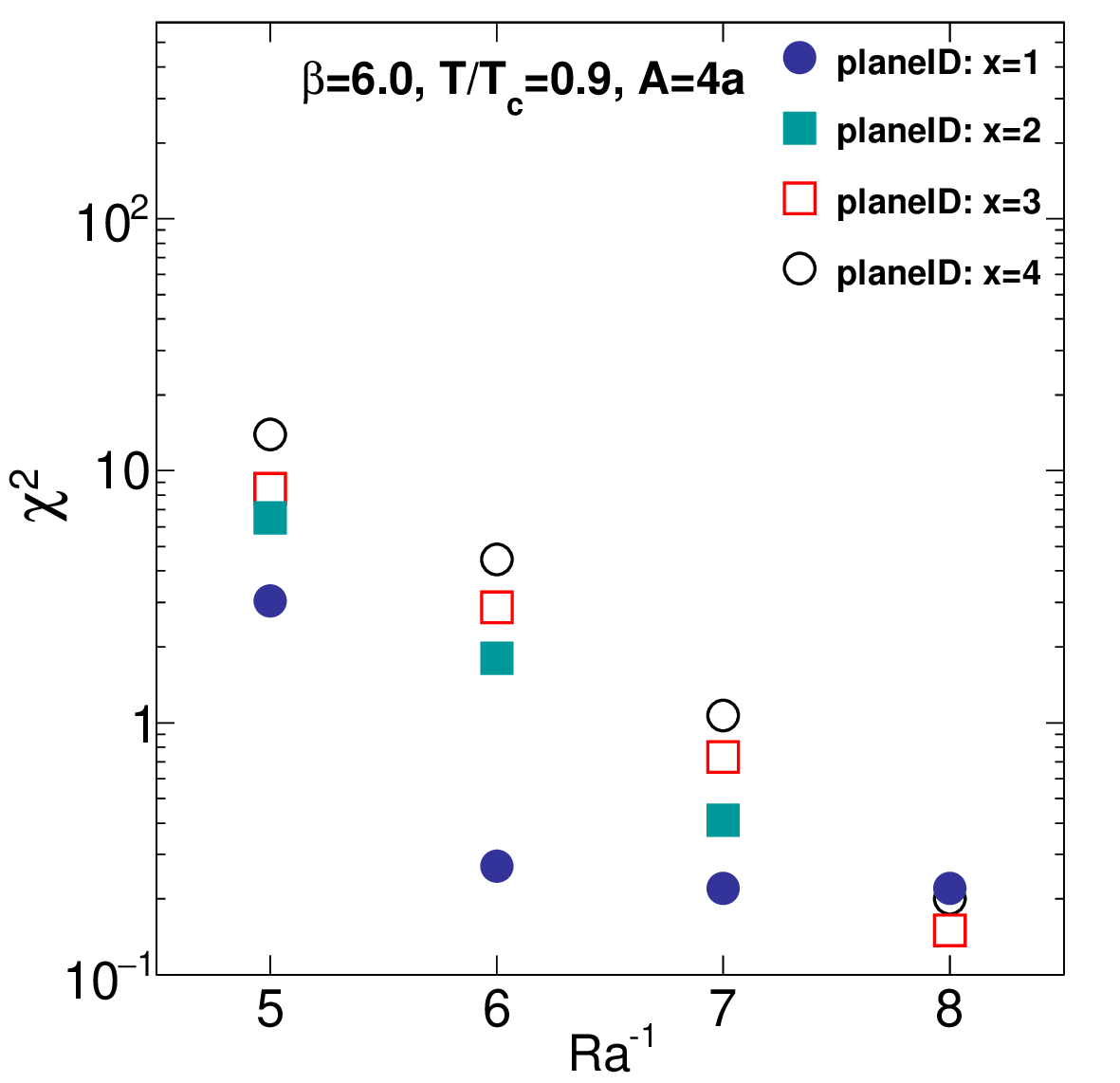}}
	\caption{(a,b) Chart of $\chi^{2}$ retrieved from the fits of
			mesonic string Eq.\eqref{sol} to MS width $W_{\parallel}^{2}(x_i)$ at $\sigma_0=0.036$ and $T/T_c=0.9$; $A=2a$ in (a) and $A=4a$ in (b). (c,d) Same as (a,b), however, $\chi^{2}$ values are from the fits of the baryonic string model Eq.\eqref{eq:inplane_StrFluct}; diquark base diameters $A=2a$ in (c) and $A=4a$ in (d).}
	\label{fig:ChiT09M}
\end{figure}
\begin{figure}[!hptb]
	\centering	 
	\subfigure[]{\includegraphics[scale=0.38]{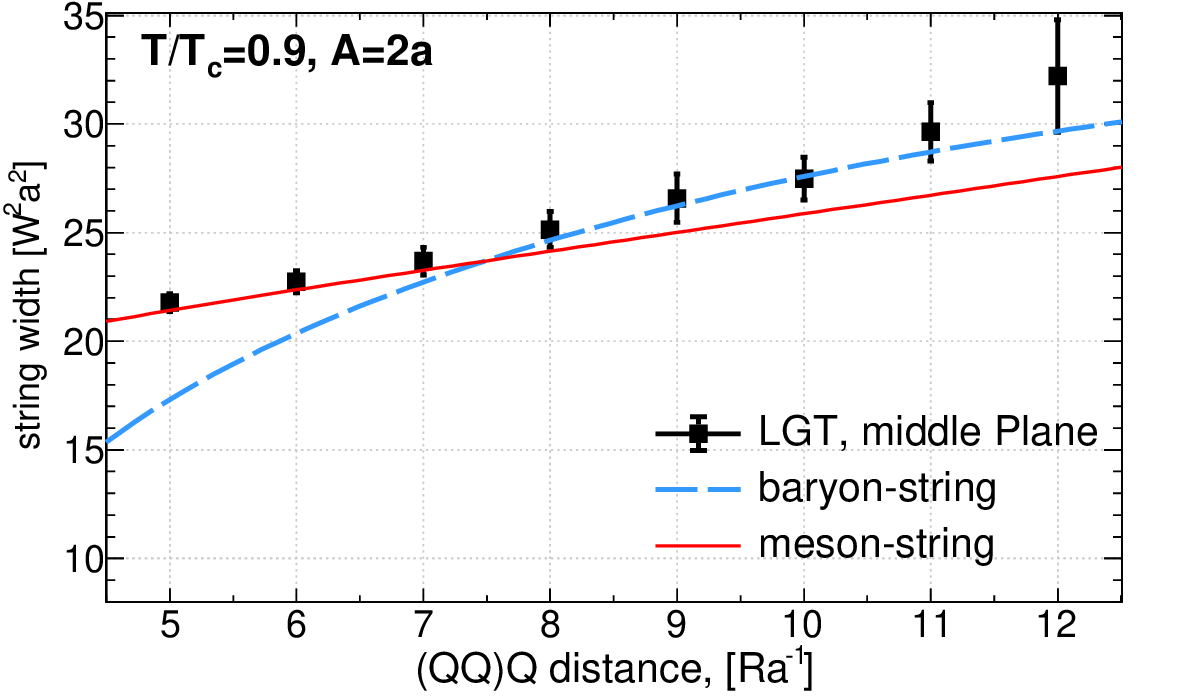}}
	\subfigure[]{\includegraphics[scale=0.38]{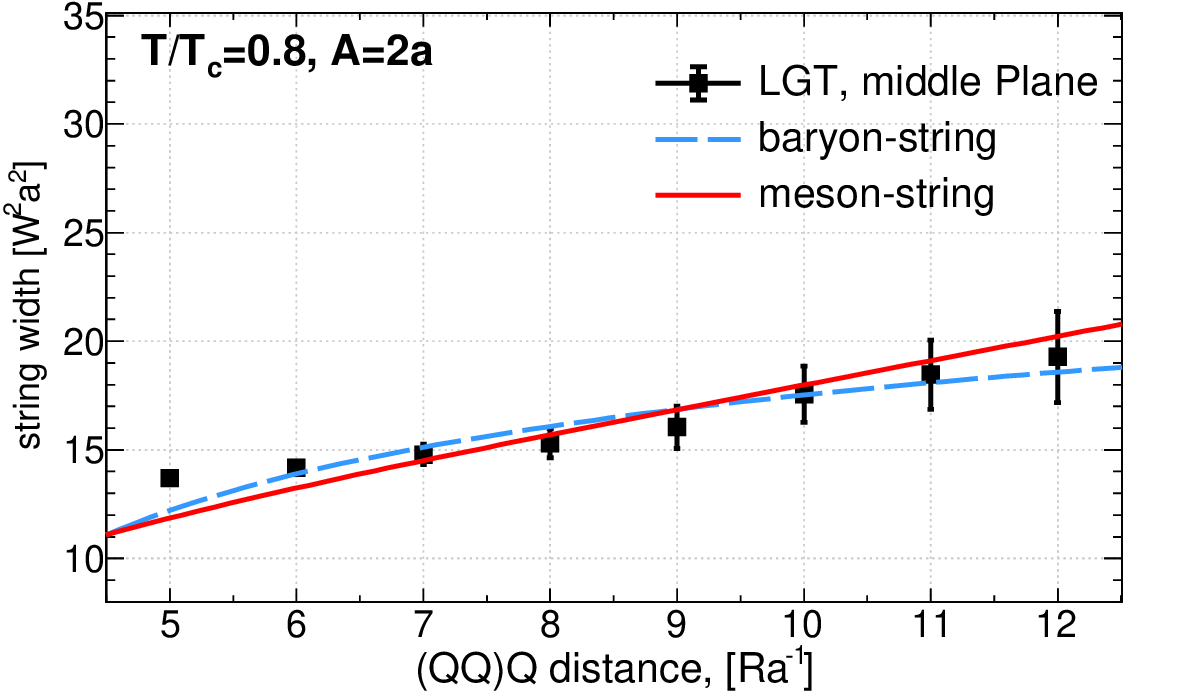}}	  
	\caption{(a) The MS width of the $(QQ)Q$ action density $(W_{\perp}^{2}(R/2)+W_{\parallel}^{2}(R/2))/2$ at the middle plane $x=R/2$ and $T/T_c=0.9$. The dashed line corresponds to fitted  to the sum of the baryonic string Eqs.\eqref{eq:perpen_StrFluct} and \eqref{eq:inplane_StrFluct}, the solid line corresponds to the mesonic string Eq.\eqref{sol}. (b) The data and fits of MS width correspond to the temperature $T/T_{c}=0.8$.}
	\label{fig:widfitmid}
\end{figure}
\begin{figure*}[t]
	\centering
	\subfigure[Plane $x=1$]{\includegraphics[scale=0.38]{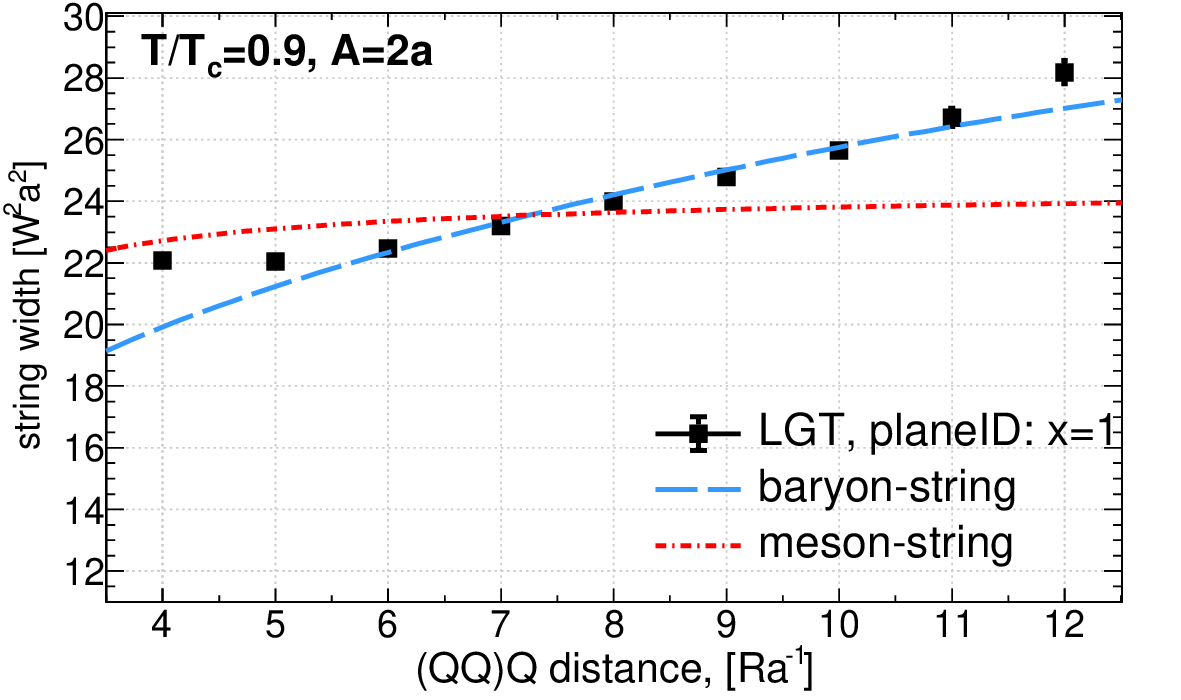}}
	\subfigure[Plane $x=2$]{\includegraphics[scale=0.38]{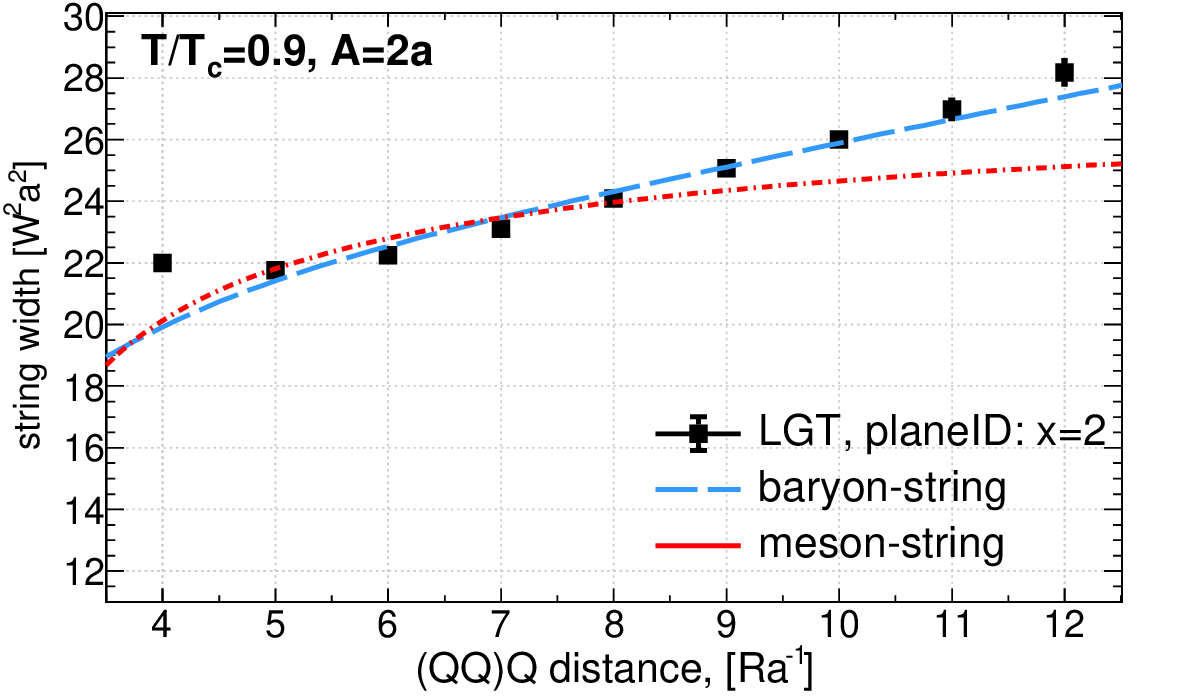}}
	\subfigure[Plane $x=1$]{\includegraphics[scale=0.38]{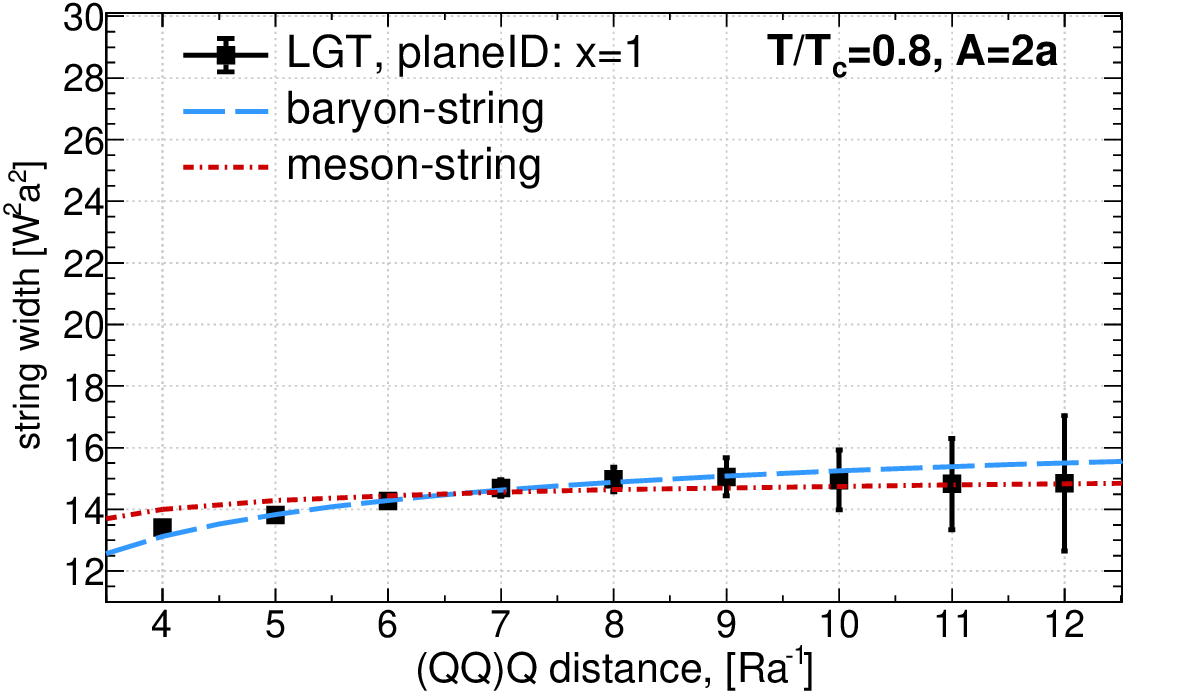}}
	\subfigure[Plane $x=2$]{\includegraphics[scale=0.38]{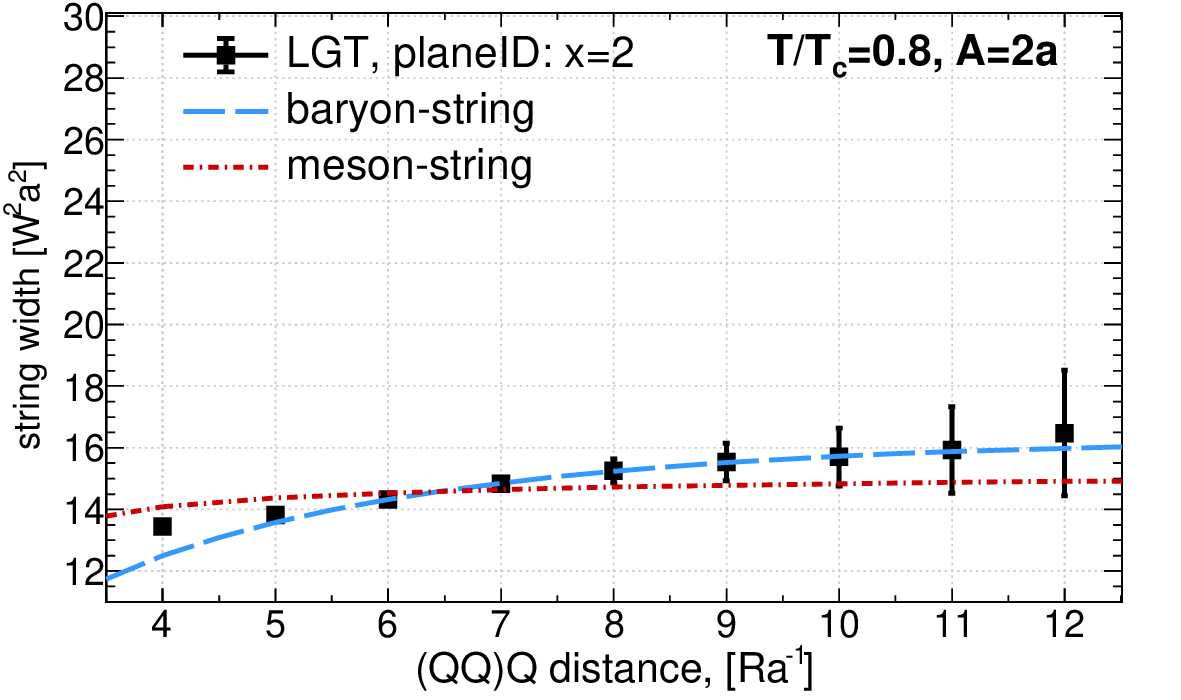}}
	\caption{
		(a) The MS width of the action density of the $(QQ)Q$  at plane $x=1$, the lines are the fit of baryonic string Eq.\eqref{eq:inplane_StrFluct} for the in-plane fluctuation of the junction and mesonic string Eq.\eqref{sol} at plane $x=1$ temperature $T/T_c=0.9$. (b) Same as (a) however at the plane $x=2$. (c,d) The data and fits of MS correspond to the temperature $T/T_{c}=0.8$.}
	\label{fig:widfitx1}
\end{figure*}
The Y-string implies perpendicular and in-plane MS width of the junction fluctuations given by  Eq.\eqref{eq:perpen_StrFluct} and Eq.\eqref{eq:inplane_StrFluct}, respectively. Since the junction's oscillations are not projected to smooth out and will likely produce a local peak, the features of the fit ought to be scrutinized at each selected transverse plane to the tube’s measured widths. 
In the same context, we recall from the discussion at the end of Sec.\ref{sec:EffectiveBosonicStringModel} that the junction must be placed in formulas Eq.\eqref{eq:perpen_StrFluct} and Eq.\eqref{eq:inplane_StrFluct} at the corresponding flux plane $x_{i}$ where the fit takes place.

Table~\ref{tab:T09P} and Table~\ref{tab:T09N} include  the outcome values of  $\chi_{\rm{d.o.f}}^{2}(x_{i})$ from the fits of Eq.\eqref{eq:inplane_StrFluct} and Eq.\eqref{eq:perpen_StrFluct} to the two width components $W_{\parallel}^{2}(x_i)$ and $W_{\perp}^{2}(x_i)$ enlisted in Table~\ref{tab:AAAw2a2T09}) for the indicated $(QQ)-Q$ distance $R \in [a,b]$ at four consecutive transverse planes $x=1$ to $x=4$, at the highest temperature $T/T_c=0.9$.

Inspection of the two tables exposes that best fits are retrieved for planes $x=1$ and $x=2$, which are one to two lattice spacings from the diquark base. This is consistent with the results in Ref.~\cite{Bakry:2014gea}, where we performed a comparative analysis with base length $A>4a$. In that analysis, it is shown that the values of $\chi_{\rm{d.o.f}}^2(x_{i})$ are optimized at the closest plane to the intersection of two distant identical Gaussians used to fit the flux profile. That is, certain planes display higher contributions received from the fluctuations in the vicinity of the junction.

The length of the base of the triangular isosceles quark configuration affects the plane at which we attain the minimum in $\chi_{\rm{d.o.f}}^2(x_{i})$~\cite{Bakry:2014gea}. The two strings of the Y-configuration connecting the diquark of base diameter $A \leq 4a$ are close enough that we obtain the best fits at the first two planes from the diquark base which are near the Fermat point of the configuration.

Actually, the occurrence of certain planes at which the lattice data best agrees with the baryonic Y-string formulas Eq.\eqref{eq:perpen_StrFluct} and Eq.\eqref{eq:inplane_StrFluct}  suggests that the junction impacts are manifesting at the highest temperatures $T/T_{c}=0.9$. The effects of the junction eventually fade away at distant planes from the diquark.

Following the same line of reasoning regarding the $3Q$ potential in the previous section, we would like to assess the mesoniclike aspects of the width of the fluctuations at $T/T_{c}=0.9$. The width of the mesonic string Eq.\eqref{sol} is fitted to the MS width $W^{2}_{\parallel}$ of the baryonic flux-tube (Table ~\ref{tab:T09P} and \ref{tab:T09N}). Similar to the baryonic string analysis, In Table~\ref{tab:T09M} the resultant $\chi_{\rm{d.o.f}}^{2}(x_{i})$ are collected from the fits corresponding to diquark-quark $(QQ)Q$ of bases diameter $A=2a$ and $A=4a$ at two values of the string tension.

The fits return  large residuals $\chi_{\rm{d.o.f}}^{2}(x_{i})$ which can readily seen from the chart of $\chi_{\rm{d.o.f}}^{2}(x_{i})$ in Fig.\ref{fig:ChiT09M}(a,b) at most planes. The consideration of string tension value $\sigma_{0}a^{2}=0.036$ reduces the residuals compared to the fits adopting $\sigma_{0}a^{2}=0.044$, even so, the poor fits are still persisting for diquark base diameter $A=2a$. This contrast with the good Y-string model's fits signals, thereof, a cross over of the flux-tube into the junction behavior.

On the other hand, at the distant planes $x=3$ and $x=4$ from the diquark of wider base $A=4a$ the fits return good $\chi_{\rm{d.o.f}}^{2}$ values at large $QQ-Q$ separation $R$. These findings would indicate that the mesonic traits of the baryonic flux tube tend to show up in the vicinity of the quark rather than the diquark at large enough $R$. Similar manifestation in general $3Q$ configurations may be anticipated when two quarks are close enough relative to the third quark.

At the temperature $T/T_c=0.9$, the lines corresponding to best fits of the baryonic string Eq.\eqref{eq:inplane_StrFluct} and mesonic string Eq.\eqref{sol} to the MS width are plotted in
Fig.\ref{fig:widfitmid}(a) and Fig.\ref{fig:widfitx1}(a,b). The figures correspond to the middle plane $R/2$ and the planes $x=1,2$, respectively. The fits  interval of the latter $R \in [5a,12a]$ and $R\in [7a,12a]$ for the former. The string tension is set to the value $\sigma_0 a^2=0.036$. 

Both figures display the poor fit of the mesonic string to the lattice data of the diquark-quark $(QQ)Q$ at this temperature scale. The plots, on the other hand, show the good correspondence of 
the baryonic string model with the data reflecting the returned $\chi^{2}_{\rm{d.o.f}}$ values in Tables~\ref{tab:T09P} and plotted in Fig.\ref{fig:ChiT09M}(c,d). This supports  that junction interactions ensue in this temperature.
\begin{table}[!hpt]
	\caption{
		$\chi^{2}(x)$ from the fit of mesonic Eq.\eqref{sol} and baryonic strings Eq.\eqref{eq:inplane_StrFluct} at  $T/T_{c}=0.8$ to the  MS width $W^{2}_{\parallel}$ of $(QQ)Q$, at planes $x=1,2$ and interval $R\in[5a,12a]$.}
	\begin{tabular}{ccc|ccc}
		\hline
		\hline
		\multicolumn{1}{c}{\bf{Meson:}}
		&\multicolumn{1}{c}{$\bf{A=2a}$}
		&\multicolumn{1}{c|}{$\bf{A=4a}$}
		&\multicolumn{1}{c}{\bf{Baryon:}}
		&\multicolumn{1}{c}{$\bf{A=2a}$}
		&\multicolumn{1}{c}{$\bf{A=4a}$}
		\\
		\hline   
		\hline    
		$\chi^{2}(1)$ &1.29  & 0.36 &&0.48 &0.06  \\
		$\chi^{2}(2)$ &0.20  & 0.18 &&0.45 &0.48  \\
		\hline		
		\hline
	\end{tabular}
\label{tab:T08PM}
\end{table}

At the lower temperature $T/T_c=0.8$, the returned $\chi^{2}(x_{i})$ from the fits to MS width $W_{\parallel}^{2}(x_{i})$ of the diquark-quark are enlisted in Table~\ref{tab:T08PM}. The two panels in the table compare the fits of both the mesonic and baryonic strings Eq.\eqref{sol} and Eq.\eqref{eq:inplane_StrFluct} fixing the string tension to the value $\sigma_{0}a^{2}=0.044$ on the fit interval $R\in[5a,12a]$, respectively.

The fits of the mesonic string returns good $\chi_{\rm{d.o.f}}^{2}(x_{i})$ at $x=1,2$ planes. This is depicted in the plots  Fig.\ref{fig:widfitx1}(c,d). Also Fig.\ref{fig:widfitmid}(b) show good match considering fit interval $R\in [7a,12a]$.  The diminishing of the deviations in Fig.\ref{fig:widfitmid}(a) and Fig.\ref{fig:widfitx1}(a,b) of the lattice data from the mesonic string's width profile is palpable. 

The findings from the analysis in this section concur with that of the potential analysis in the preceding section. In addition, we stress on the close analogy between the presented results at the temperature $T/T_{c}=0.8$, which is near the end of the plateau region of QCD phase diagram~\cite{QCDPhase}, with around 10\% reduction in the string tension~\cite{Bakry:2020flt,Kac,PhysRevD.85.077501}, and the analysis utilizing the Wilson-loop overlap formalism~\cite{Bissey:2009gw} or three Polyakov-loops ~\cite{Koma:2017hcm} at low temperature.

\section{Conclusion and prospect}  
\label{sec:Conclusions}
In this work, we inspect the similarity between the gluon flux tubes for the quark-antiquark $Q\bar{Q}$ and three quark systems at finite temperature. We approximate the baryonic quark-diquark $(QQ)Q$ configuration by constructing the two quarks at a small separation distance of at least $0.2$ fm.

The potential and energy-density characteristics of the $Q\bar{Q}$ and $(QQ)Q$ systems are examined. Both the potential and the action-density correlator provide the same almost identical structure up to temperatures near the end of the plateau region of the QCD phase diagram~\cite{QCDPhase}. 
 
However, when the temperature gets close to the deconfinement point, the similarity between the two systems breakdown. The gluonic characteristics display splitting for $(QQ)-Q$ distance $R < 1.0$ fm.

The numerical data of the $(QQ)Q$ potential are investigated in light of the fits of mesonic and baryonic string models. The baryonic string model approaches the free mesonic string reproducing the same value for the string tension. However, near the critical point $T/T_ c=0.9$, in contrast to the free mesonic string the Y-baryonic string model's fit of the potential yields a decent $\chi^{2}$ value with a string tension value same as the corresponding $Q\bar{Q}$ arrangement.

Similarly, the analysis of the MS width of the energy profile indicates baryonic-like aspects consistent with the Goldstone modes of Y-bosonic string at $T/T_c=0.9$ at all considered transverse planes. The mesonic string profile displays large deviations from the diquark-quark data at planes close to the diquark system. At the lower temperature, the $(QQ)Q$ baryon displays a broadening profile consistent with both mesonic and Y-string models with the same string tension as the quark-antiquark $Q\bar{Q}$ system.

These findings limit the validity that, in the quenched approximation, the $(QQ)Q$ precisely share many properties in common with the $Q\bar{Q}$ only to temperatures corresponding to the plateau region of the QCD phase diagram~\cite{Koma:2017hcm, Bissey:2009gw}; otherwise, excited baryonic states can manifest around small neighborhoods of the QCD critical point signaling a cross over into the junction behavior.

It would be intriguing to perform the computations afterward while using smaller lattices and taking greater temperatures or dynamical quarks into consideration. It is justifiable that the meson-baryon similarity would be questioned in the context of an excited spectrum ~\cite{ExBar, Juge:2002br, Bicudo:2018yhk}, or in the presence of strong magnetic fields ~\cite{Bali:2012zg, SBstrongMagnetic, Bonati:2018uwh, refId0, MarquesValois:2021kvf}. Future work ought to probe these arrangements, which are likely to be of substantial importance to phenomenological models of hadron structure.

\begin{acknowledgments}
The authors thank the Yukawa Institute for Theoretical Physics, Kyoto University. Discussions during the YITP workshop YITP-T-14-03 on “Hadrons and Hadron Interactions in QCD" were useful in completing this work. We thank Philippe de Forcrand for suggesting the investigation of the mesonic aspects of the (3Q) systems. We also acknowledge the suggestions and careful revision provided by the PRD reviewers. This work has been funded by the Chinese Academy of Sciences President's International Fellowship Initiative grants No.2015PM062 and No.2016PM043, the Recruitment Program of Foreign Experts, NSFC grants (No.~11035006, No.~11175215, No.~11175220) and the Hundred Talent Program of the Chinese Academy of Sciences (Y101020BR0).
\end{acknowledgments}

\appendix

\label{sec:Appendix_A}
\section{Flux tubes in 2D-plane} 

\begin{figure*}[!hptb]
	\centering 
	\subfigure[\tiny{$\parallel$-plane,$R=4a$}]{\includegraphics[scale=0.16]{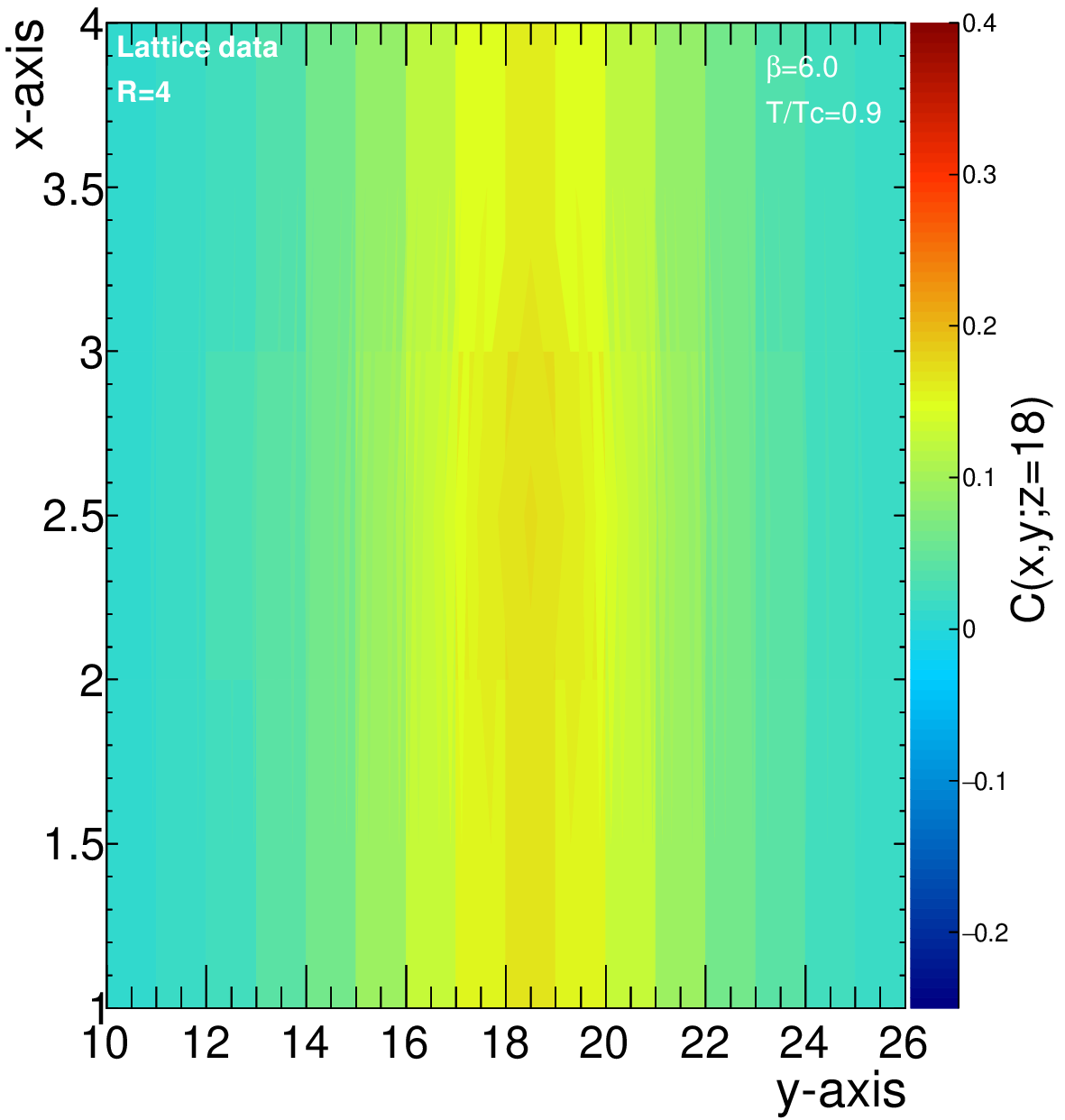}}
	\subfigure[\tiny{$\parallel$-plane,$R=5a$}]{\includegraphics[scale=0.16]{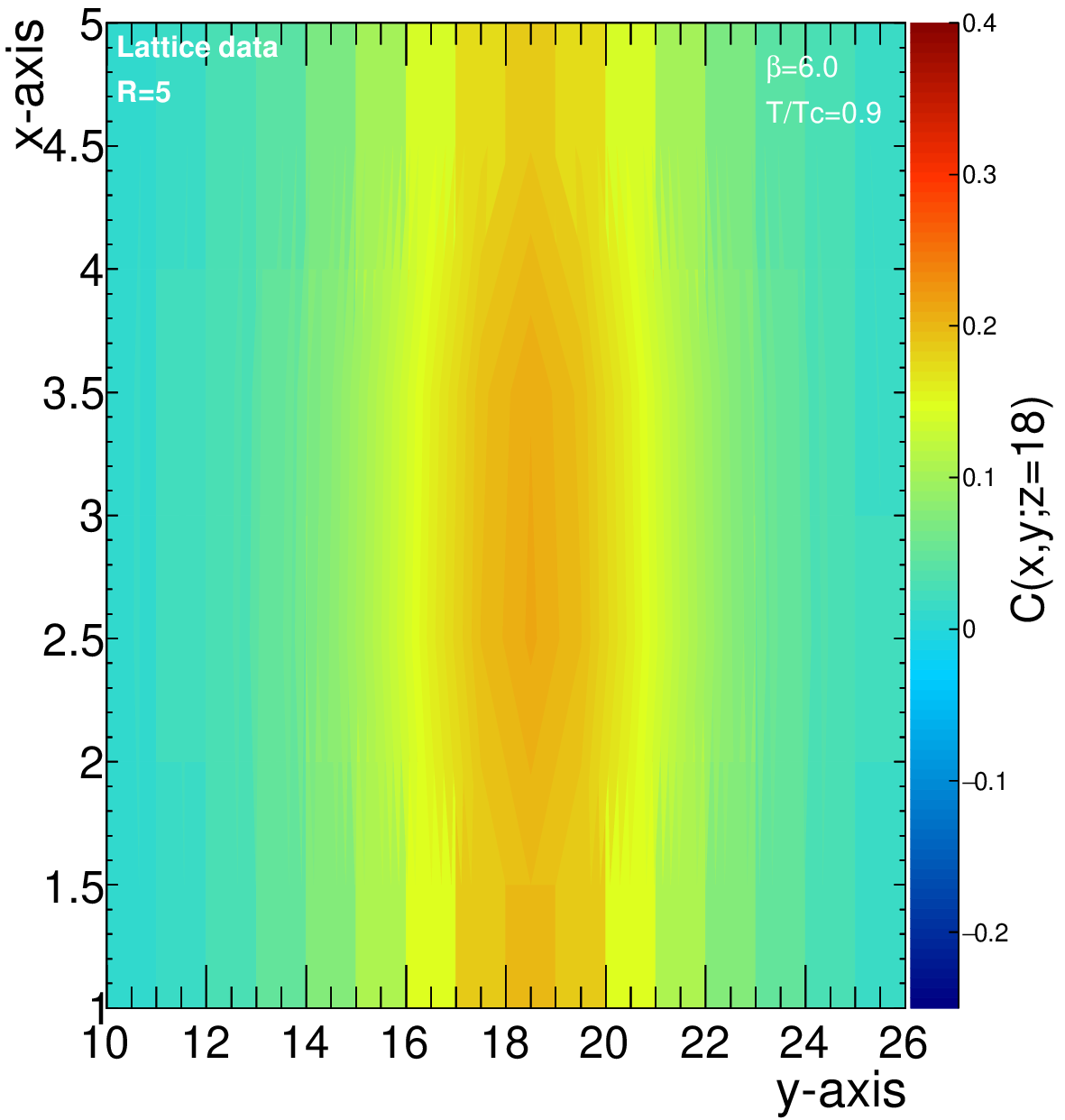}}
	\subfigure[\tiny{$\parallel$-plane,$R=6a$}]{\includegraphics[scale=0.16]{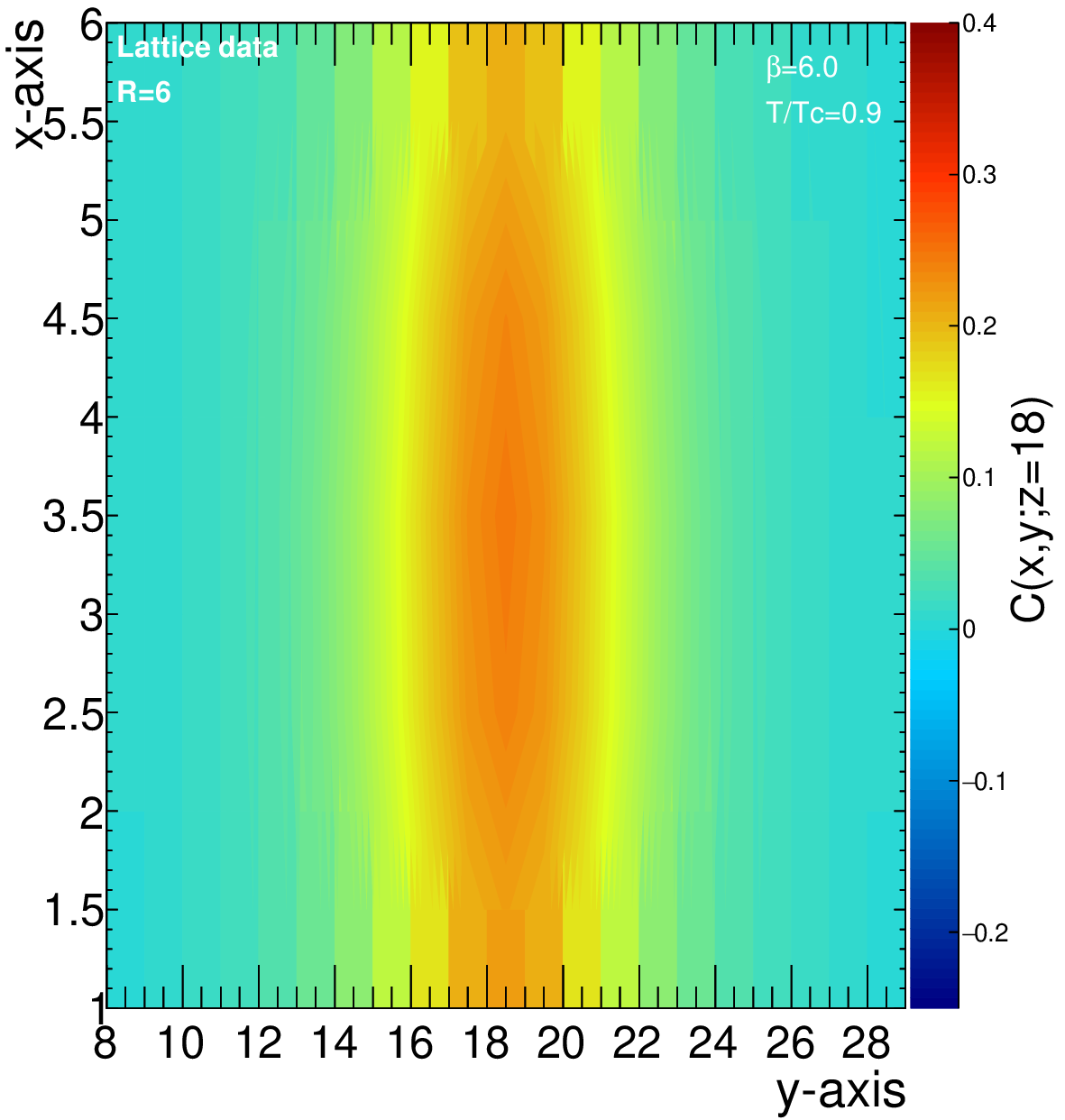}}
	\subfigure[\tiny{$\parallel$-plane,$R=9a$}]{\includegraphics[scale=0.16]{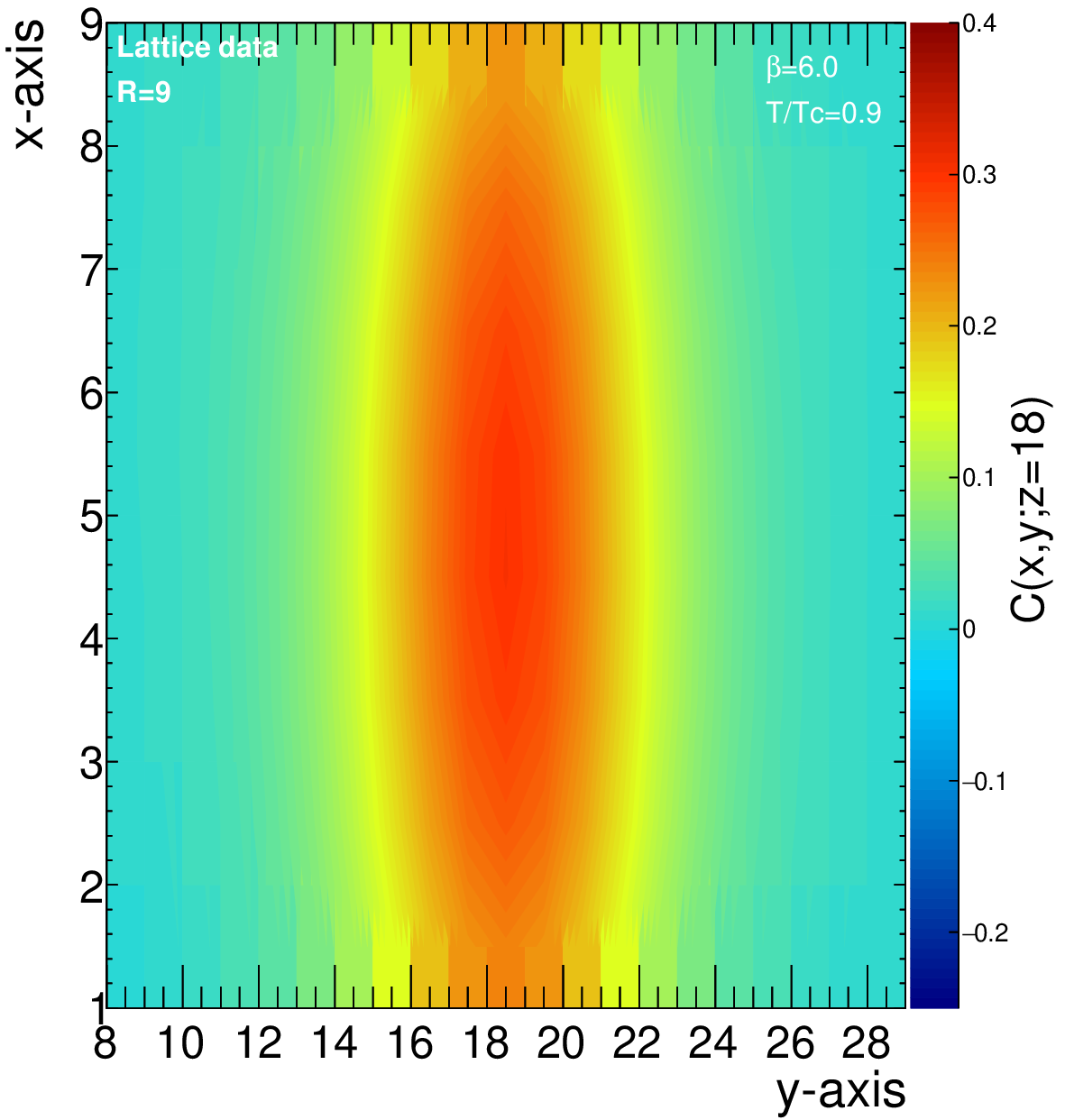}}
	\subfigure[\tiny{$\parallel$-plane,$R=12a$}]{\includegraphics[scale=0.16]{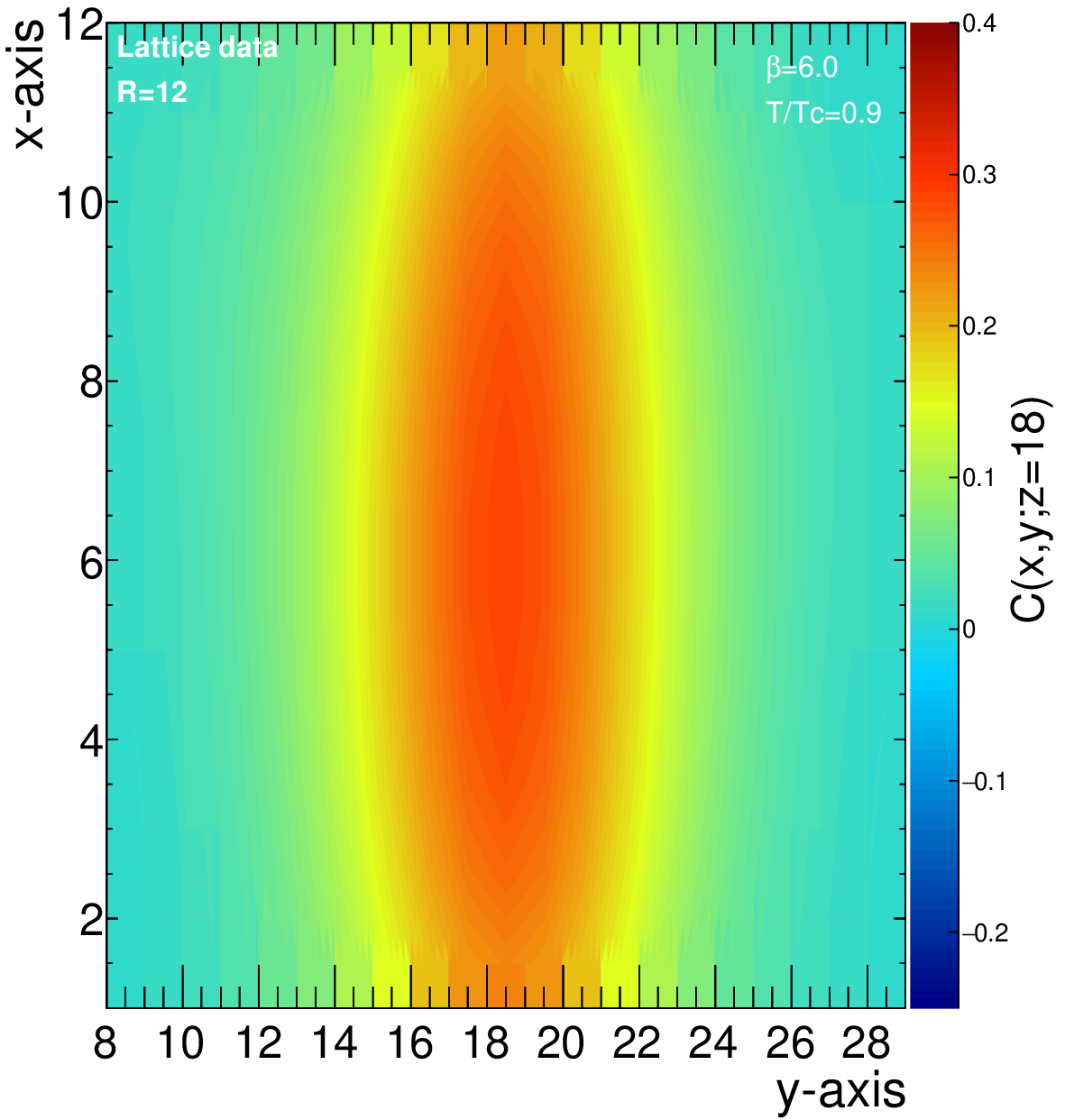}}\\
	
	\subfigure[\tiny{$\parallel$-plane,$R=4a$}]{\includegraphics[scale=0.16]{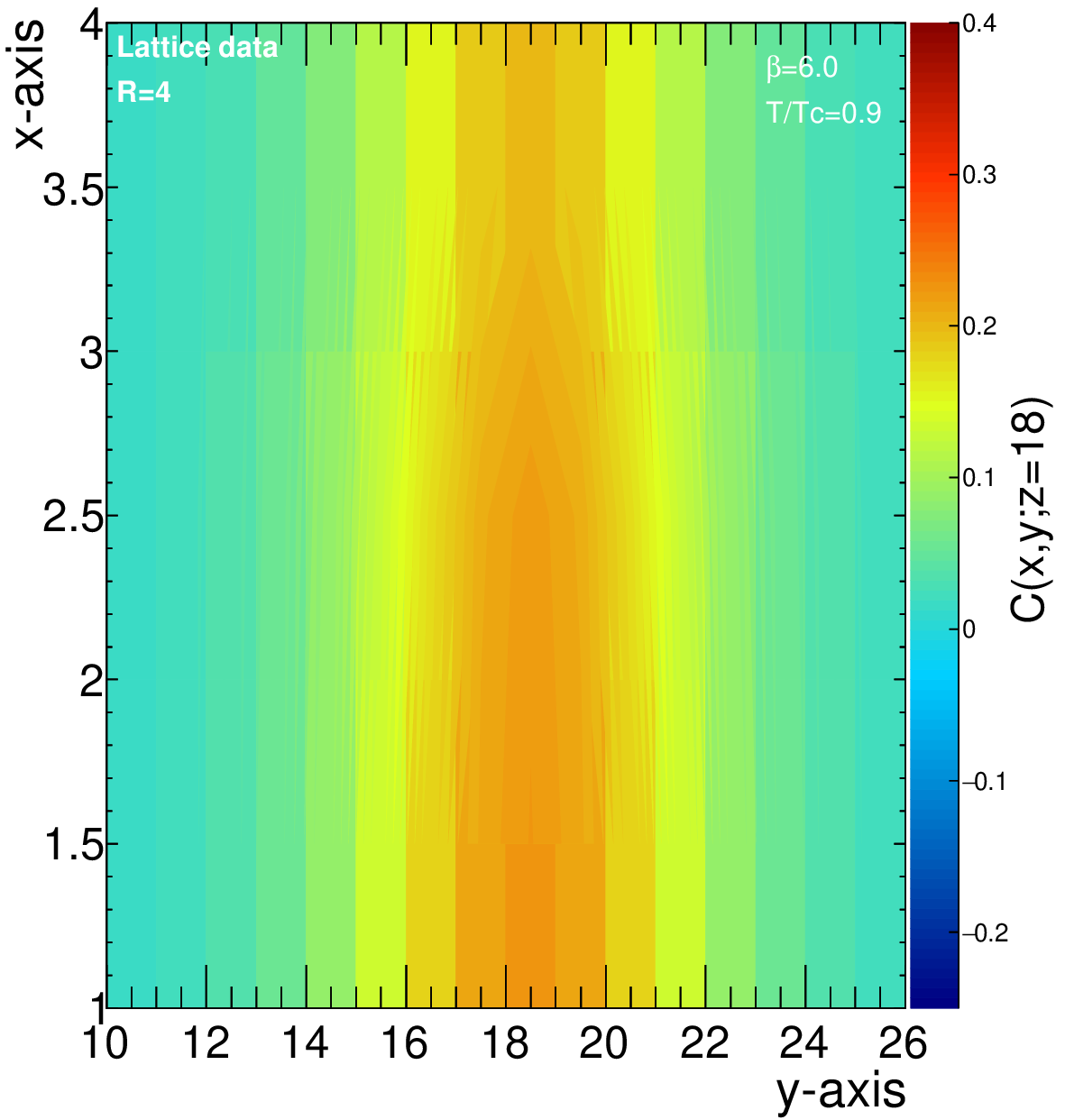}}
	\subfigure[\tiny{$\parallel$-plane,$R=5a$}]{\includegraphics[scale=0.16]{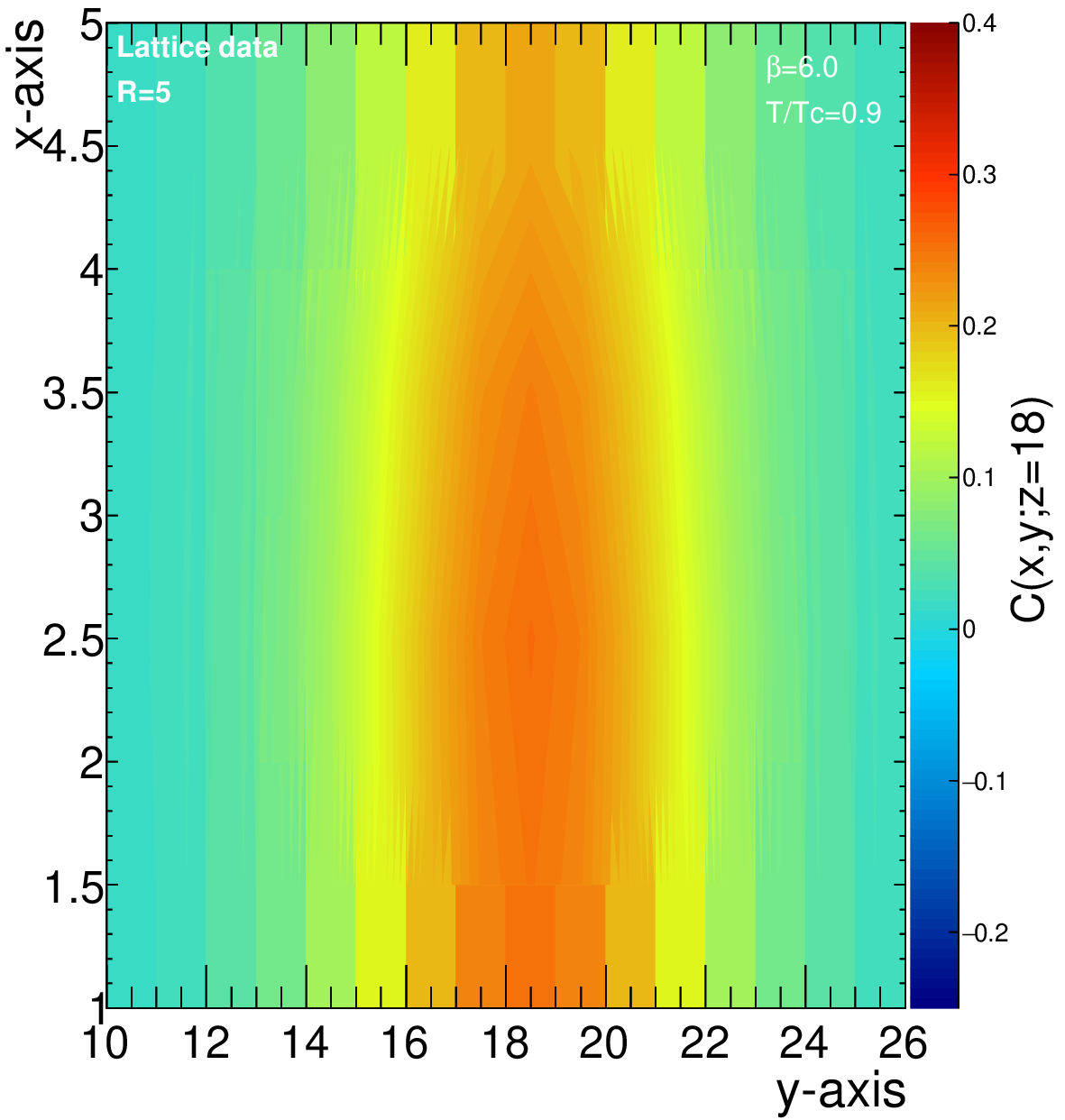}}
	\subfigure[\tiny{$\parallel$-plane,$R=6a$}]{\includegraphics[scale=0.16]{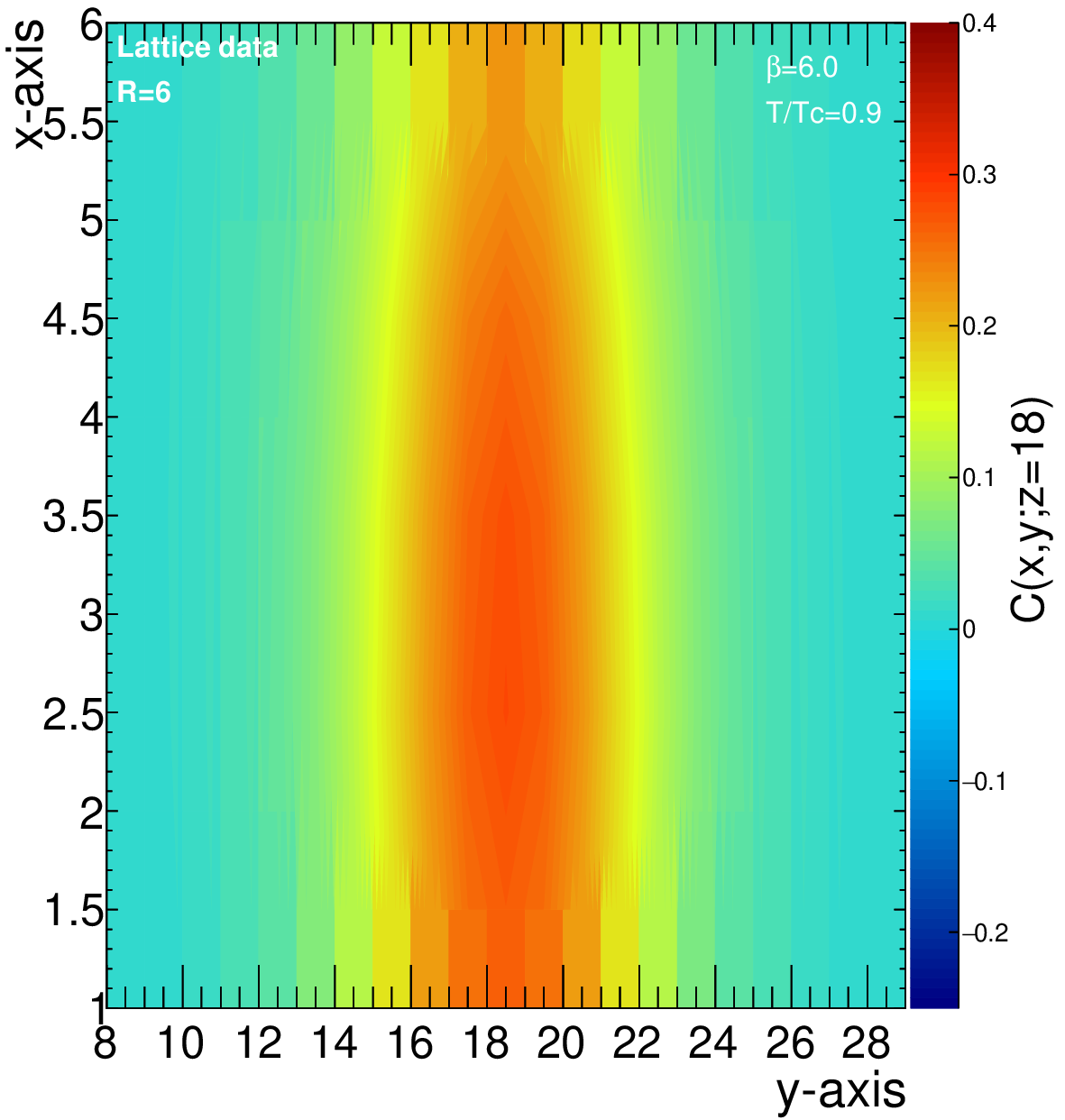}}
	\subfigure[\tiny{$\parallel$-plane,$R=9a$}]{\includegraphics[scale=0.16]{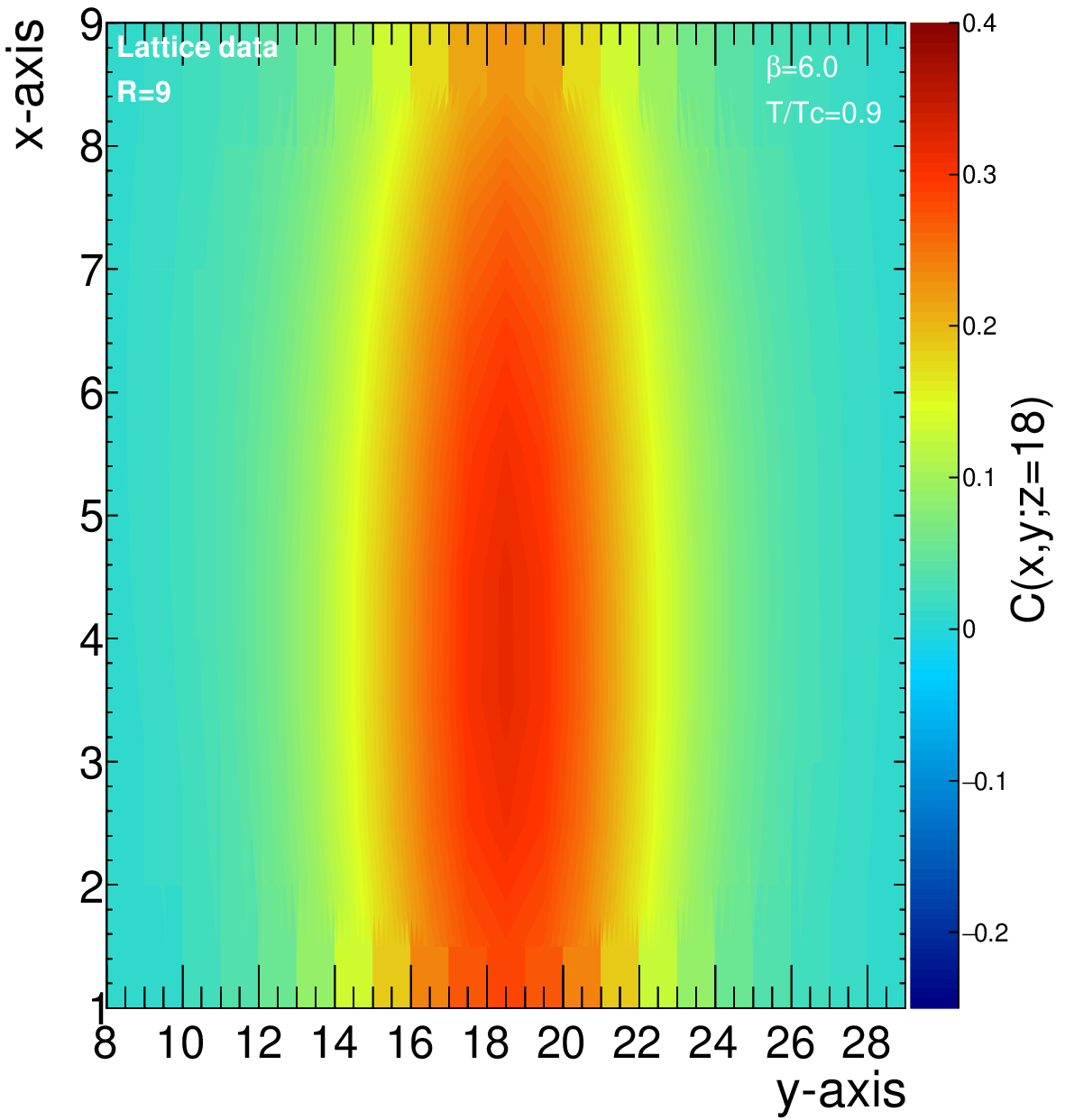}}
	\subfigure[\tiny{$\parallel$-plane,$R=12a$}]{\includegraphics[scale=0.16]{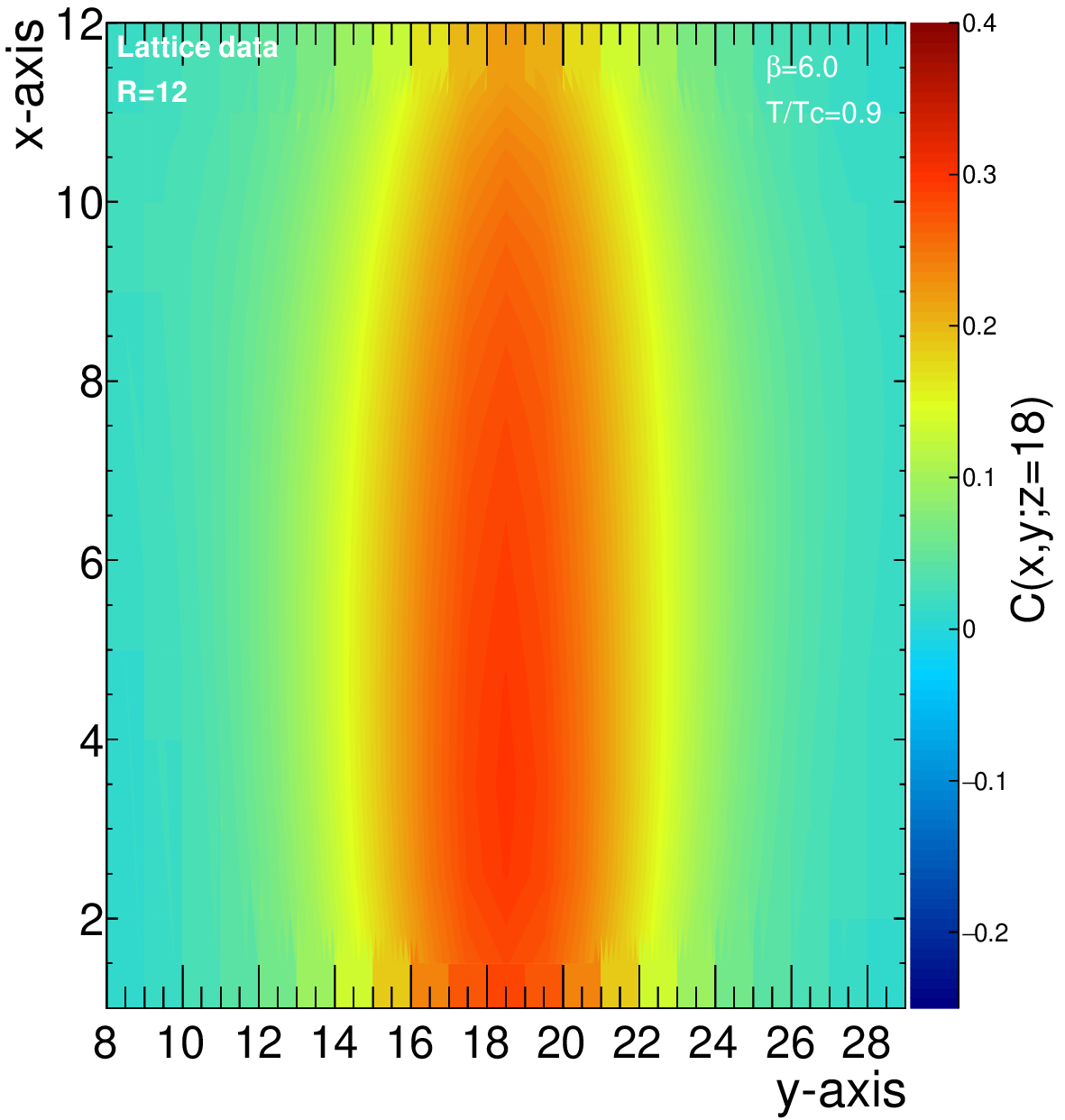}}
	\caption{Flux tubes in 2D-plane for $(QQ)-Q$ distance $R=4a, 5a, 6a, 9a, 12a$ and temperature $T/T_{c} = 0.9$. The baryon system the flux tube is shown for two base length $A=2a, 4a$ in the first and second row, respectively.}   
	\label{fig:2DFluxTubes_TTc09}
\end{figure*}

\begin{figure*}[!hpt]
	\centering
	\subfigure[\tiny{$\parallel$-plane,$R=4a$}]{\includegraphics[scale=0.16]{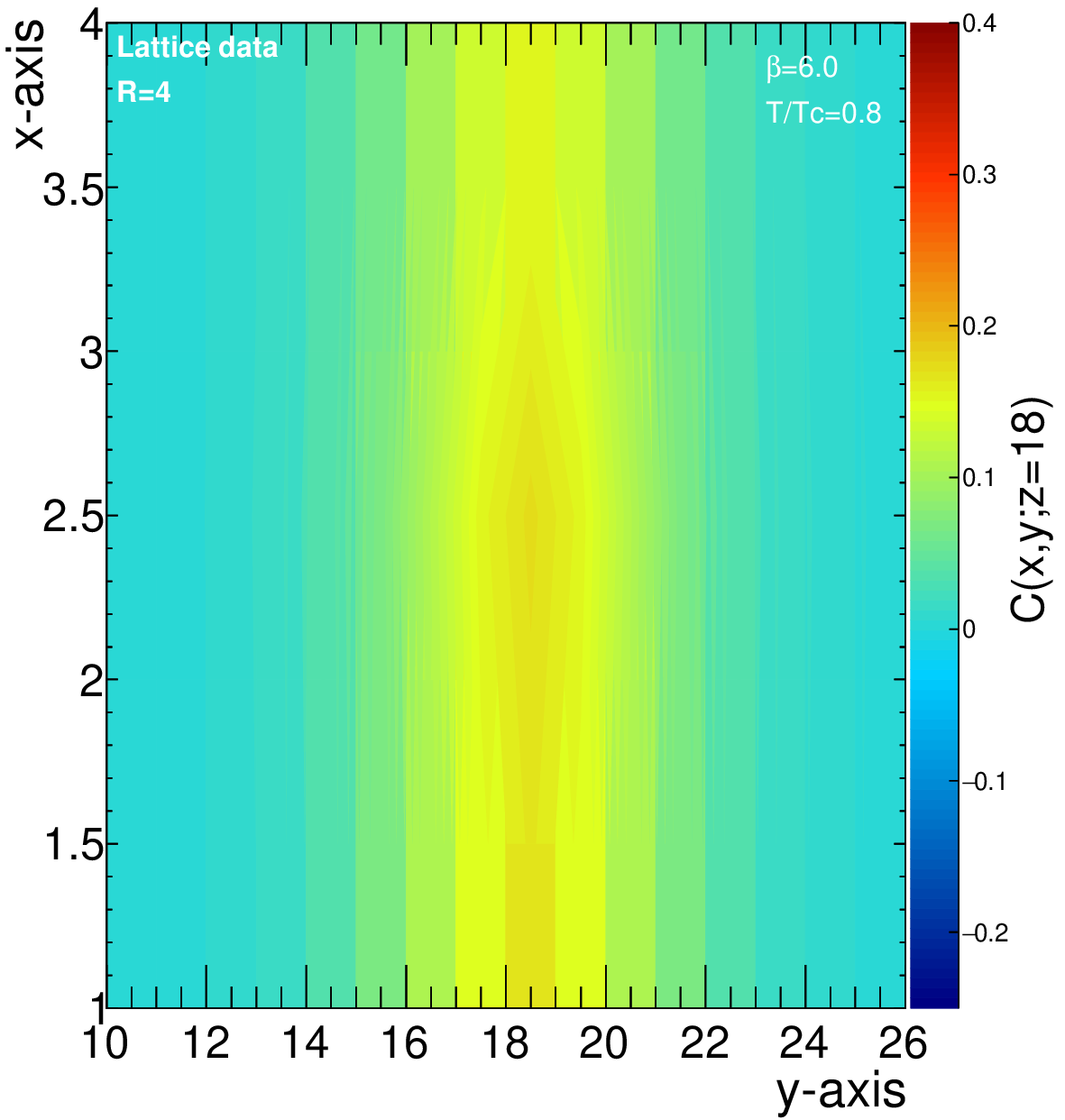} }
	\subfigure[\tiny{$\parallel$-plane,$R=5a$}]{\includegraphics[scale=0.16]{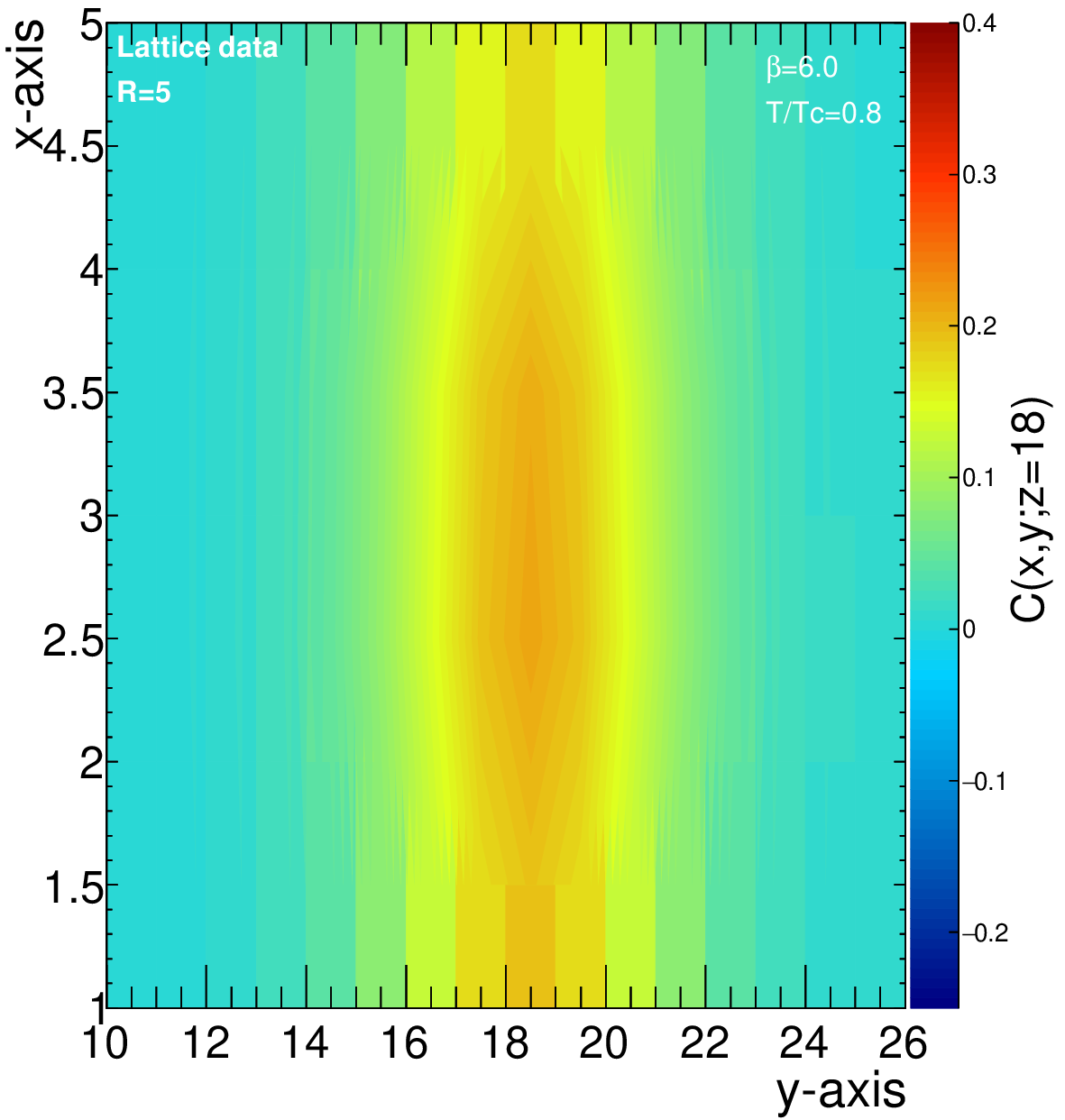} }
	\subfigure[\tiny{$\parallel$-plane,$R=6a$}]{\includegraphics[scale=0.16]{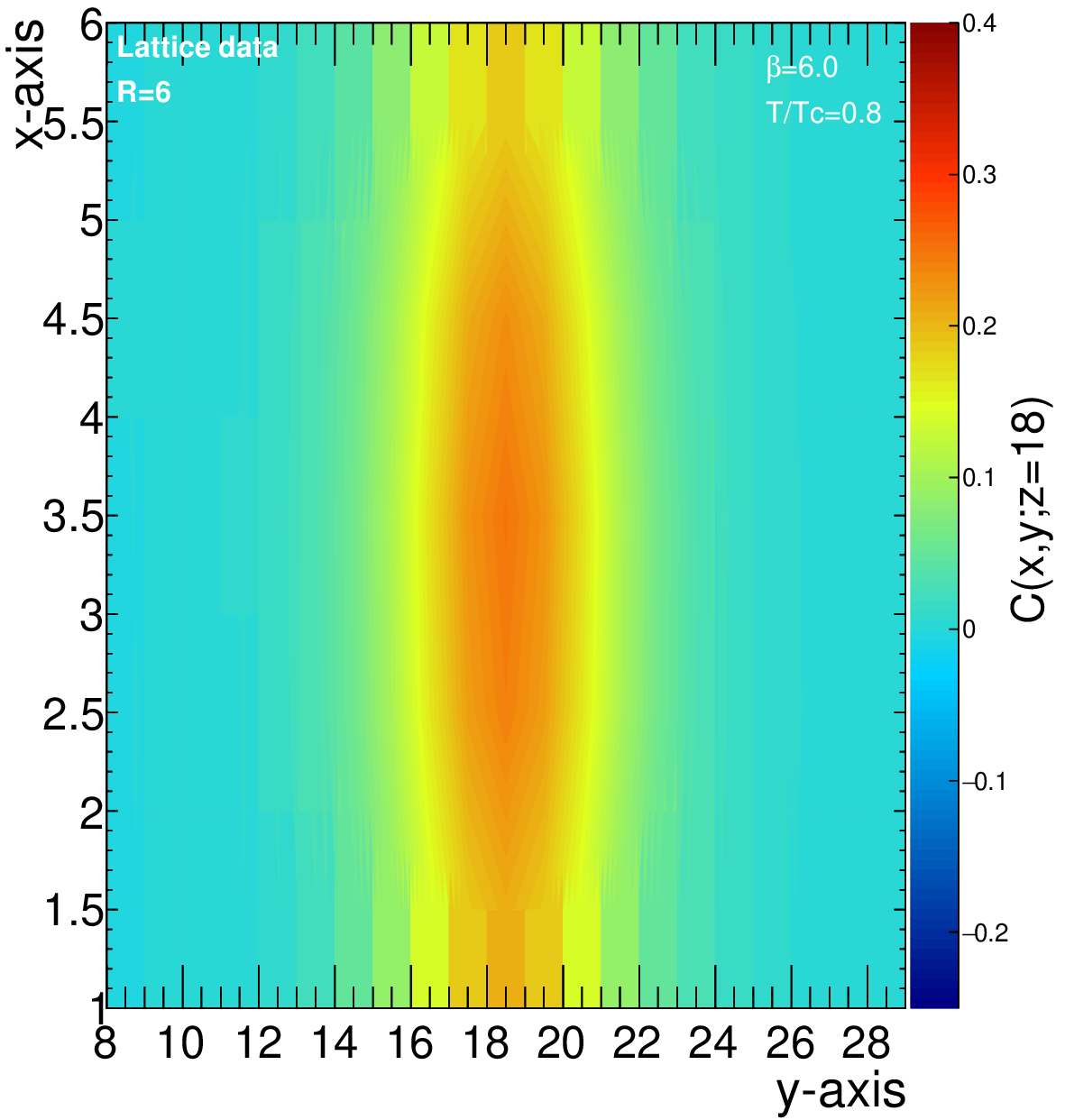} }
	\subfigure[\tiny{$\parallel$-plane,$R=9a$}]{\includegraphics[scale=0.16]{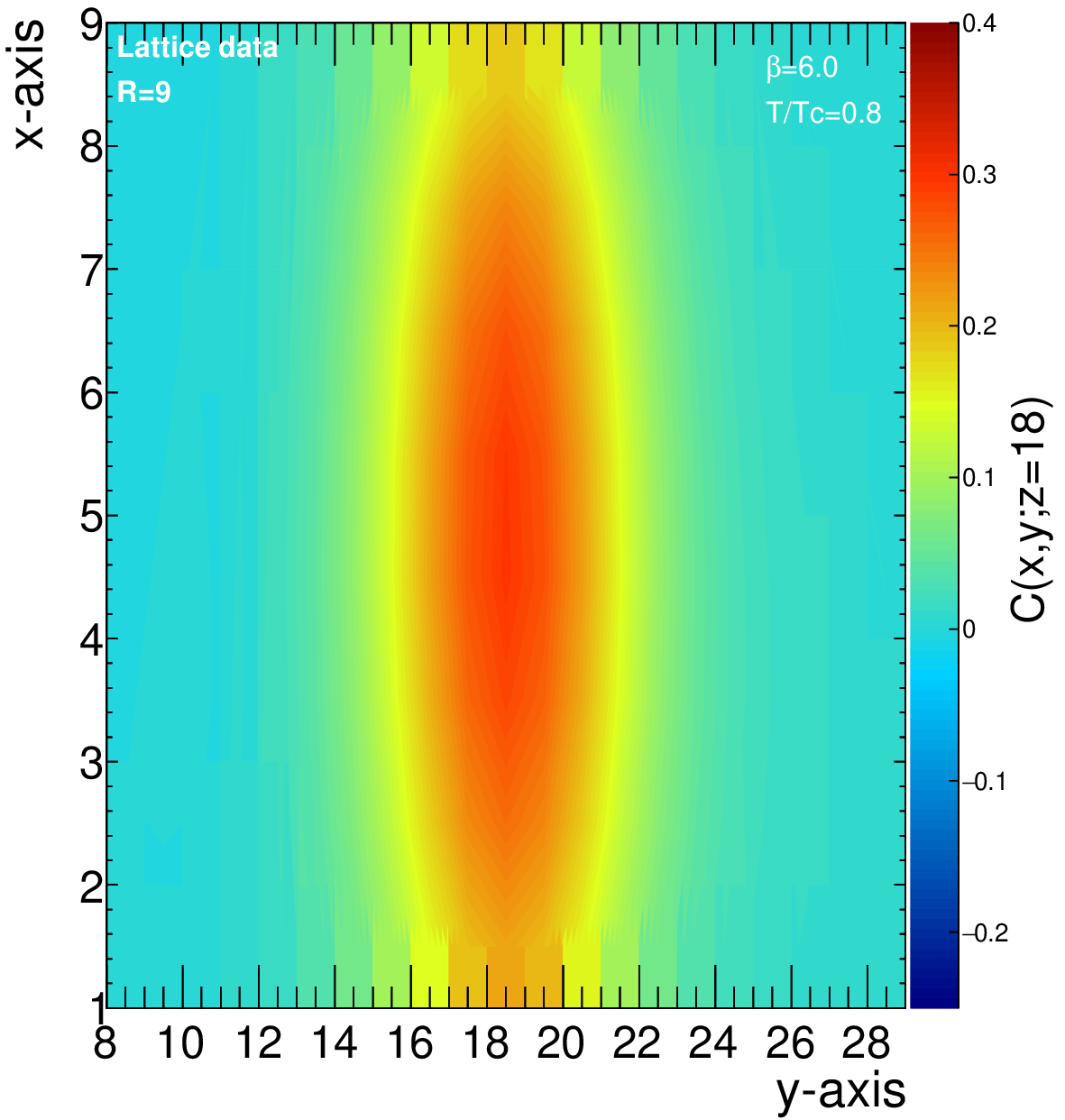} }
	\subfigure[\tiny{$\parallel$-plane,$R=12a$}]{\includegraphics[scale=0.16]{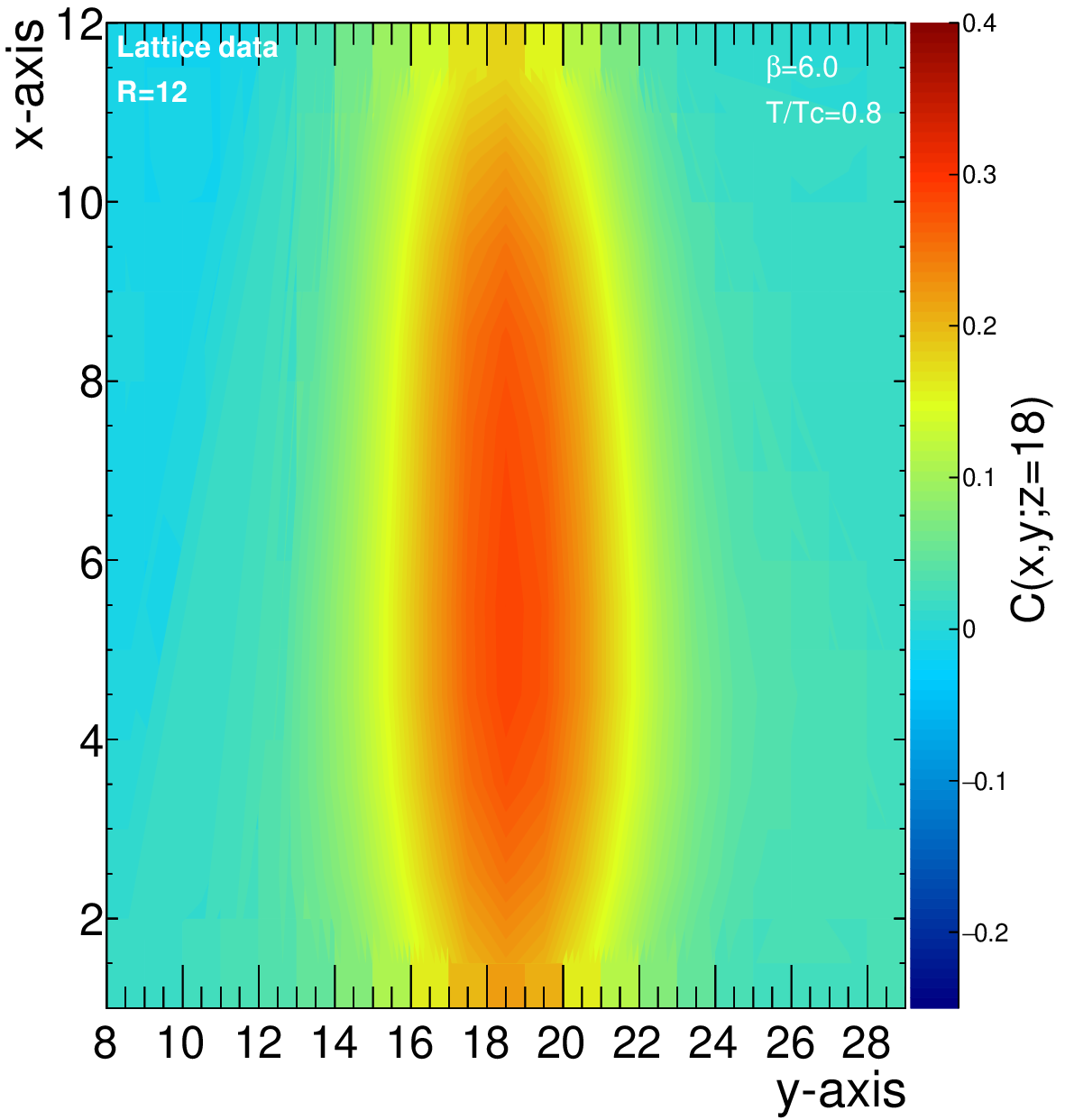} }\\
	
	\subfigure[\tiny{$\parallel$-plane,$R=4a$}]{\includegraphics[scale=0.16]{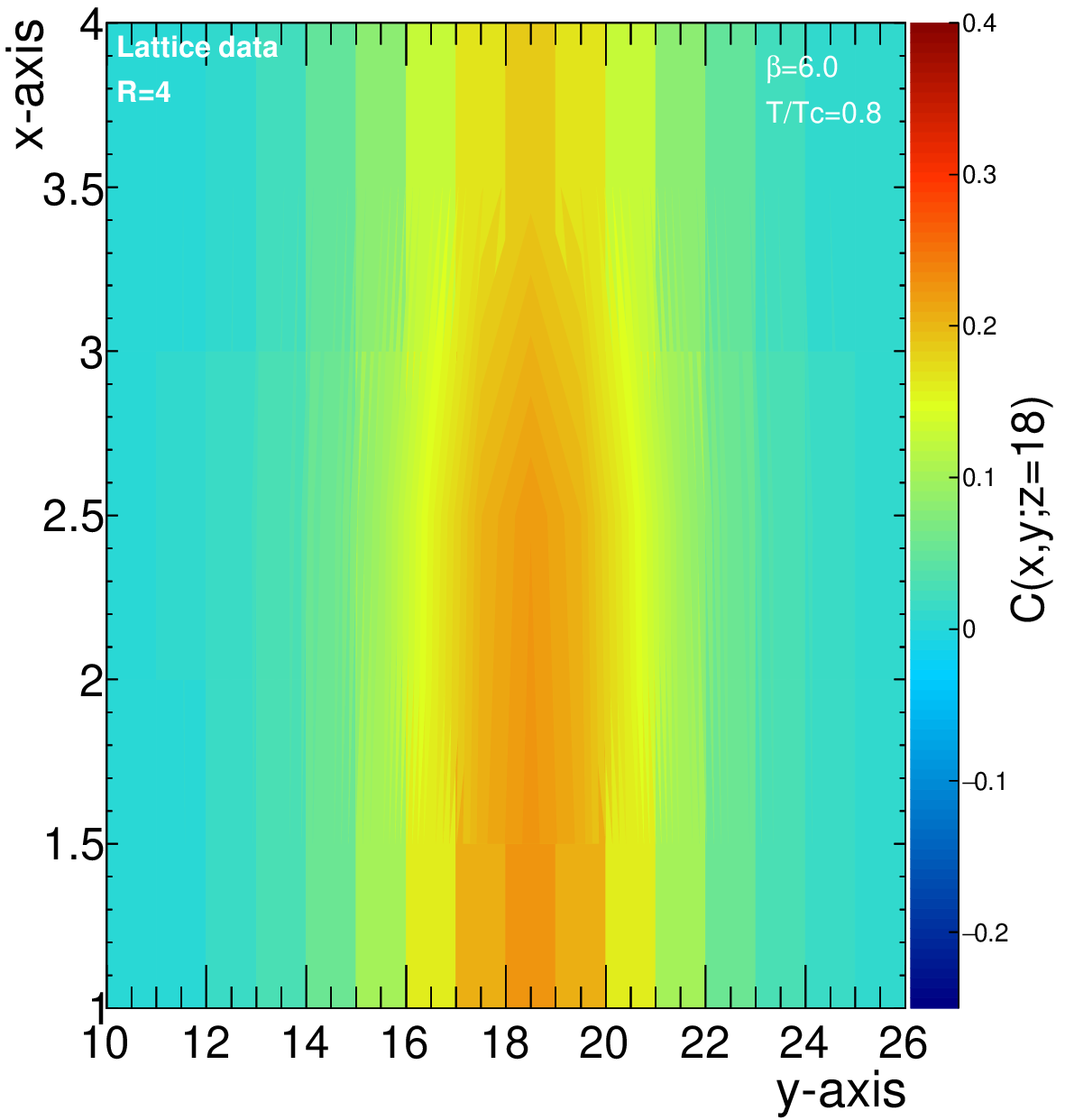} }
	\subfigure[\tiny{$\parallel$-plane,$R=5a$}]{\includegraphics[scale=0.16]{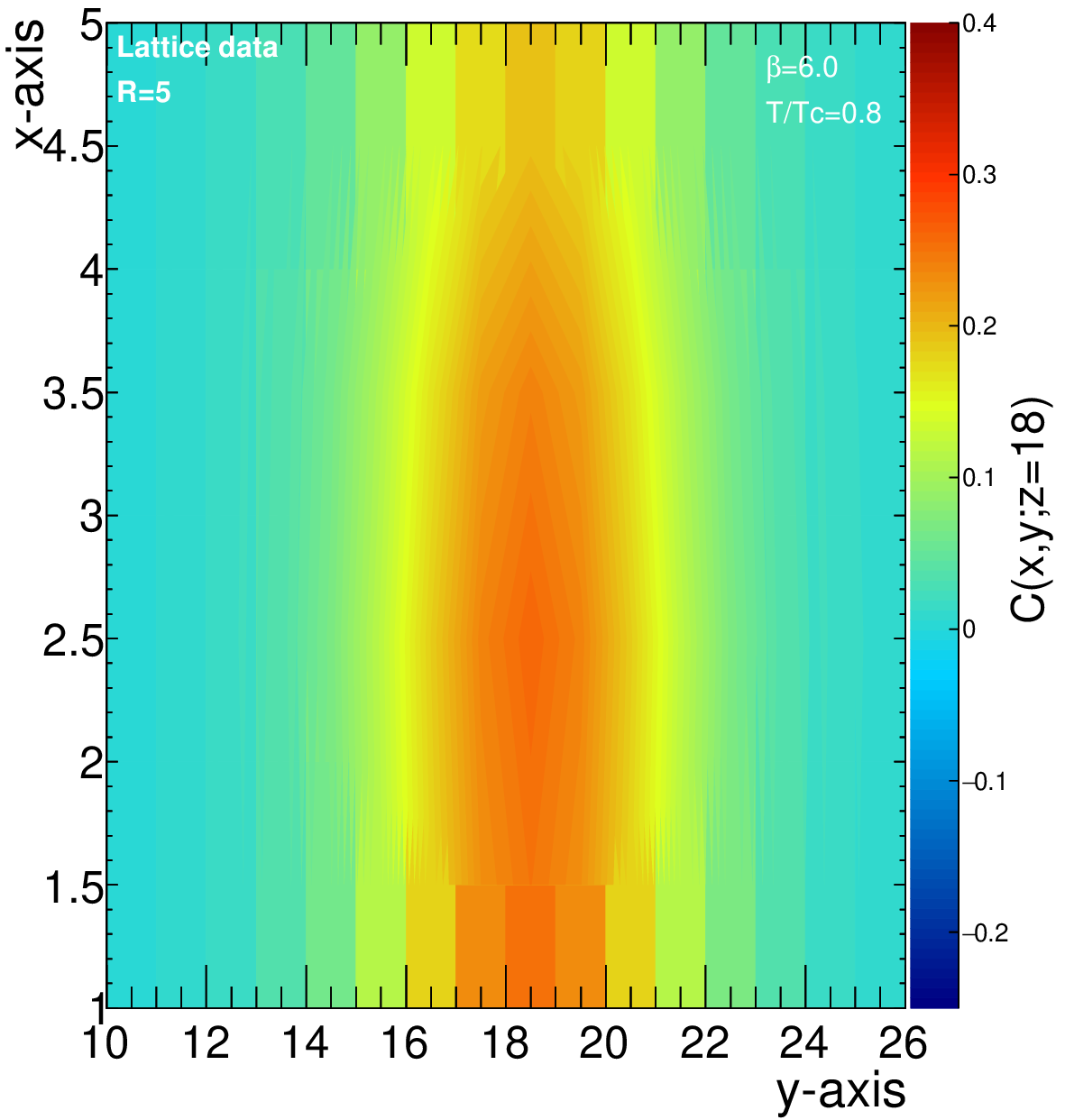} }
	\subfigure[\tiny{$\parallel$-plane,$R=6a$}]{\includegraphics[scale=0.16]{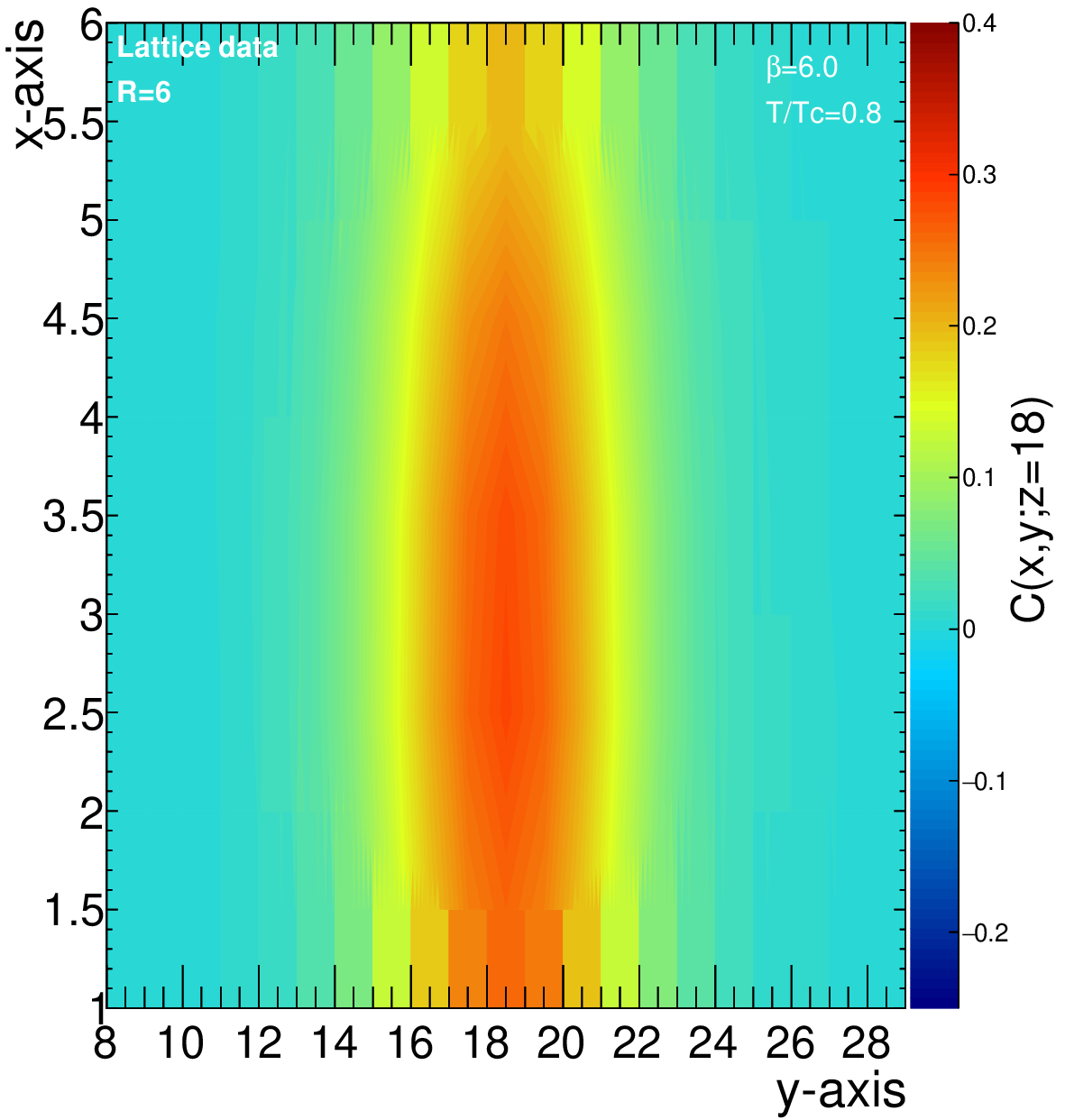} }
	\subfigure[\tiny{$\parallel$-plane,$R=9a$}]{\includegraphics[scale=0.16]{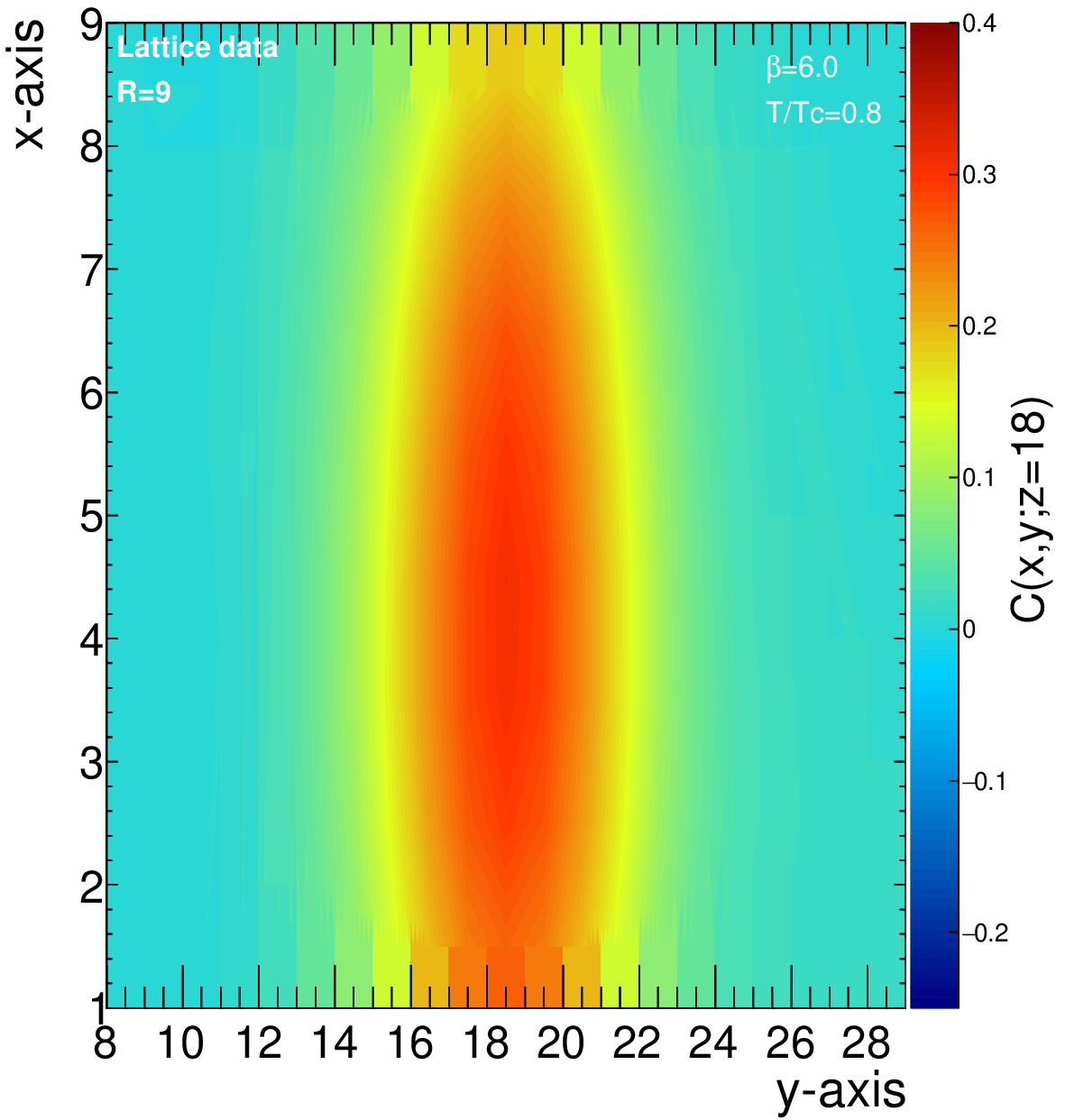} }
	\subfigure[\tiny{$\parallel$-plane,$R=12a$}]
	{\includegraphics[scale=0.16]{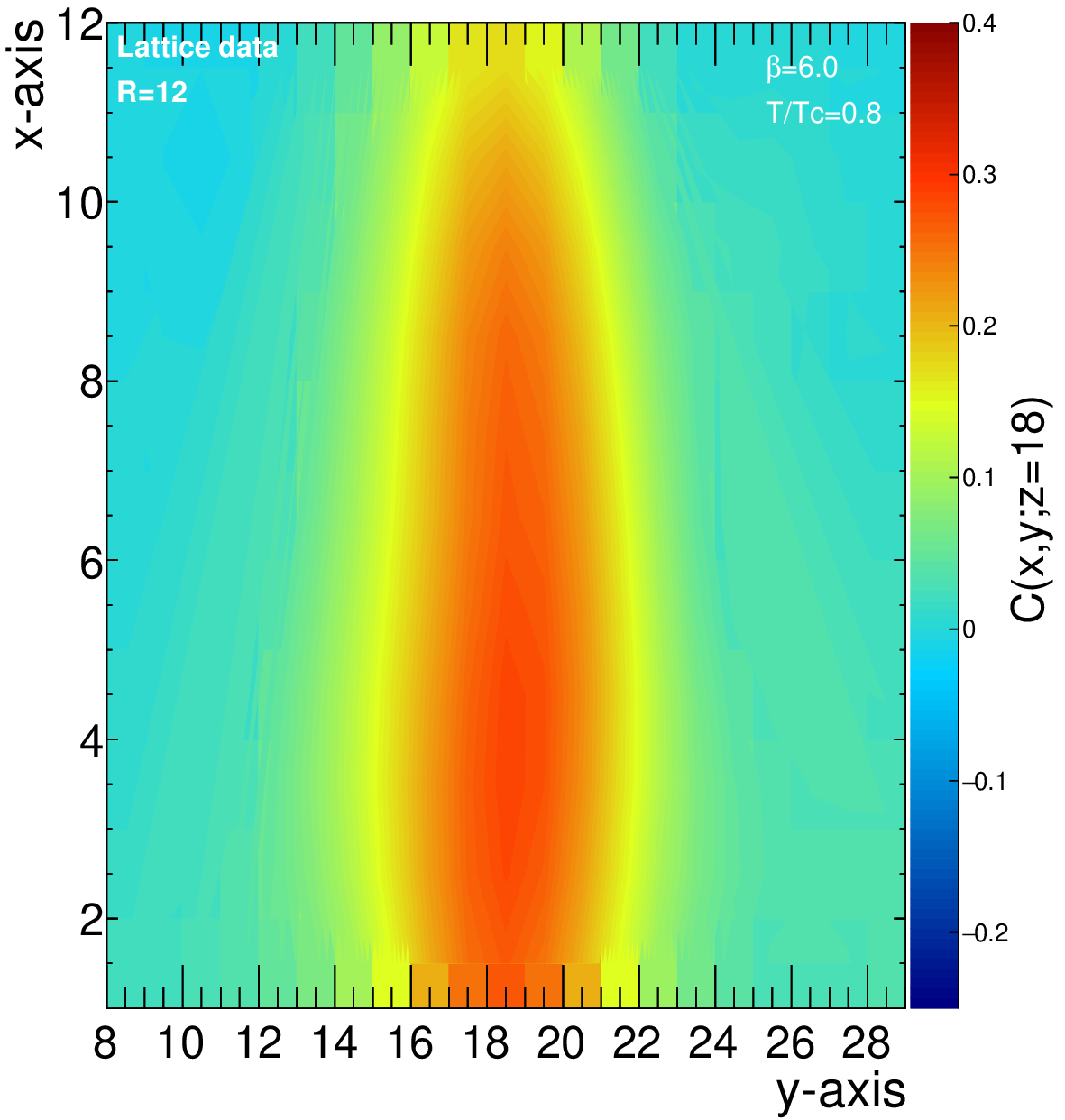} }   
	\caption{
		Flux tubes in 2D-plane for $(QQ)-Q$ distance $R=4a, 5a, 6a, 9a, 12a$ and temperature $T/T_{c} = 0.8$. The first and second row correspond to the two base length $A=2a, 4a$, respectively.  
	}  
	\label{fig:2DFluxTubes_TTc08}
\end{figure*}

\begin{figure*}[!hpt]
	\subfigure[\tiny{$Q\bar{Q}$-system:$R=4a$}]{\includegraphics[scale=0.16]{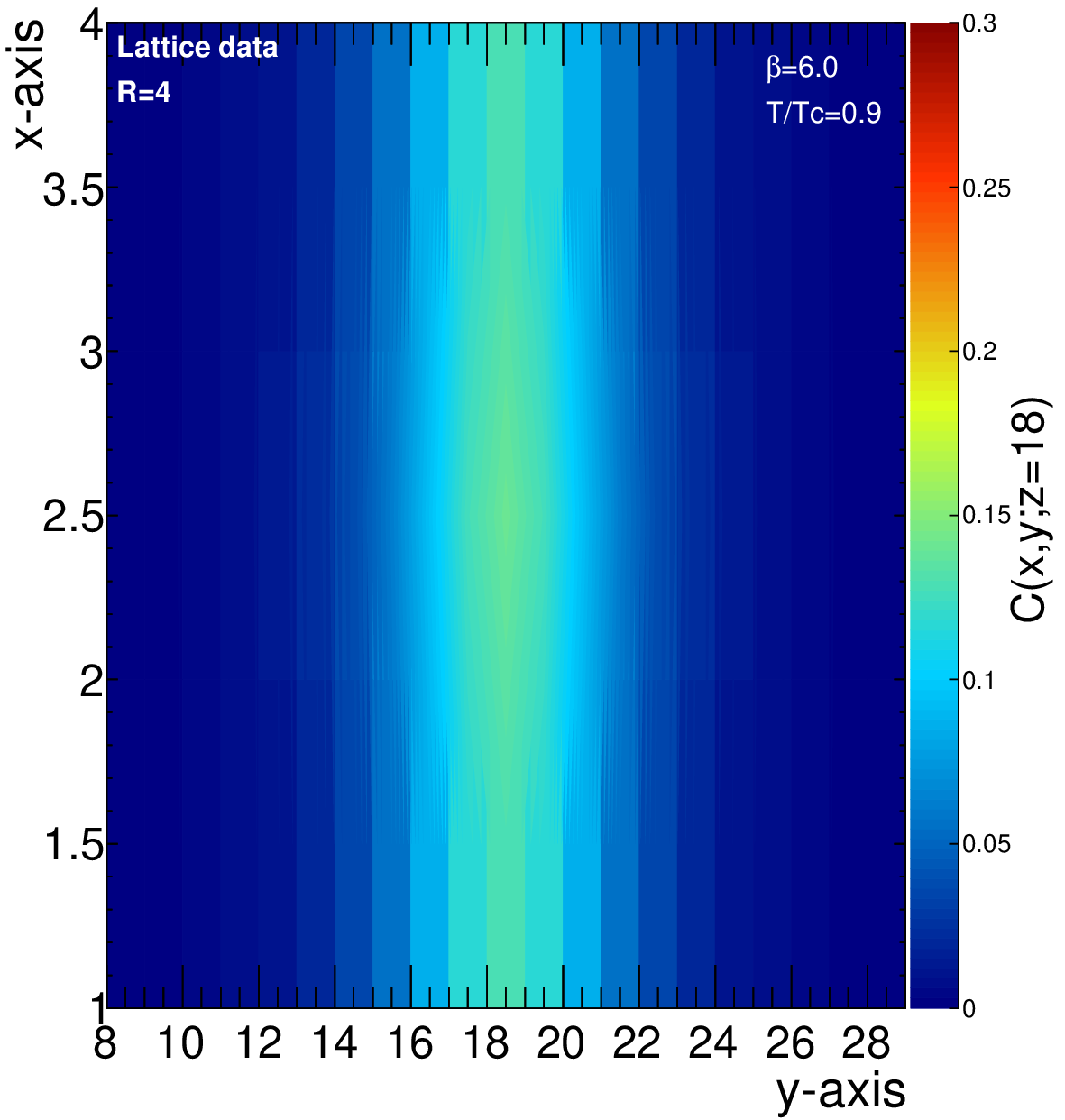} }
	\subfigure[\tiny{$Q\bar{Q}$-system:$R=5a$}]{\includegraphics[scale=0.16]{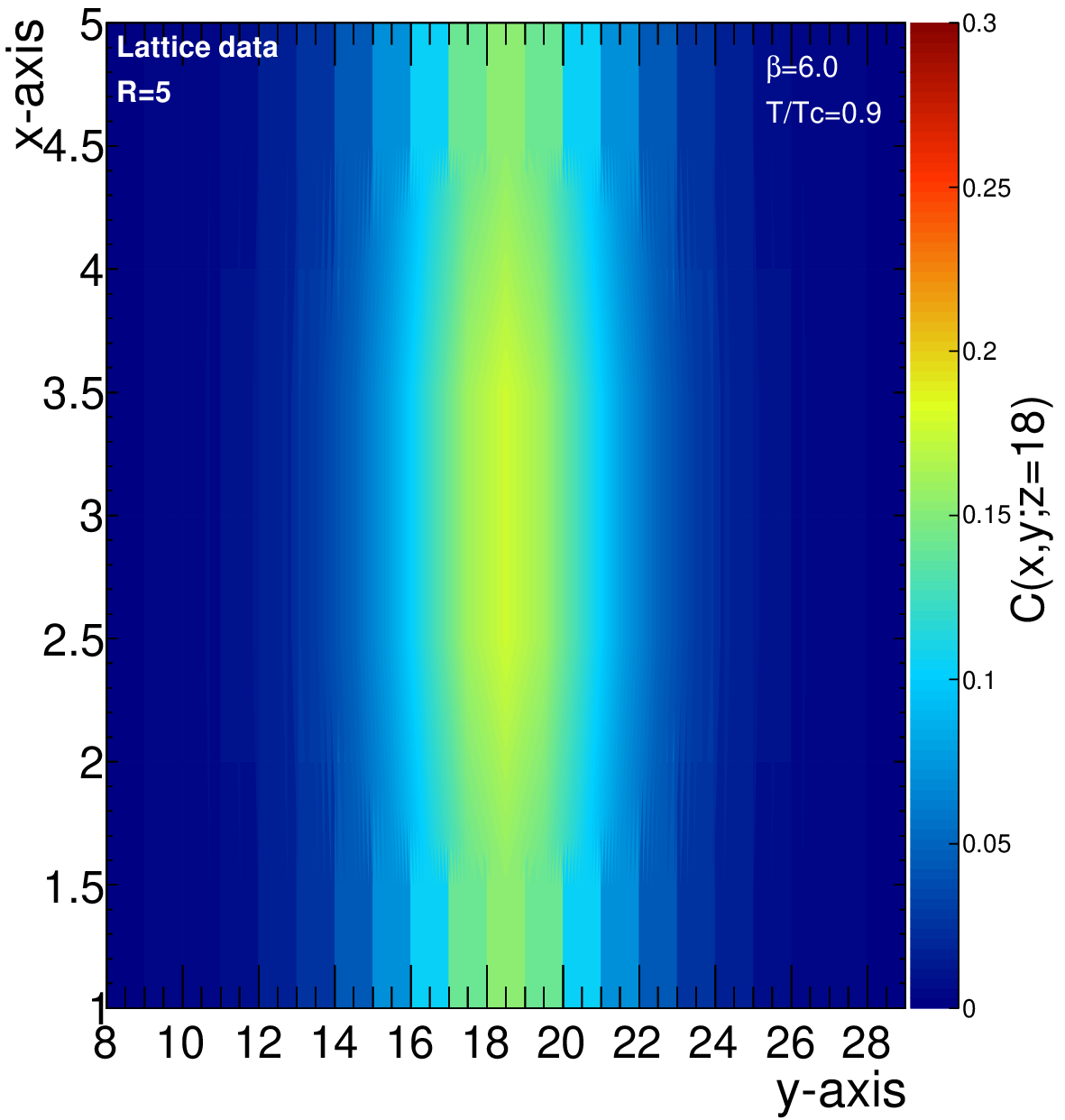} } 
	\subfigure[\tiny{$Q\bar{Q}$-system:$R=6a$}]{\includegraphics[scale=0.16]{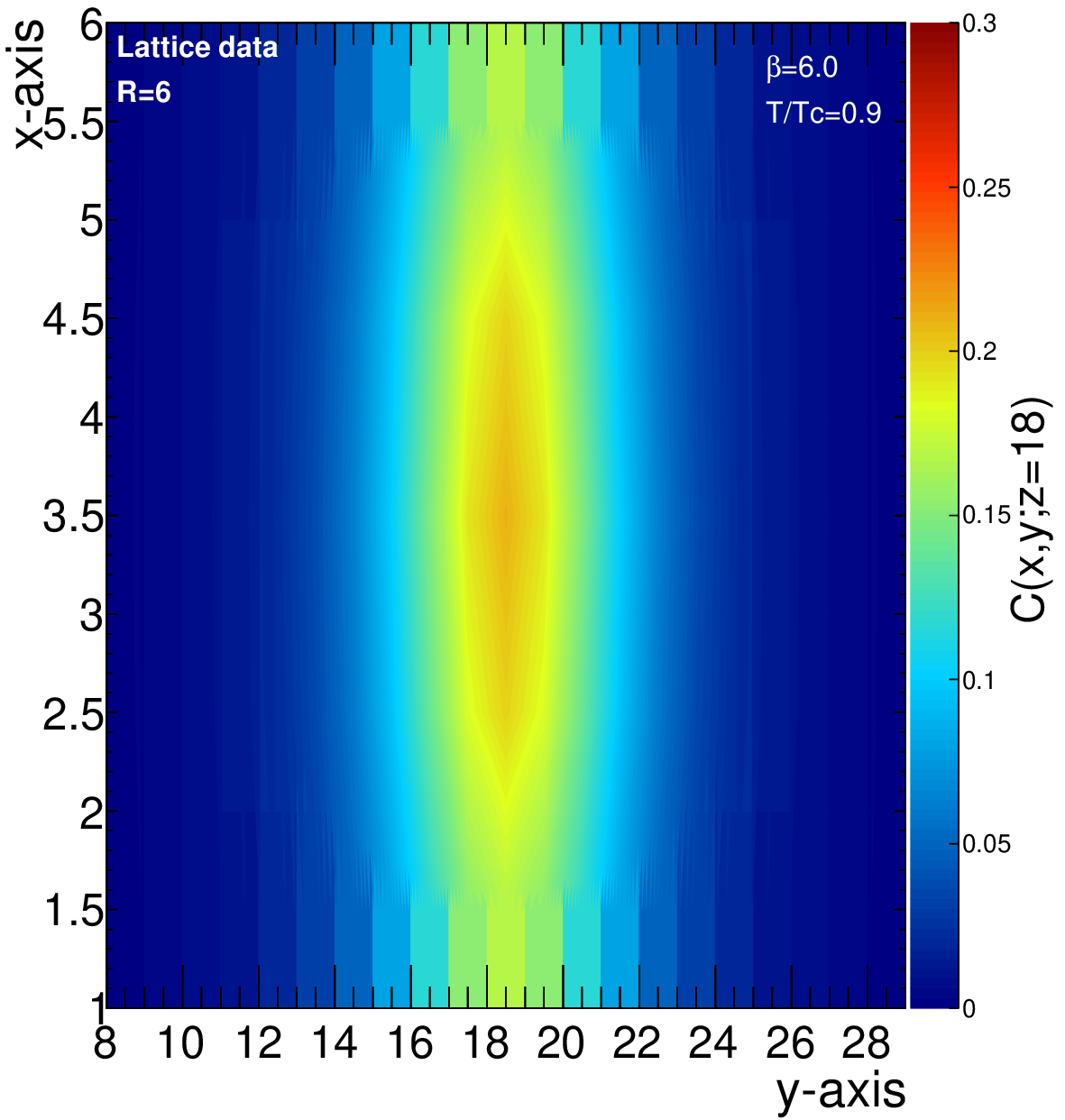} }
	\subfigure[\tiny{$Q\bar{Q}$-system:$R=9a$}]{\includegraphics[scale=0.16]{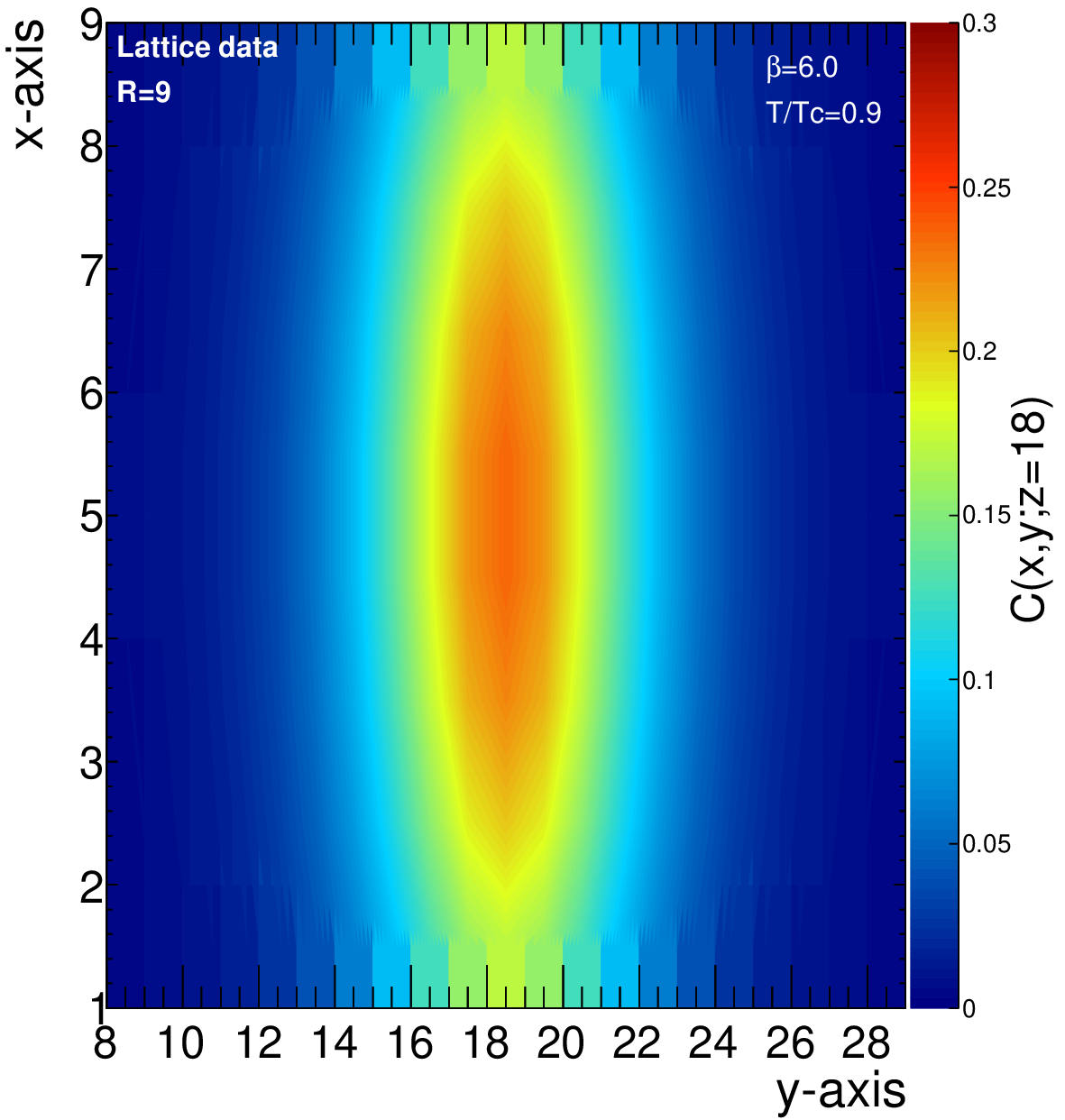} }  
	\subfigure[\tiny{$Q\bar{Q}$-system:$R=12a$}]{\includegraphics[scale=0.16]{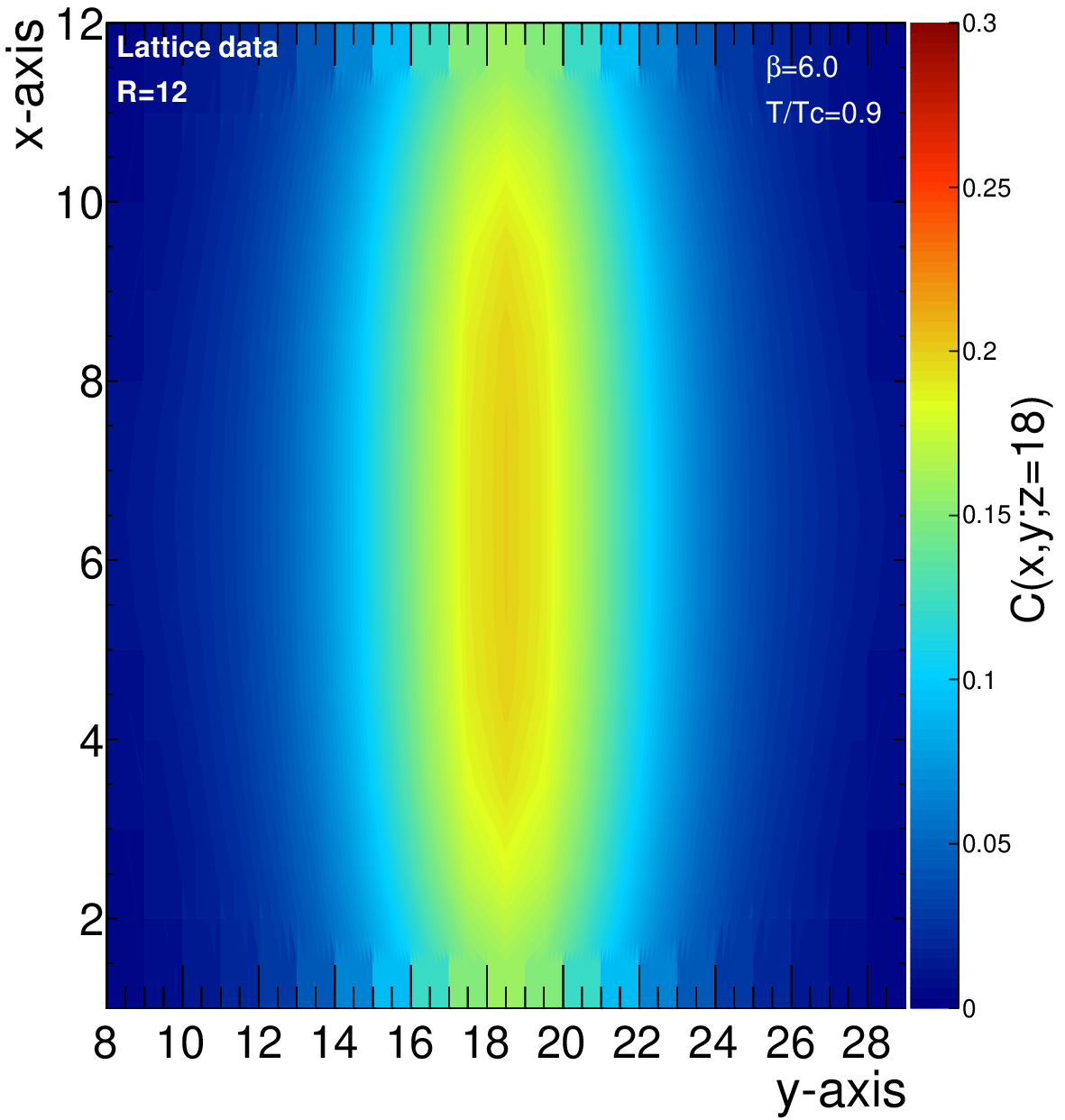} }
	\caption{Flux tubes of the $Q\bar{Q}$ quark-antiquark in 2D-plane for source separations $R=4a, 5a, 6a, 9a, 12a$ and temperature $T/T_{c} = 0.9$.
	}
	\label{QQT09} 
\end{figure*}

\begin{figure*}[!hpt]
	\subfigure[\tiny{$Q\bar{Q}$-system:$R=4a$}]{\includegraphics[scale=0.16]{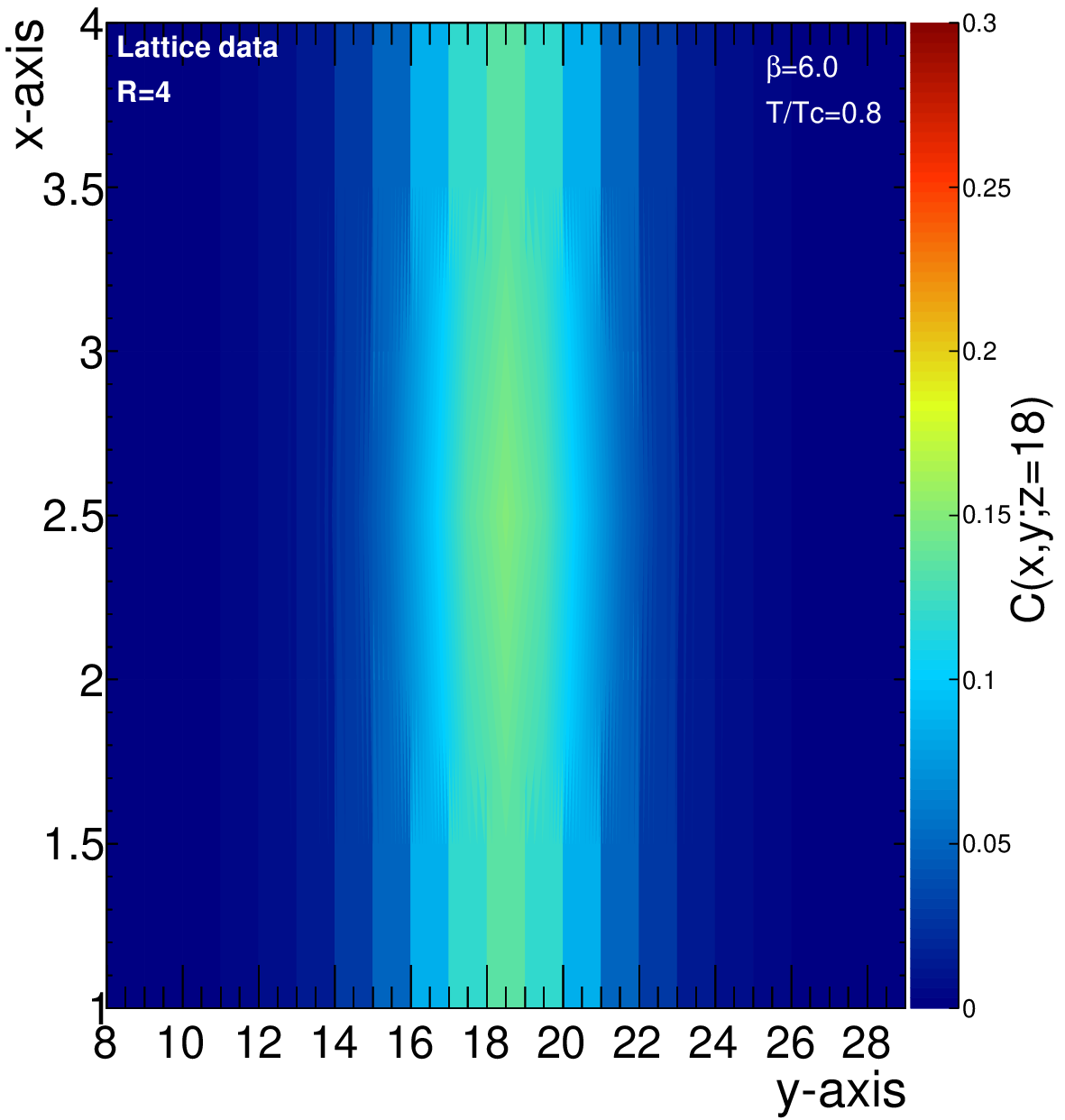} }
	\subfigure[\tiny{$Q\bar{Q}$-system:$R=5a$}]{\includegraphics[scale=0.16]{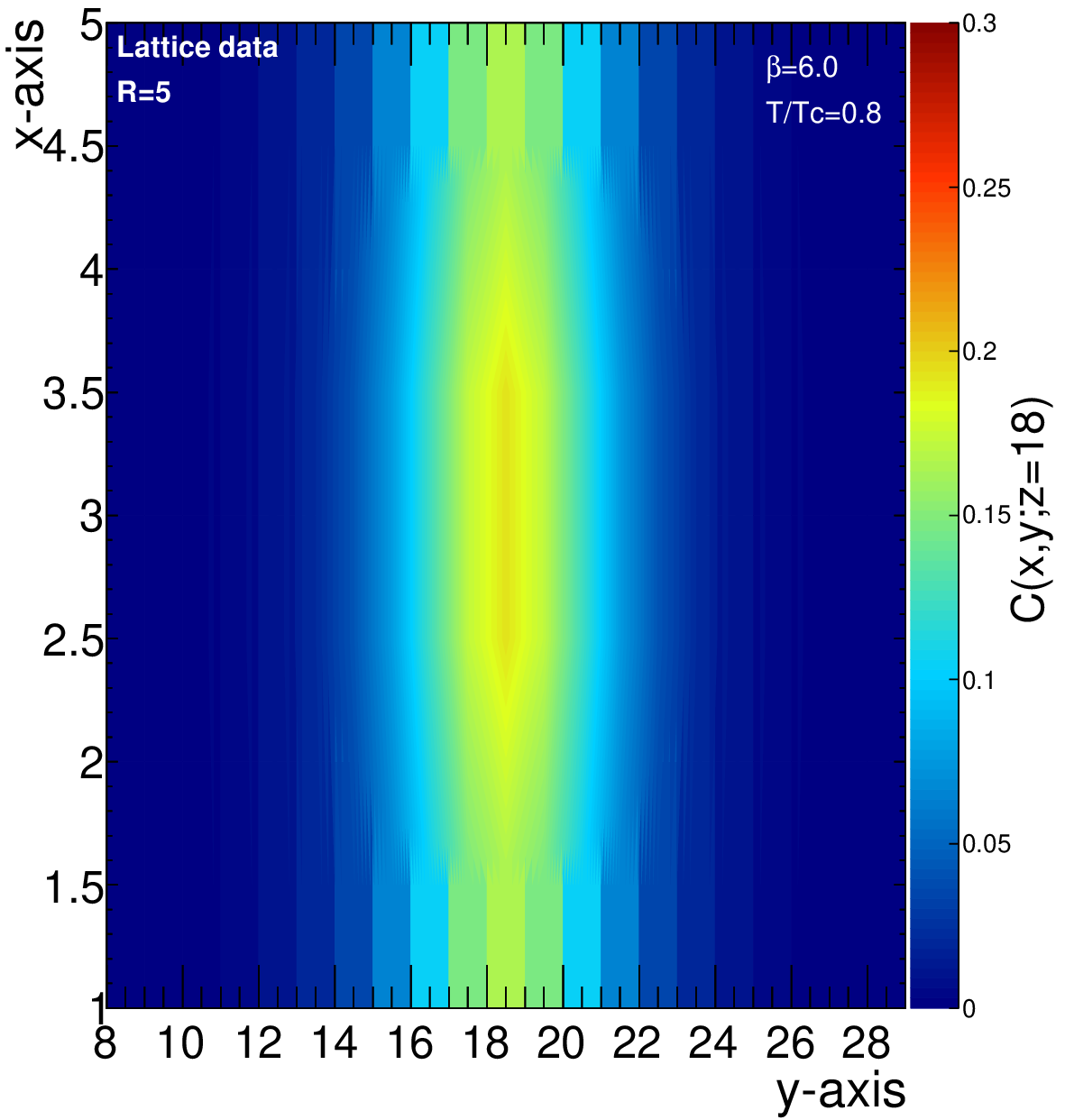} }    
	\subfigure[\tiny{$Q\bar{Q}$-system:$R=6a$}]{\includegraphics[scale=0.16]{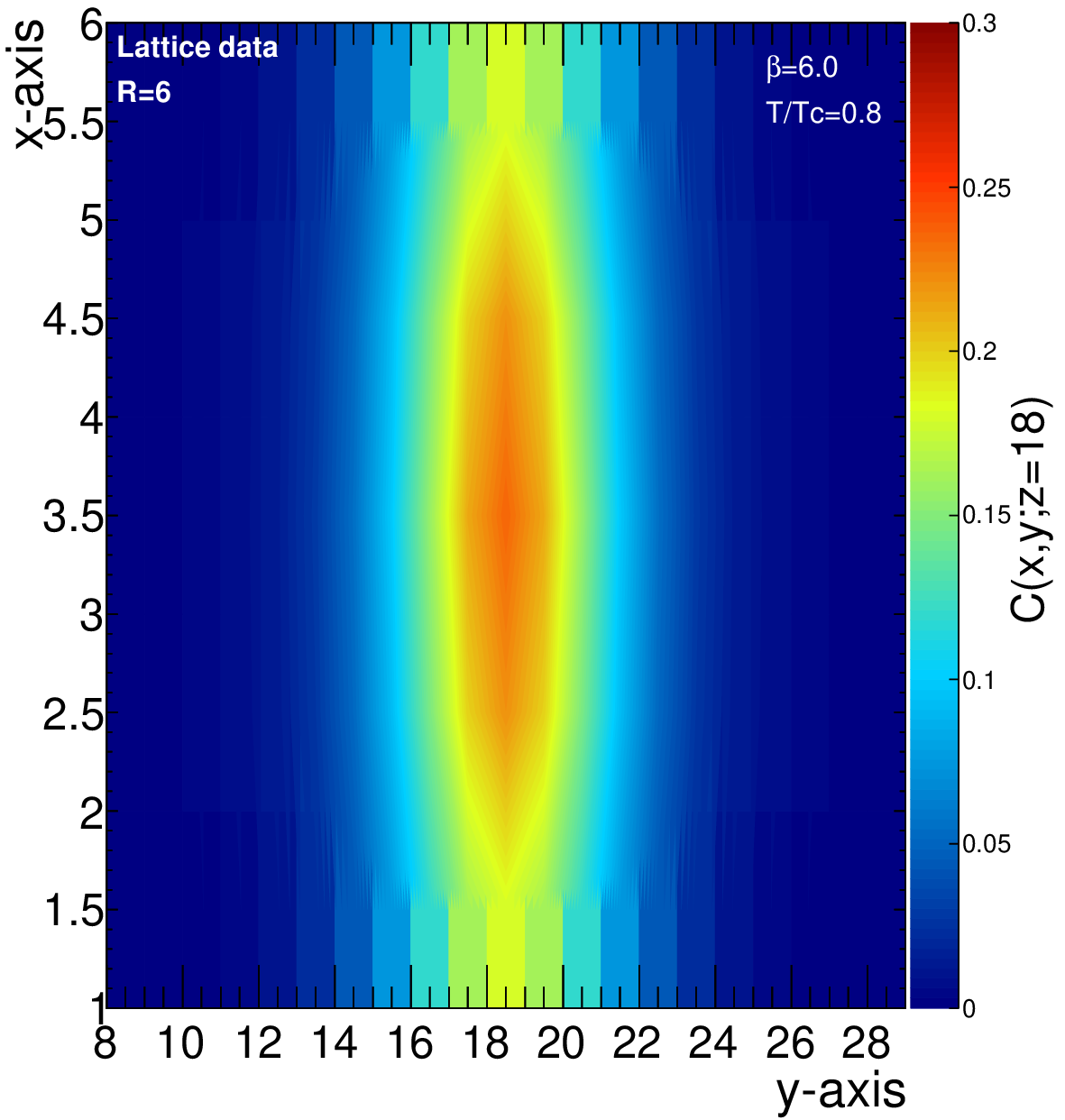} }
	\subfigure[\tiny{$Q\bar{Q}$-system:$R=9a$}]{\includegraphics[scale=0.16]{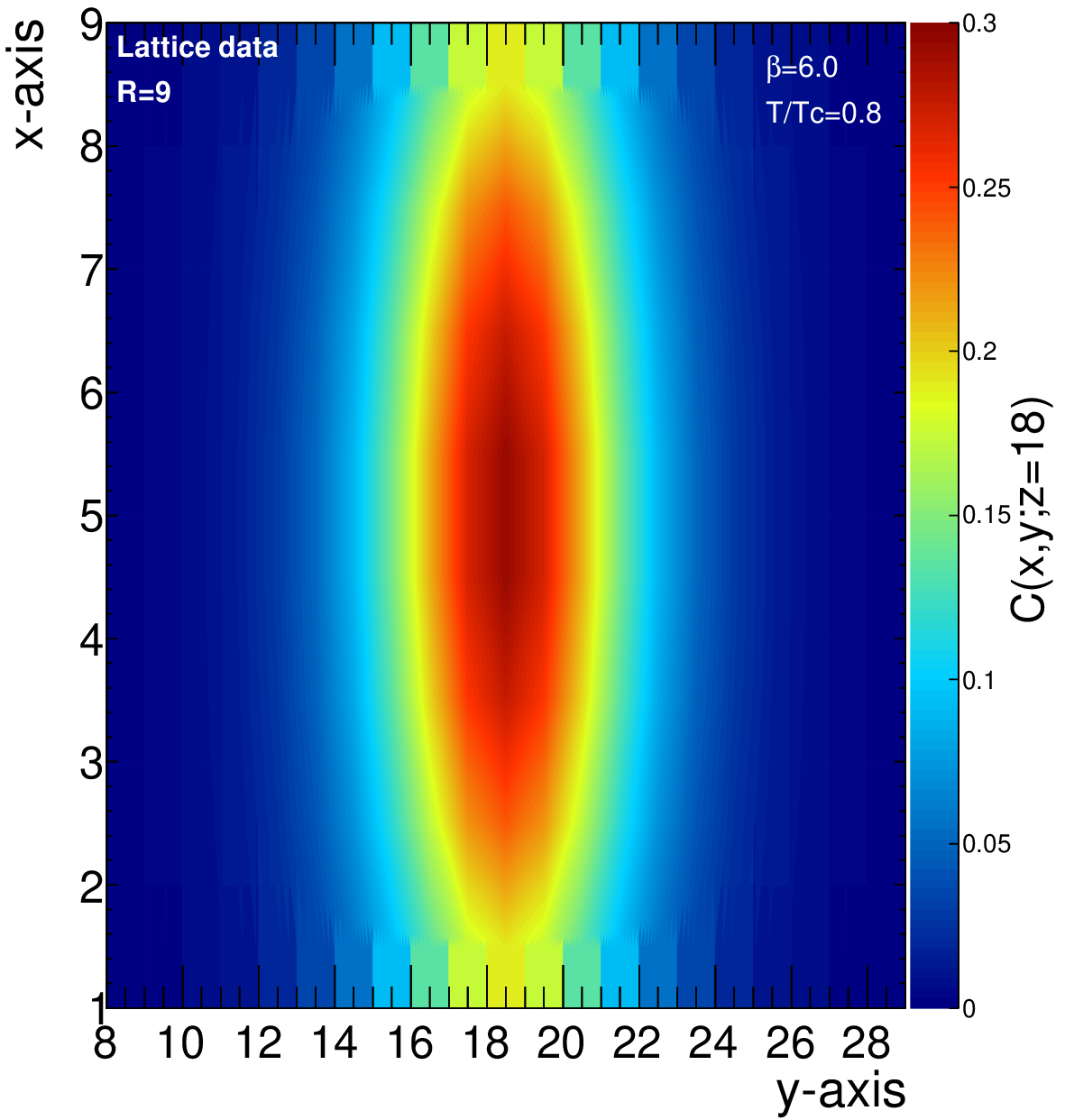} }  
	\subfigure[\tiny{$Q\bar{Q}$-system:$R=12a$}]{\includegraphics[scale=0.16]{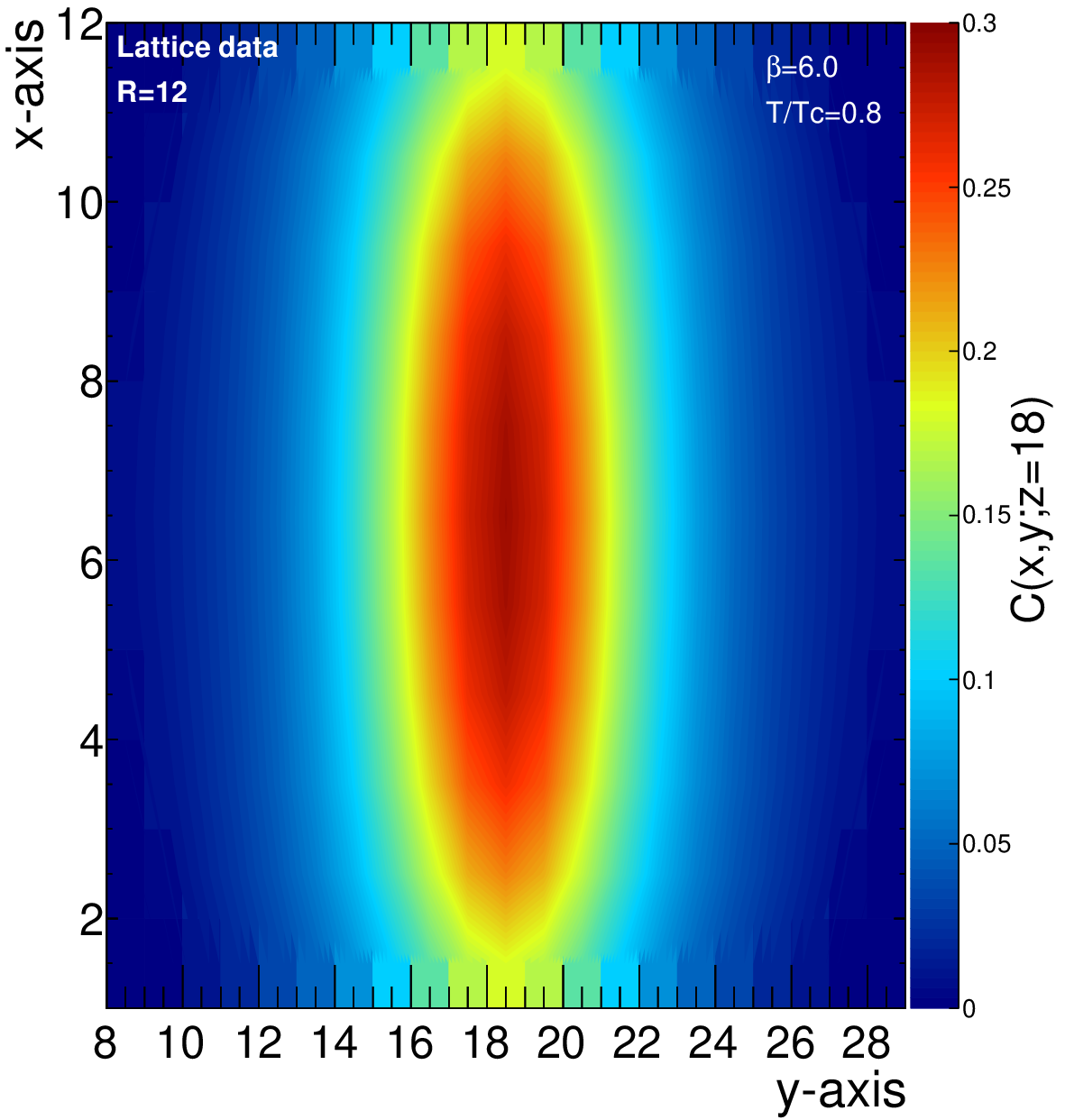} }
	\caption{Flux tubes of the $Q\bar{Q}$ quark-antiquark in 2D-plane for source separations $R=4a, 5a, 6a, 9a, 12a$ and temperature $T/T_{c} = 0.8$.
	}
	\label{QQT08} 
\end{figure*}

The in-plane action density maps at various separation distances, specifically $R=4a; 5a; 6a; 9a$, and $R=12a$, for baryon and meson systems. Figures \ref{fig:2DFluxTubes_TTc09} and \ref{fig:2DFluxTubes_TTc08} assimilate the $(QQ)Q$ system at the temperature scales $T/T_{c}= 0.9$ and $T/T_{c}= 0.8$, respectively. Each row corresponds to the action density in the $xy$ plane at different base lengths, which are $A=2a$ and $A=4a$. In contrast, Figs.~\ref{QQT09} and \ref{QQT08} depict the density map of the $Q\bar{Q}$ system at the corresponding separation distances, $R$, and temperature scales. 

The $\chi^{2}$ values obtained from the fits of the $Q\bar{Q}$ and $(QQ)Q$ systems are presented in Tables~\ref{tab:AAAw2a2T09} and \ref{tab2:AAAw2a2T08}. These tables provide width measurements for the first four planes of both the $Q\bar{Q}$ and $(QQ)Q$ systems. The width measurements in Table \ref{tab:AAAw2a2T09}, at $T/T_{c}= 0.9$, depict equal MS width components of the orthogonal $W^{2}_{\perp}$ and in-plane action density $W^{2}_{\parallel}$. Table \ref{tab2:AAAw2a2T08}, at temperature $T/T_{c}= 0.8$, similarly contains averaged width values for $x_{i}=1,2,3,4$.

\begin{table*}[!htb]
		\begin{tabular}{cc|ccc|ccc|ccc|ccc}
			\hline
			\hline
			\multicolumn{2}{c|}{plane}   
			&\multicolumn{3}{c|}{$x=1$}   
			&\multicolumn{3}{c|}{$x=2$}
			&\multicolumn{3}{c|}{$x=3$} 
			&\multicolumn{3}{c}{$x=4$}\\
			\multicolumn{2}{c|}{$n=R/a$}   
			&\multicolumn{1}{c}{$A$}  &\multicolumn{1}{c}{$w^{2}a^{-2}$}&\multicolumn{1}{c|}{$\chi^2_{\rm{dof}}$}
			&\multicolumn{1}{c}{$A$}  &\multicolumn{1}{c}{$w^{2}a^{-2}$}&\multicolumn{1}{c|}{$\chi^2_{\rm{dof}}$}
			&\multicolumn{1}{c}{$A$}  &\multicolumn{1}{c}{$w^{2}a^{-2}$}&\multicolumn{1}{c|}{$\chi^2_{\rm{dof}}$}
			&\multicolumn{1}{c}{$A$}  &\multicolumn{1}{c}{$w^{2}a^{-2}$}&\multicolumn{1}{c}{$\chi^2_{\rm{dof}}$}\\ 
			\hline
			\multirow{3}{*}{\begin{turn}{90} \scriptsize{$R=4a$} \end{turn}}
			&$QQ$         &0.0654(1) &15.7(1)  &4.3  &0.0711(1) &15.4(1)  &4.40 &0.0654(1) &15.7(1)   &4.3\\
			&$(QQ)Q_{\parallel}$ &0.0837(2) &22.1(1) &5.8  &0.0864(2) &22.0(1) &5.74 &0.0810(2) &22.7(1) &5.84\\ 
			&$(QQ)Q_{\perp}$ &0.0833(2) &22.2(1) &8.13 &0.0862(2) &22.9(1) &8.57 &0.0810(2) &22.2(1) &10.51\\
			\hline
			\multirow{3}{*}{\begin{turn}{90} \scriptsize{$R=5a$} \end{turn}}
			&$QQ$                &0.0777(2) & 16.7(5)  & 3.98& 0.0894(2)& 16.1(1) &4.03&0.0894(2)&16.1(1)  &4.3& 0.0777(2)&16.7(1)  &3.98\\
			&$(QQ)Q_{\parallel}$ &0.1(2)    &22.13(2)&2.39& 0.1064(2)&21.8(1)&4.87& 0.1053(2)&21.7(1)&5.10& 0.0951(2)&22.0(1)&5.33 \\ 
			&$(QQ)Q_{\perp}$     &0.0986(2) &22.1(1)&7.09&0.1062(2)&27.7(7)&6.83&0.1052(2)&21.6(1)&7.68&0.0952(2)&22.0(1)&9.45 \\
			\hline
			\multirow{3}{*}{\begin{turn}{90} \scriptsize{$R=6a$} \end{turn}}
			&$QQ$       &    0.0842(2)&18.0(2)    & 3.27 &  0.1003(2) & 17.3(2) & 3.2& 0.1062(3) & 17.1(2)  &3.1&0.0843(2)& 17.3(2)    & 3.21 \\
			&$(QQ)Q_{\parallel}$&0.1097(3) &22.5(2) &2.98& 0.1212(3)&22.3(2)&3.51&  0.1245(3)&22.1(2)&3.86 &0.1190(3)&22.1(2)&4.30 \\
			&$(QQ)Q_{\perp}$& 0.1092(3)& 22.7(2)&5.10& 0.1209(3)&22.2(2)&5.19&0.1243(3)&22.0(2)&5.17&0.1190(3)&22.1(2)&5.98\\
			\hline
			\multirow{3}{*}{\begin{turn}{90} \scriptsize{$R=7a$} \end{turn}}
			&$QQ$       & 0.0867(3) & 19.4(2)  & 2.5& 0.1048(3) & 18.8(2) & 2.47 & 0.1147(4)& 18.6(2)    & 2.3 & 0.0867(3)& 18.6(2)     & 2.3 \\
			&$(QQ)Q_{\parallel}$ &0.1161(3)&23.2(2)&1.6& 0.1306(4)&23.2(2)&2.15& 0.1449(6)&23.0(2)&1.44& 0.1367(5)&22.9(2)&3.02 \\ 
			&$(QQ)Q_{\perp}$ &0.1156(3)&23.7(2)&5.20&0.1303(4)&23.2(2)&4.02&0.1448(6)&23.0(2)&2.35& 0.1366(5)&22.9(2)&3.48 \\
			\hline
			\multirow{3}{*}{\begin{turn}{90} \scriptsize{$R=8a$} \end{turn}}
			&$QQ$       &0.9(2) & 20.9(3)  & 1.94& 0.1049(4) & 20.4(3) & 1.91 & 0.1165(5) & 20.3(3)  &1.71 & 0.1206(5) &  20.3(3)    & 1.62 \\
			&$(QQ)Q_{\parallel}$ &0.1195(4) &24.0(3)  &0.74& 0.1354(5)&24.1(2) &1.10& 0.1449(6)&24.2(2)&1.44&0.1476(6)&24.2(2)&1.83\\ 
			&$(QQ)Q_{\perp}$    & 0.1190(4)&24.9(3)  &4.67& 0.1351(5)&24.6(1) &3.32&0.1448(6)&24.3(2) &2.35&0.1475(7)&24.2(3)&2.00 \\
			\hline	
			\multirow{3}{*}{\begin{turn}{90} \scriptsize{$R=9a$} \end{turn}}
			&$QQ$ & 0.0856(4)& 22.3(4) & 1.38 & 0.1025(5)& 22.0(4)  & 1.43 & 0.1140(6)& 22.1(4)  & 1.30& 0.1199(7)& 22.3(4)    & 1.17     \\
			&$(QQ)Q_{\parallel} $    & 0.1209(5)&24.8(4) &0.33 &0.1368(6)&25.1(4)&0.49& 0.1474(8)&25.4(4)&0.48& 0.1524(8)&25.6(4)&0.96\\ 
			&$(QQ)Q_{\perp}$    &0.1204(5)&26.0(4)&4.11& 0.1366(7)&25.7(4)&2.92&0.1473(8)&25.7(4)&1.79&0.1523(9)&25.7(4)&1.21\\
			\hline	        
			\multirow{3}{*}{\begin{turn}{90} \scriptsize{$R=10a$} \end{turn}}
			&$QQ$       & 0.0840(5) & 23.6(5)    & 0.91 & 0.0992(6) & 23.5(5) &0.98& 0.1094(8)& 24.0(5)  & 0.91  & 0.1151(9)& 24.5(6)   &0.83        \\
			&$(QQ)Q_{\parallel}$      &0.1210(6)&25.6(5)&0.17& 0.1361(8)&26.0(4)&0.20& 0.1464(9)&26.5(4)&0.20& 0.152(1)&27.0(6)&0.46 \\ 
			&$(QQ)Q_{\perp}$  &0.1205(7)&27.1(4)&3.27&0.1359(8)&27.0(4)&2.52&0.146(1)&27.1(4)&1.5&0.152(1)&27.3(5)&0.80\\
			\hline
			\multirow{3}{*}{\begin{turn}{90} \scriptsize{$R=11a$} \end{turn}}
			&$QQ$ & 0.0821(7) & 24.7(7)   & 0.57 & 0.0953(8) & 25.0(7)  & 0.6& 0.1038(9) &  25.8(7)   & 0.55 & 0.1084(1)  &  26.8(8)    & 0.50    \\
			&$(QQ)Q_{\parallel}$&0.1205(8)&26.7(6)&0.09  & 0.134(1)&27.0(6)&0.09& 0.143(1)&27.6(6)  &0.14   & 0.149(1)&28.4(6)&0.22 \\
			&$(QQ)Q_{\perp}$   & 0.1199(8)&28.2(6)&2.21 & 0.134(1)&28.1(6)&1.95&0.143(1)& 28.5(6)   &1.27  &0.149(1) &28.9(6)&0.61 \\
			\hline
			\multirow{3}{*}{\begin{turn}{90} \scriptsize{$R=12a$} \end{turn}}
			&$QQ$              & 0.0799(8) &  25.4(9) & 0.34 &  0.0910(9)  & 26.1(9) & 0.34    & 0.097(1)   & 27.5(1.0)    & 0.29  & 0.101(1) & 29.1(1.1)  & 0.24    \\
			&$(QQ)Q_{\parallel}$ &0.119(1)&28.2(8)&0.08& 0.131(1)&28.2(8)&0.06 & 0.138(1)&28.7(8)  &0.07   & 0.143(2)&29.7(1.0)&0.12  \\ 
			&$(QQ)Q_{\perp}$     &0.119(1)&29.3(8)&1.21&0.131(1) &29.2(8)&1.28 &0.138(1) &29.7(8)  &0.97   &0.143(2) &30.5(8)&0.49  \\  
			\hline
			\hline			
		\end{tabular}
	\caption{
		The MS width $W_{\parallel}^2,W_{\perp}^2$ and amplitude of the in-plane and perpendicular action-density of $(QQ)Q$ measured at the corresponding planes using fits to a double-Gaussian ansatz Eq.\eqref{eq:conGE} at temperature $T/T_{c}=0.9$.
	}
	\label{tab:AAAw2a2T09}	
\end{table*}

\begin{table*}[!hpt]
		\begin{tabular}{cc|ccc|ccc|ccc|ccc}
			\hline
			\hline
			\multicolumn{2}{c|}{plane}   
			&\multicolumn{3}{c|}{$x=1$}   
			&\multicolumn{3}{c|}{$x=2$}
			&\multicolumn{3}{c|}{$x=3$} 
			&\multicolumn{3}{c}{$x=4$}\\
			\multicolumn{2}{c|}{$n=R/a$}   
			&\multicolumn{1}{c}{$A$}  &\multicolumn{1}{c}{$w^{2}a^{-2}$}&\multicolumn{1}{c|}{$\chi^2_{\rm{dof}}$}
			&\multicolumn{1}{c}{$A$}  &\multicolumn{1}{c}{$w^{2}a^{-2}$}&\multicolumn{1}{c|}{$\chi^2_{\rm{dof}}$}
			&\multicolumn{1}{c}{$A$}  &\multicolumn{1}{c}{$w^{2}a^{-2}$}&\multicolumn{1}{c|}{$\chi^2_{\rm{dof}}$}
			&\multicolumn{1}{c}{$A$}  &\multicolumn{1}{c}{$w^{2}a^{-2}$}&\multicolumn{1}{c}{$\chi^2_{\rm{dof}}$}\\ 
			\hline
			\multirow{2}{*}{\begin{turn}{90} \tiny{$4a$} \end{turn}}
			&$QQ$&0.075(1)&13.0(1)&0.84&0.0812(2)&12.9(1)&0.87&0.0752(1)&13.0(1)&0.84\\
			&$(QQ)Q$&0.0904(2)&13.4(1)&1.73&0.0936(1)&13.3(1)&1.60&0.0903(5)&12.9(2)&1.91\\
			\hline
			\multirow{2}{*}{\begin{turn}{90} \tiny{$5a$} \end{turn}}
			&$QQ$&0.093(2)&13.5(1)&0.45&0.1066(2)&13.4(1)&0.51&0.1066(2)&13.4(1)    &0.51&0.0935(2)&13.5(1)&0.5\\
			&$(QQ)Q$&0.1081(2)&13.8(1)&1.82&0.1183(2)&13.7(1)&1.43&0.1155(3)&13.7(2)&1.22&0.1081(2)&13.8(1)&1.82\\
			\hline
			\multirow{2}{*}{\begin{turn}{90} \tiny{$6a$} \end{turn}}
			&$QQ$       &0.106(3)&14.2(1)&0.15&0.1262(3)&14.0(1)   &0.20&0.1262(3)&14.0(1)    &0.20&0.1262(3)&14.0(1)&0.2\\
			&$(QQ)Q$&0.1206(3)&14.3(2)&2.04&0.1373(3)&14.3(2)&1.51&0.1413(4)&14.2(2)&1.06&0.1322(3)&14.1(2)&0.87\\
			\hline
			\multirow{2}{*}{\begin{turn}{90} \tiny{$7a$} \end{turn}}
			&$QQ$ &0.115(4)&15.0(2)&0.04&0.1397(5)&14.8(2)&0.05&0.1397(5)&14.7(2)&0.05&0.1530(5)&14.7(2)       &0.07\\
			&$(QQ)Q$&0.1283(4)&14.9(2)&2.14&0.1501(5)&14.9(2)&1.68&0.1604(5)&14.8(2)&1.09&0.1604(6)&14.7(2)&1.09\\
			\hline
			\multirow{2}{*}{\begin{turn}{90} \tiny{$8a$} \end{turn}}	&$QQ$&0.116(5)&16.7(1)      &0.17&0.1478(7)&15.7(3)&0.01&0.1478(7)&15.6(3)&0.01&0.1721(9)&15.6(3)&0.02\\
			&$(QQ)Q$&0.1323(6)&15.5(3)&2.01&0.1573(7)&15.5(4)&1.80&0.1724(8)&15.5(3)&1.22&0.1764(8)&15.3(3)&0.75\\
			\hline	
			\multirow{2}{*}{\begin{turn}{90} \tiny{$9a$} \end{turn}}
			&$QQ$ &0.122(9)&17.2(4)&0.01&0.1478(7)&16.9(4)   &0.01&0.151(1) & 16.7(4)&0.01&0.183(2)&16.7(4)&0.01\\
			&$(QQ)Q $&0.1338(9)&16.1(5)&1.81&0.160(1)&16.1(5)&1.81&0.179(1)&16.2(5)&	1.36&0.187(1)&16.0(5)&0.88\\
			\hline	        
			\multirow{2}{*}{\begin{turn}{90} \tiny{$10a$} \end{turn}}
			&$QQ$&0.12(1)&18.9(6)&0.06&0.152(2) &18.5(6)&0.02&0.152(2) &18.2(6)&0.02&0.189(2)&18.1(7)&0.01\\
			&$(QQ)Q$&0.134(1)&16.5(8)&1.68&0.160(1)&16.8(7)&	1.69&0.181(2)&16.9(8)&	1.36&0.189(2)&16.1(5)&1.15\\
			\hline
			\multirow{2}{*}{\begin{turn}{90} \tiny{$11a$} \end{turn}}
			&$QQ$&0.12(2)&20.5(9)&0.20&0.151(3)&20.4(8)&0.07&0.151(3)&20.2(9)&0.07&0.189(4)&20.0(9)&0.01\\
			&$(QQ)Q$&0.133(2)&17(1)&1.84&0.160(2)&18(1.0)&1.57&0.181(2)&18(1.1)&	1.19&0.193(3)&18(1.2)&0.94\\
			\hline
			\multirow{2}{*}{\begin{turn}{90} \tiny{$12a$} \end{turn}}
			&$QQ$&0.11(3)&21.7(1.6)&0.38&0.147(4)&22.1(1.3)&0.21&0.147(4)&22.4(1.3)  &0.21&0.186(5)&22.4(1.3)&0.01\\
			&$(QQ)Q$&0.133(2)&19(1.6)&2.70&0.159(3)&19(1.5)&1.75&0.180(3)&20(1.6)&	1.01&0.192(4)&19.5(1.8)&0.77\\
			\hline
			\hline			
		\end{tabular}
	\caption{The symmetrized MS width $W^2=(W^2_{\parallel}+W^2_{\perp})/2$ and amplitude of the $(QQ)Q$ action-density measured at the corresponding planes using fits to a double-Gaussian ansatz Eq.\eqref{eq:conGE} at temperature $T/T_{c}=0.8$.}
	\label{tab2:AAAw2a2T08}
\end{table*}


\end{document}